\documentclass[epsf,amssymb,aps,prd,nofootinbib,floatfix,superscriptaddress,secnumarabic,preprintnumbers,longbibliography]{revtex4-2}

\usepackage{graphicx}
\usepackage{slashed}
\usepackage{amsmath}
\usepackage{bm}
\usepackage{epsf}
\usepackage{epsfig}
\usepackage{epstopdf}
\usepackage{hyperref}
\usepackage[dvipsnames]{xcolor}
\usepackage{tikz}
\usepackage{float}
\usepackage{mathtools}
\usetikzlibrary{positioning}
\DeclareFontFamily{U}{BOONDOX-calo}{\skewchar\font=45 }
\DeclareFontShape{U}{BOONDOX-calo}{m}{n}{
  <-> s*[1.05] BOONDOX-r-calo}{}
\DeclareFontShape{U}{BOONDOX-calo}{b}{n}{
  <-> s*[1.05] BOONDOX-b-calo}{}
\DeclareMathAlphabet{\mathcalboondox}{U}{BOONDOX-calo}{m}{n}
\SetMathAlphabet{\mathcalboondox}{bold}{U}{BOONDOX-calo}{b}{n}
\DeclareMathAlphabet{\mathbcalboondox}{U}{BOONDOX-calo}{b}{n}
\usepackage[normalem]{ulem}
\usepackage{subcaption}
\numberwithin{equation}{subsection}

\newcommand{\be}{\begin{equation}}
\newcommand{\ee}{\end{equation}}
\newcommand{\bea}{\begin{eqnarray}}
\newcommand{\eea}{\end{eqnarray}}

\newcommand{\MSbar}{{\overline{\rm MS}}}

\newcommand{\bmb}[1]{{\color{Blue} \bm{#1}}}
\newcommand{\bmp}[1]{{\color{Plum} \bm{#1}}}
\newcommand{\bmr}[1]{{\color{Red} \bm{#1}}}

\usepackage{stackengine,xcolor}

\begin{document}
\vspace*{1cm}

\title{Renormalization of asymmetric staple-shaped Wilson-line operators in lattice and continuum perturbation theory}

\author{G.~Spanoudes}
\email[]{gspano01@ucy.ac.cy}
\affiliation{Department of Physics, University of Cyprus,  \\
  P.O. Box 20537, 1678 Nicosia, Cyprus\bigskip}

\author{M.~Constantinou}
\email{marthac@temple.edu}
\affiliation{Department of Physics, Temple University, \\
  Philadelphia, Pennsylvania 19122 - 1801, USA\bigskip}

\author{H.~Panagopoulos\bigskip}
\email[]{haris@ucy.ac.cy}
\affiliation{Department of Physics, University of Cyprus,  \\
  P.O. Box 20537, 1678 Nicosia, Cyprus\bigskip}

\begin{abstract}
  In this work, we study the renormalization of nonlocal quark bilinear operators containing an asymmetric staple-shaped Wilson line at the one-loop level in both lattice and continuum perturbation theory. These operators enter the first-principle calculation of transverse momentum-dependent parton distribution functions (TMDPDFs) in lattice QCD using the formulation of Large Momentum Effective Theory. We provide appropriate RI$'$-type conditions that address the power and logarithmic divergences, as well as the mixing among staple operators of different Dirac structures, using a number of different possible projectors. A variant of RI$'$, including calculations of rectangular Wilson loops, which cancel the pinch-pole singularities of the staple operators at infinite length and reduce residual power divergences, is also employed. We calculate at one-loop order the conversion matrix, which relates the quasi-TMDPDFs in the RI$'$-type schemes to the reference scheme $\MSbar$ for arbitrary values of the renormalization momentum scale and of the dimensions of the staple. 
  
\end{abstract}

\maketitle

\section{Introduction}
\label{Introduction}
One of the directions of research in lattice QCD, which has shown rapid progress in the last decade, is the first-principle study of a family of nonperturbative distribution functions (DFs) that describe the internal structure of hadrons: parton distribution functions (PDFs), generalized parton distribution functions (GPDs) and transverse-momentum dependent parton distribution functions (TMDPDFs). All three types of DFs are crucial for a comprehensive understanding of the three-dimensional hadron picture. Calculating DFs from first principles has long been a challenge in Hadron Physics due to their nonperturbative and light-cone nature. The latter does not allow for a direct nonperturbative computation of DFs on a Euclidean lattice. In the last decade, a groundbreaking approach by X. Ji~\cite{Ji:2013dva} has overcome this issue. The approach connects Euclidean equal-time correlation functions (referred to as quasi-DFs), which are accessible by lattice simulations, to the physical light-cone DFs using the framework of Large Momentum Effective Theory (LaMET)~\cite{Ji:2014gla}. This breakthrough has paved the way for extracting, for the first time, DFs from lattice simulations.

Several groups have successfully employed this approach in the calculation of various quark and gluon DFs on the lattice: quark PDFs and GPDs in Refs.~\cite{Lin:2014zya,Alexandrou:2015rja,Chen:2016utp,Alexandrou:2016jqi,Zhang:2017bzy,Alexandrou:2017huk,Zhang:2017zfe,Alexandrou:2018pbm,Alexandrou:2018eet,LatticeParton:2018gjr,Zhang:2018nsy,Alexandrou:2019lfo,Izubuchi:2019lyk,Cichy:2019ebf,Chai:2020nxw,Zhang:2020gaj,Bhattacharya:2020xlt,Bhattacharya:2020jfj,Alexandrou:2020zbe,Alexandrou:2020uyt,Alexandrou:2020qtt,Lin:2020ssv,Fan:2020nzz,Gao:2020ito,Bringewatt:2020ixn,Hua:2020gnw,Alexandrou:2021oih,Alexandrou:2021bbo, Bhattacharya:2021moj,Gao:2021dbh,Hua:2022kcm,Gao:2022iex,Bhattacharya:2022aob,LatticeParton:2022xsd,Bhattacharya:2023nmv,Orginos:2017kos,Karpie:2018zaz,Karpie:2019eiq,Joo:2019jct,Joo:2019bzr,Joo:2020spy,Bhat:2020ktg,DelDebbio:2020rgv,Karpie:2021pap,Egerer:2021ymv,HadStruc:2021qdf,Bhat:2022zrw,HadStruc:2022nay}, quark TMDPDFs and soft-function in Refs.~\cite{Ji:2014hxa, Engelhardt:2015xja, Radyushkin:2016hsy, Radyushkin:2017ffo, Yoon:2017qzo, Broniowski:2017gfp, Ji:2018hvs,Shanahan:2019zcq,Ebert:2022fmh,LatticeParton:2020uhz,Shanahan:2020zxr,Li:2021wvl,Schlemmer:2021aij,Shanahan:2021tst,Ji:2021uvr,Zhang:2022xuw,LatticePartonLPC:2023pdv,Alexandrou:2023ucc}, gluon PDFs in Refs.~\cite{Fan:2018dxu,Zhang:2018diq,Fan:2020cpa,Fan:2021bcr,Salas-Chavira:2021wui,HadStruc:2021wmh,HadStruc:2022yaw,Khan:2022vot,Fan:2022kcb,Delmar:2023agv} and hadronic light-cone distribution amplitudes (DA) in Refs.~\cite{Jia:2015pxx, Radyushkin:2017gjd, Zhang:2017bzy, Broniowski:2017wbr, Chen:2017gck}. An overview of recent progress in the research of DFs on the lattice can be found in Refs.~\cite{Cichy:2018mum,Ji:2020ect,Constantinou:2020pek,Cichy:2021lih,Cichy:2021ewm}. The goal of these studies is to complement the planned experimental programs for investigating the 3D tomography of the nucleon in major experimental facilities, such as the electron ion colliders of USA~\cite{Accardi:2012qut,AbdulKhalek:2021gbh} and China~\cite{Anderle:2021wcy}. Theoretical studies can significantly complement the experimental investigations, especially when there are limitations either in the experimental programs or in the phenomenological models used for the analysis of the experimental data.

The computation of DFs using Ji's approach (also called the ``quasi-PDFs'' approach) involves a three-step process. Firstly, nonperturbative calculations are performed to determine hadron matrix elements of gauge-invariant quark or gluon nonlocal operators. These operators contain path-ordered Wilson lines with specific shapes, such as straight lines for PDFs and staple-shaped lines for TMDPDFs. Secondly, the nonlocal operators are renormalized to establish a connection with physically measurable quantities. This task is challenging compared to the case of local operators, as explained below. Lastly, the renormalized lattice DFs are perturbatively matched to the corresponding physical light-cone DFs using LaMET. One-loop matching formulae have been extracted in the literature for several quasi DFs~\cite{Wang:2019tgg,Wang:2017qyg}; in some cases a two-loop formula is also available~\cite{Chen:2020arf,Li:2020xml}. 

In this study, we focus on the implementation of the second step in the quasi-PDFs approach regarding the renormalization of nonlocal operators using one-loop perturbation theory in both continuum and lattice regularizations. While a nonperturbative calculation of the renormalization functions in numerical simulations of lattice QCD is desirable, perturbation theory can give important feedback for the development of an appropriate nonperturbative renormalization prescription, which addresses all kinds of divergences, as well as possible mixing with operators of equal or lower dimension allowed by global symmetries. Most importantly, perturbation theory gives us the matching functions (at a given order) between nonperturbative renormalization schemes and continuum perturbative schemes -- primarily $\MSbar$ -- employed in phenomenology. 

Studies of nonlocal operators with Wilson lines in continuum perturbation theory go back decades, including seminal work~\cite{Mandelstam:1968hz,Polyakov:1979gp,Makeenko:1979pb,Dotsenko:1979wb,Craigie:1980qs,Brandt:1981kf,Stefanis:1983ke,Knauss:1984rx,Dorn:1986dt,Korchemsky:1987wg} for the renormalization of open and closed (loops) Wilson lines, with and without singular points (cusps, self-intersections), and having quark or gluon fields at the endpoints. 
Lattice studies of nonlocal operators have emerged only in the last decade after the development of the quasi-PDFs approach. The first perturbative lattice calculation of Wilson-line operators was made by our group in Ref.~\cite{Constantinou:2017sej}, to one loop for massless
quarks, using the Wilson/clover fermion action and a variety of Symanzik-improved gluon actions. A straight Wilson line with quark fields at the endpoints was employed in order to investigate the renormalization of quark quasi-PDFs. This study showed that the lattice formulation introduces several new complications, such as mixing among operators of equal dimension and different twists, and power-law divergences (even in the absence of mixing with lower-dimensional operators, in contrast to the case of local operators). A number of extensions of this study have been followed by our group regarding the presence of finite quark mass~\cite{Spanoudes:2018zya} and the calculation of one-loop artifacts to all orders in the lattice spacing $a$~\cite{Constantinou:2022aij}. In a different extension of our study, we have considered nonlocal operators with a symmetric staple-shaped Wilson line~\cite{Constantinou:2019vyb} in order to investigate the renormalization of quark quasi-TMDPDFs. In our current work, we extend further the latter calculation by employing the more general case of an asymmetric staple-shaped Wilson line, where all three segments of the staple can have different lengths. Studies by other groups along these lines have appeared in Refs.~\cite{Chen:2016fxx,Ishikawa:2016znu,Carlson:2017gpk,Xiong:2017jtn}.

Staple-shaped nonlocal operators have a wide range of applicability. They enter the analysis of semi-inclusive deep inelastic scattering (SIDIS) processes, as well as the Drell-Yan (DY) processes,
in a kinematic region where the photon virtuality is large and the measured transverse momentum of the produced hadron is of the order of $\Lambda_{\rm QCD}$~\cite{Bomhof:2006dp}. In these analyses, the segments of the staple that are parallel to the direction in which the hadron is boosted have an infinite length. Thus, while our study focuses on finite lengths of the staple segments, an extrapolation to infinite limit must be taken in the renormalized matrix elements of these operators. The presence or not of an asymmetry in the shape of the staple operators affects their renormalization. Besides the presence of an additional scale in the renormalization conditions, the mixing pattern is different between symmetric and asymmetric staple operators (see Sec. \ref{Sym}). However, this difference is not visible in one-loop lattice perturbation theory, as concluded by the present calculation.

As stated before, there are a number of challenges to address in order to renormalize the nonlocal Wilson-line operators of arbitrary shape: 
\begin{enumerate}
    \item Power divergences arise for cutoff regularized theories, such as lattice QCD~\cite{Dotsenko:1979wb}. The divergences depend on the total length ($L$) of the Wilson line and can be absorbed in an exponential factor of the form $e^{- \delta m L}$, where $\delta m$ is a dimensionful regularization-dependent quantity whose magnitude diverges linearly with the regulator.
    \item Logarithmic divergences arise not only from contact terms but also from the singular points of the Wilson line. In the case of a straight line, singular points are just the endpoints. For operators of different shapes, there can also be cusp divergences~\cite{Brandt:1981kf}, which depend on the angle and number of cusps present. In the case of the staple line, there are two cusps of angle $\pi / 2$. These divergences can be addressed by using typical renormalization schemes, such as RI$'$ or ratio schemes. 
    \item Finite operator mixing arises between Wilson-line operators $\mathcal{O}_\Gamma$ with different (products of) Dirac matrices $\Gamma$ depending on the regularization. In the case of straight-line operators, the one-loop computation~\cite{Constantinou:2017sej} shows mixing in pairs between nonlocal fermion bilinears of the form ($\mathcal{O}_\Gamma$, $\mathcal{O}_{\{\Gamma, \gamma_{\nu_1}\}/2}$), where $\hat{\nu}_1$ is the direction of the straight line, for chirality-breaking actions. Symmetry arguments (reflections, charge conjugation)~\cite{Chen:2017mie} confirm that the one-loop mixing pattern is also valid nonperturbatively. In the case of (symmetric and asymmetric) staple operators, the one-loop computation~\cite{Constantinou:2019vyb} shows mixing in pairs with a different pattern: ($\mathcal{O}_\Gamma$, $\mathcal{O}_{[\Gamma, \gamma_{\nu_2}]/2}$), where $\hat{\nu}_2$ is the direction of the sides of the staple line. However, symmetry arguments~\cite{Alexandrou:2023ucc} show a wider mixing pattern: quadruplets of operators emerge in the asymmetric case and triplets in the symmetric case when nonchiral fermions are employed.
    \item In the case of (symmetric and asymmetric) staple operators, a pinch-pole singularity~\cite{Ji:2018hvs} arises when one takes the infinite limit of lateral sizes of the staple (this limit is included in the definition of TMDPDFs). This singularity -- a linear divergence -- comes from the gluon exchange between the two parallel (``longitudinal'') segments of the staple. 
    \item The renormalization functions are (in general) complex in nonminimal subtraction schemes\footnote{The presence of an imaginary part in the renormalization functions depends on the exact definition of the renormalization scheme. For example, one can define purely real renormalization conditions by using only the real part of ``projected'' Green's functions; in this way, both the real and the imaginary part of the renormalized Green's functions will be rendered finite.}.
\end{enumerate}

Nonperturbative renormalization prescriptions~\cite{GHP,Alexandrou:2017huk,Chen:2017mzz,Shanahan:2019zcq,Ji:2021uvr,Alexandrou:2023ucc} have already been employed in lattice simulations for both straight-line and staple operators using RI$'$-type schemes. There are also studies of alternative prescriptions, such as the ratio (or short-distance ratio) scheme~\cite{Zhang:2022xuw}, Wilson-line-mass-subtraction scheme~\cite{Musch:2010ka,Chen:2016fxx}, RI-xMOM scheme (using the auxiliary-field approach~\cite{Dorn:1986dt,Chen:2016fxx, Ji:2017oey, Green:2017xeu, Wang:2017eel}) and a hybrid renormalization scheme~\cite{Ji:2020brr}. 

While a standard RI$'$-type scheme (extended in the presence of mixing) can treat the various types of divergences and mixing of the nonlocal operators at one loop, recent nonperturbative examinations~\cite{Zhang:2020rsx,Zhang:2022xuw} at different lattice spacings have provided compelling evidence indicating the existence of residual linear divergences in both straight-line and staple operators. In this regard, an alternative version of RI$'$, as suggested in Refs.~\cite{Ebert:2019tvc,Ji:2021uvr}, which includes the computation of vacuum expectation value for rectangular Wilson loops, is also employed in our current work. This prescription (described in Sec. \ref{Renormalization conditions}) is expected to suppress the residual power divergences from the Green's functions of the staple operators (see Sec. \ref{ResultsLR}).

The paper is organized as follows: In Sec.~\ref{Calculation setup}, we provide the setup of our calculation, including the definition of the operators under study and the renormalization conditions that we employ throughout, corresponding to four different variants of the RI$'$ scheme. Sections~\ref{ResultsDR}  and~\ref{ResultsLR} present our main results in dimensional and lattice regularization, respectively. This includes both the renormalization functions and conversion matrices between the RI$'$-type and $\MSbar$ schemes. In Sec.~\ref{Conclusions}, we summarize and give some future plans. We have also included three appendices. In Appendix~\ref{IntegrationMethod}, we provide a list of formulae for the calculation, in dimensional regularization, of one-loop Feynman integrals appropriate to nonlocal operators. Appendix~\ref{IntegralList} contains the definitions of Feynman-parameter integrals appearing in our results. In Appendix~\ref{ap.C}, we collect one-loop results for the renormalization of the quark field in both dimensional and lattice regularizations. 

\section{Calculation setup}
\label{Calculation setup}

\subsection{Definition of asymmetric staple-shaped Wilson-line operators}

\begin{figure}[H]
    \centering
    \includegraphics[width=0.28\textwidth]{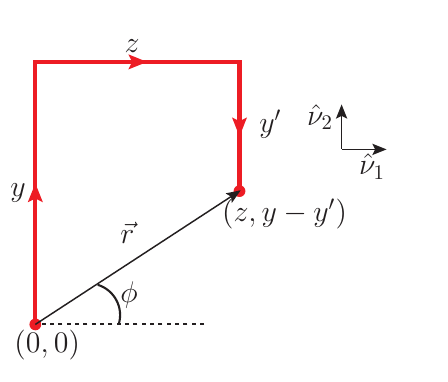}
    \caption{The shape of the asymmetric staple line, as defined in the operator $\mathcal{O}_\Gamma (x,z,y,y')$. Here, the staple is placed at $x=0$.}
    \label{fig:staple_line}
  \end{figure}

First, we define the operators under study along with our conventions\footnote{A number of different conventions can be found in the literature. For example, one can match our convention to Ref.~\cite{Alexandrou:2023ucc} through $y \rightarrow l$, $z \rightarrow b$, and $(y - y') \rightarrow -z$.}. The asymmetric staple-shaped Wilson-line fermion bilinear operators are defined in Euclidean space as follows:
\begin{equation}
    \mathcal{O}_\Gamma (x,z,y,y') \equiv \Bar{\psi}(x) \ \Gamma \ \mathcal{W} (x, z, y, y') \ \psi (x + z \hat{\nu}_1 + (y-y') \hat{\nu}_2),
    \label{O_Gamma}
\end{equation}
where $\mathcal{W} (x,z,y,y')$ denotes the staple-shaped Wilson line as given schematically in Fig.~\ref{fig:staple_line} and defined as:
\begin{eqnarray}
  \mathcal{W} (x,z,y,y') &\equiv& \mathcal{U} (x,y \, \hat{\nu}_2) \ \mathcal{U} (x+y \, \hat{\nu}_2,z \, \hat{\nu}_1) \ \mathcal{U}^\dagger (x + z \, \hat{\nu}_1 + (y-y') \, \hat{\nu}_2,y' \, \hat{\nu}_2), \\
  \mathcal{U} (r,\ell \, \hat{\mu}) &\equiv& \mathcal{P} \exp\bigg[ig\int_0^\ell d\bar{\ell} \ \mathcal{A}_{\mu}(r+\bar{\ell} \, \hat{\mu})\bigg].
\end{eqnarray}
$\mathcal{U} (r,\ell \, \hat{\mu})$ denotes the straight-line path-ordered ($\mathcal{P}$) exponential (Wilson line), expressed in terms of the gluon field $\mathcal{A}_\mu$, which connects the points $r$ and $r + \ell \hat{\mu}$. A lattice discretization of $\mathcal{U} (r,\ell \ \hat{\mu})$ in terms of gluon links $U_\mu (x)$ which connect points $x$ and $x+a \hat{\mu}$ is given below:
\begin{equation}
    \mathcal{U} (r,\ell \ \hat{\mu}) = \prod_{\bar{\ell}=0}^{\ell \mp 1} U_{\pm \mu} (r + \bar{\ell} a \hat{\mu}),
\end{equation}
where $U_{-\mu}(x) \equiv U_\mu^\dagger(x -a \hat\mu)$ and upper (lower) signs correspond to $\ell>0$ ($\ell<0$). Other discretizations involve smeared gluon links, e.g., stout, HYP, Wilson flow. We plan to investigate the impact of stout smearing at the one-loop level in future work. There is a total of 16 different operators which can be extracted from Eq.~\eqref{O_Gamma} depending on the choice of the Dirac matrix $\Gamma$ inserted between the fermion and antifermion fields: $\Gamma = \openone$ (scalar $S$), $\gamma_5 \equiv \gamma_1 \gamma_2 \gamma_3 \gamma_4$ (pseudoscalar $P$), $\gamma_\mu$ (vector $V_\mu$), $\gamma_5 \gamma_\mu$ (axial vector $A_\mu$), $\sigma_{\mu \nu} \equiv [\gamma_\mu, \gamma_\nu]/2$ (tensor $T_{\mu\nu}$). Of particular interest is the study of vector, axial-vector, and tensor operators, which correspond to the unpolarized, helicity, and transversity types of TMDPDFs, respectively. 

Since we are working in one-loop perturbation theory with massless fermions, the flavor content of the fermion fields $\psi$ and $\Bar{\psi}$ is irrelevant for our calculations up to this order. Thus, our one-loop results are valid for both flavor singlet and nonsinglet operators, given that there is no additional mixing with gluon nonlocal operators~\cite{Green:2017xeu,Zhang:2018diq}.  

\subsection{Symmetry properties of staple-shaped operators}
\label{Sym}

We examine below the properties of staple operators under symmetry transformations. Similar investigations can be found in~\cite{Ji:2021uvr,Alexandrou:2023ucc}. Operators with the same behavior under all symmetries can mix among themselves. In this way, one can identify the mixing pattern of the staple operators based on nonperturbative arguments. 
Since the staple operators depend on two special directions ($\hat{\nu}_1$ and $\hat{\nu}_2$) in which the staple is defined, we consider appropriate versions of the symmetry transformations of the QCD action (in Euclidean space) with respect to the special directions. 
\begin{enumerate}
    \item {\bf Translational symmetry:} The operators are covariant under translations in Euclidean space. Similarly to the case of local operators ${\cal O}(x)$ which cannot mix with translated versions of themselves (${\cal O}(y)$), mixing among nonlocal operators involving different paths and shapes for the Wilson line joining the fermion-antifermion pair cannot occur~\cite{Dotsenko:1979wb,Brandt:1981kf,Dorn:1986dt}.
    \item {\bf Two-dimensional (2D) rotational (or square, on the lattice) symmetry:} The operators are covariant under rotations over the two-dimensional (2D) plane transverse to the plane in which the staple is defined. The 16 staple operators are classified into two representations of the 2D rotational group: (1) scalar: $\{ S, P, V_{\nu_1}, V_{\nu_2}, A_{\nu_1}, A_{\nu_2}, T_{\nu_1 \nu_2}, T_{\nu_3 \nu_4} \}$, (2) vector: $\{(V_{\nu_3}, V_{\nu_4}), (A_{\nu_3}, A_{\nu_4}), (T_{\nu_3 \nu_1}, T_{\nu_4 \nu_1}), (T_{\nu_3 \nu_2},  T_{\nu_4 \nu_2}) \}$, where $(\nu_1, \nu_2, \nu_3, \nu_4)$ correspond to different orthogonal directions in the 4D Euclidean space. The first representation is one-dimensional while the second is two-dimensional reducible (e.g., $(V_{\nu_3}, V_{\nu_4})$ splits into two {\it different} one-dimensional representations: $V_{\nu_3}+ i V_{\nu_4}$ and $V_{\nu_3}- i V_{\nu_4}$).  Operator mixing can only occur among operators that support the same irreducible representation. 
    The fact that an operator such as 
    $V_{\nu_3}+ i V_{\nu_4}$ can mix with $A_{\nu_3}+ i A_{\nu_4}$ but not with $A_{\nu_3}- i A_{\nu_4}$ implies certain relations among the corresponding renormalization and mixing coefficients.
    Thus, in contrast to the local operators, which are covariant under 4D rotations, the residual rotational symmetry cannot completely prevent mixing between scalar, pseudoscalar, vector, axial-vector, and tensor operators. 
    Also, operators which support the same representation of the 4D rotational group but different representations of the 2D rotational group, e.g. $V_{\nu_1}$ (or $V_{\nu_2}$) and $V_{\rho \neq (\nu_1, \nu_2)}$, will not share the same renormalization factor (in contrast to the case of local operators); at least, the 2D rotational symmetry prevents the mixing among these operators.
    \item {\bf Parity ($\mathcal{P}$):} In Euclidean space, temporal and spatial directions are not distinguished. Thus, parity can be generalized in any direction. The generalized parity transformations $\mathcal{P}_{\mu}$ for the fermion and gluon fields with respect to the direction $\mu$ are defined below~\cite{Gattringer:2010zz}. Here, $\vec{x}$ is the 3-vector, which is perpendicular to the $\mu$ direction.
\begin{eqnarray}
 \psi(\vec{x}, x_{\mu}) &\xrightarrow[]{\mathcal{P}_{\mu}}& \gamma_{\mu} \psi (-\vec{x},x_{\mu}), \\
      \Bar{\psi}(\vec{x}, x_{\mu}) &\xrightarrow[]{\mathcal{P}_{\mu}}& \Bar{\psi} (-\vec{x},x_{\mu}) \gamma_{\mu}, \\
      \mathcal{A}_{\mu} (\vec{x},x_{\mu}) &\xrightarrow[]{\mathcal{P}_{\mu}}& \mathcal{A}_{\mu} (-\vec{x},x_{\mu}), \
      U_{\mu}(\vec{x},x_{\mu}) \xrightarrow[]{\mathcal{P}_{\mu}} U_{\mu} (-\vec{x},x_{\mu}), \\
      \mathcal{A}_{\nu \neq \mu}(\vec{x},x_{\mu}) &\xrightarrow[]{\mathcal{P}_{\mu}}& -\mathcal{A}_{\nu} (-\vec{x},x_{\mu}), \       U_{\nu \neq \mu}(\vec{x},x_{\mu}) \xrightarrow[]{\mathcal{P}_{\mu}} U^{\dagger}_{\nu} (-\vec{x}-\hat{\nu},x_{\mu}).
      \end{eqnarray}    

The transformation of the staple-shaped operators under generalized parity is:
\begin{equation}
\mathcal{O}_{\Gamma} (x,z,y,y') \xrightarrow[]{\mathcal{P}_{\mu}} \mathcal{O}_{\gamma_{\mu} \Gamma \gamma_{\mu}} (x, (-1)^{\delta_{\mu \nu_1} + 1} \ z, (-1)^{\delta_{\mu \nu_2} + 1} \ y, (-1)^{\delta_{\mu \nu_2} + 1} \ y').
\end{equation}
In the above relation it is understood that, after the parity transformation ${\cal{P}}_\mu$, there follows a translation $T_{2\,\vec x}$ by an amount $2\,\vec x$\,; such a translation is clearly allowed by translational invariance, and it also does not affect the ``relative coordinates" $y, y', z$. Thus:
\begin{equation}\vec x \xrightarrow[]{{\cal{P}}_\mu} -\vec x \xrightarrow[]{T_{2\,\vec x}} + \vec x\end{equation}
and $\vec x$ remains unchanged. Due to parity, mixing between operator $\mathcal{O}_\Gamma$ with operators $\mathcal{O}_{\Gamma \gamma_5}$, $\mathcal{O}_{\Gamma \gamma_5 \gamma_{\nu_1}}$, $\mathcal{O}_{\Gamma \gamma_5 \gamma_{\nu_2}}$, $\mathcal{O}_{\Gamma \gamma_5 \gamma_{\nu_1} \gamma_{\nu_2}}$ is prevented, as shown below (see table \ref{tab:CPT}). 
\item {\bf Time reversal ($\mathcal{T}$):} As in the case of parity, time reversal in Euclidean space is generalized in any direction. The generalized time-reversal transformations $\mathcal{T}_{\mu}$ for the fermion and gluon fields with respect to the direction $\mu$ are defined below. We use the same notation as in parity.
\begin{eqnarray}
    \psi(\vec{x}, x_{\mu}) &\xrightarrow[]{\mathcal{T}_{\mu}}& \gamma_{\mu} \gamma_5 \psi (\vec{x},-x_{\mu}), \\
      \Bar{\psi}(\vec{x}, x_{\mu}) &\xrightarrow[]{\mathcal{T}_{\mu}}& \Bar{\psi} (\vec{x},-x_{\mu}) \gamma_5 \gamma_{\mu}, \\
      \mathcal{A}_{\mu}(\vec{x},x_{\mu}) &\xrightarrow[]{\mathcal{T}_{\mu}}& -\ \mathcal{A}_{\mu}(\vec{x},-x_{\mu}), \ 
      U_{\mu}(\vec{x},x_{\mu}) \xrightarrow[]{\mathcal{T}_{\mu}} U^{\dagger}_{\mu} (\vec{x},-x_{\mu} - \hat{\mu}), \\
      \mathcal{A}_{\nu \neq \mu}(\vec{x},x_{\mu}) &\xrightarrow[]{\mathcal{T}_{\mu}}& \mathcal{A}_{\nu} (\vec{x},-x_{\mu}), \
      U_{\nu \neq \mu}(\vec{x},x_{\mu}) \xrightarrow[]{\mathcal{T}_{\mu}} U_{\nu} (\vec{x},-x_{\mu}).
\end{eqnarray}
The transformation of the staple-shaped operators under generalized time reversal is (a translation $T_{2 \, x_\mu}$ has also been applied):
\begin{equation}
    \mathcal{O}_{\Gamma} (x,z,y,y') \xrightarrow[]{\mathcal{T}_{\mu}} \mathcal{O}_{\gamma_5 \gamma_{\mu} \Gamma \gamma_{\mu} \gamma_5} (x,(-1)^{\delta_{\mu \nu_1}} \ z, (-1)^{\delta_{\mu \nu_2}} \ y, (-1)^{\delta_{\mu \nu_2}} \ y').
\end{equation}
Time reversal does not provide any additional information to the mixing compared to the residual rotational symmetry and generalized parity.
\item {\bf Charge conjugation ($\mathcal{C}$):} The transformations of fermion and gluon fields under charge conjugation are given below:
\begin{eqnarray}
    \psi(x) &\xrightarrow[]{\mathcal{C}}& C^{-1} \Bar{\psi}^{T} (x), \\
      \Bar{\psi}(x) &\xrightarrow[]{\mathcal{C}}& -\psi^{T} (x) \ C, \\
      \mathcal{A}_{\mu}(x) &\xrightarrow[]{\mathcal{C}}& -\mathcal{A}_{\mu}^T (x), \ U_{\mu}(x) \xrightarrow[]{\mathcal{C}} \left(U^{\dagger}_{\mu} (x)\right)^{T},
\end{eqnarray}
where $C$ is the charge conjugation matrix satisfying $C \gamma_\mu C^{-1} = - \gamma^T_\mu$. The transformation of the staple-shaped operators under charge conjugation is:
 \begin{equation}
   \mathcal{O}_{\Gamma} (x,z,y,y') \xrightarrow[]{\mathcal{C}} \mathcal{O}^\dagger_{\gamma_4 (C \Gamma C^{-1})^{\ast} \gamma_4} (x,z,y,y'),
 \end{equation} 
where $\gamma_4$ is the Dirac matrix in the temporal direction. Charge conjugation does not provide any additional information on the mixing of asymmetric staple operators compared to the previously mentioned symmetries. However, in the case of symmetric staple operators ($y' = y$), charge conjugation forbids the mixing between operator $\mathcal{O}_\Gamma$ with operators $\mathcal{O}_{\Gamma \gamma_{\nu_1}}$ or $\mathcal{O}_{\Gamma \gamma_{\nu_2}}$ or $\mathcal{O}_{\Gamma \gamma_{\nu_1} \gamma_{\nu_2}}$, when $[\Gamma, \Gamma \gamma_{\nu_1}] = 0$ or $[\Gamma, \Gamma \gamma_{\nu_2}] = 0$ or $[\Gamma, \Gamma \gamma_{\nu_1} \gamma_{\nu_2}] = 0$, respectively, as shown below (see table \ref{tab:CPT}).

\item {\bf Chiral transformations:}
Under chiral transformations of fermion fields:
\begin{eqnarray}
    \psi(x) &\xrightarrow[]{\alpha}& e^{i \alpha \gamma_5} \psi (x), \\
      \Bar{\psi}(x) &\xrightarrow[]{\alpha}& \Bar{\psi} (x) \ e^{i \alpha \gamma_5},
\end{eqnarray}
the staple-shaped operators are invariant only for $\Gamma = \gamma_{\mu}, \gamma_5 \gamma_{\mu}$. Thus, mixing between operator $\mathcal{O}_\Gamma$ with operators $\mathcal{O}_{\Gamma \gamma_{\nu_1}}$, $\mathcal{O}_{\Gamma \gamma_{\nu_2}}$ is eliminated in chirality-preserving actions. 
\end{enumerate}

Since $\mathcal{C}$, $\mathcal{P}$, and $\mathcal{T}$ transformations can flip the direction of one or more staple segments or can give a hermitian conjugate operator, it is useful to consider the following basis of eight operators for each $\Gamma$:
\begin{eqnarray}
\mathcal{O}_\Gamma^{++\phantom{\leftrightarrow}} &\equiv& \mathcal{O}_{\Gamma} (x,\phantom{+}z,\phantom{+}y,\phantom{+}y'), \\
\mathcal{O}_\Gamma^{-+\phantom{\leftrightarrow}} &\equiv&
\mathcal{O}_{\Gamma} (x,-z,\phantom{+}y,\phantom{+}y'), \\
\mathcal{O}_\Gamma^{+-\phantom{\leftrightarrow}} &\equiv&\mathcal{O}_{\Gamma} (x,\phantom{+}z,-y,-y'), \\
\mathcal{O}_\Gamma^{--\phantom{\leftrightarrow}} &\equiv&\mathcal{O}_{\Gamma} (x,-z,-y,-y'), 
\end{eqnarray}
and their hermitian conjugates. The action of the three discrete symmetries is manifest by taking linear combinations of the eight operators~\cite{Alexandrou:2023ucc}, which are odd/even under $\mathcal{C}$, $\mathcal{P}$, $\mathcal{T}$:
\begin{eqnarray}
    \widetilde{\mathcal{O}}_1 (\Gamma) &\equiv& [ \mathcal{O}_\Gamma^{++} + \mathcal{O}_\Gamma^{-+} + \mathcal{O}_\Gamma^{+-} + \mathcal{O}_\Gamma^{--} + {(\mathcal{O}_\Gamma^{++})}^\dagger + {(\mathcal{O}_\Gamma^{-+})}^\dagger + {(\mathcal{O}_\Gamma^{+-})}^\dagger + {(\mathcal{O}_\Gamma^{--})}^\dagger]/8, \label{Otilde1} \\
    \widetilde{\mathcal{O}}_2 (\Gamma) &\equiv& [ \mathcal{O}_\Gamma^{++} - \mathcal{O}_\Gamma^{-+} - \mathcal{O}_\Gamma^{+-} + \mathcal{O}_\Gamma^{--} + {(\mathcal{O}_\Gamma^{++})}^\dagger - {(\mathcal{O}_\Gamma^{-+})}^\dagger - {(\mathcal{O}_\Gamma^{+-})}^\dagger + {(\mathcal{O}_\Gamma^{--})}^\dagger]/8, \\
    \widetilde{\mathcal{O}}_3 (\Gamma) &\equiv& [ \mathcal{O}_\Gamma^{++} - \mathcal{O}_\Gamma^{-+} + \mathcal{O}_\Gamma^{+-} - \mathcal{O}_\Gamma^{--} + {(\mathcal{O}_\Gamma^{++})}^\dagger - {(\mathcal{O}_\Gamma^{-+})}^\dagger + {(\mathcal{O}_\Gamma^{+-})}^\dagger - {(\mathcal{O}_\Gamma^{--})}^\dagger]/8, \\
    \widetilde{\mathcal{O}}_4 (\Gamma) &\equiv& [ \mathcal{O}_\Gamma^{++} + \mathcal{O}_\Gamma^{-+} - \mathcal{O}_\Gamma^{+-} - \mathcal{O}_\Gamma^{--} + {(\mathcal{O}_\Gamma^{++})}^\dagger + {(\mathcal{O}_\Gamma^{-+})}^\dagger - {(\mathcal{O}_\Gamma^{+-})}^\dagger - {(\mathcal{O}_\Gamma^{--})}^\dagger]/8, \\
    \widetilde{\mathcal{O}}_{5} (\Gamma) &\equiv& [ \mathcal{O}_\Gamma^{++} + \mathcal{O}_\Gamma^{-+} + \mathcal{O}_\Gamma^{+-} + \mathcal{O}_\Gamma^{--} - {(\mathcal{O}_\Gamma^{++})}^\dagger - {(\mathcal{O}_\Gamma^{-+})}^\dagger - {(\mathcal{O}_\Gamma^{+-})}^\dagger - {(\mathcal{O}_\Gamma^{--})}^\dagger]/8, \\
    \widetilde{\mathcal{O}}_{6} (\Gamma) &\equiv& [ \mathcal{O}_\Gamma^{++} - \mathcal{O}_\Gamma^{-+} - \mathcal{O}_\Gamma^{+-} + \mathcal{O}_\Gamma^{--} - {(\mathcal{O}_\Gamma^{++})}^\dagger + {(\mathcal{O}_\Gamma^{-+})}^\dagger + {(\mathcal{O}_\Gamma^{+-})}^\dagger - {(\mathcal{O}_\Gamma^{--})}^\dagger]/8, \\
    \widetilde{\mathcal{O}}_{7} (\Gamma) &\equiv& [ \mathcal{O}_\Gamma^{++} - \mathcal{O}_\Gamma^{-+} + \mathcal{O}_\Gamma^{+-} - \mathcal{O}_\Gamma^{--} - {(\mathcal{O}_\Gamma^{++})}^\dagger + {(\mathcal{O}_\Gamma^{-+})}^\dagger - {(\mathcal{O}_\Gamma^{+-})}^\dagger + {(\mathcal{O}_\Gamma^{--})}^\dagger]/8, \\
    \widetilde{\mathcal{O}}_{8} (\Gamma) &\equiv& [ \mathcal{O}_\Gamma^{++} + \mathcal{O}_\Gamma^{-+} - \mathcal{O}_\Gamma^{+-} - \mathcal{O}_\Gamma^{--} - {(\mathcal{O}_\Gamma^{++})}^\dagger - {(\mathcal{O}_\Gamma^{-+})}^\dagger + {(\mathcal{O}_\Gamma^{+-})}^\dagger + {(\mathcal{O}_\Gamma^{--})}^\dagger]/8. \label{Otilde8}
\end{eqnarray}
In table \ref{tab:CPT}, we provide the action of the symmetry transformations $\mathcal{C}$, $\mathcal{P}$, $\mathcal{T}$ on the operators $\widetilde{\mathcal{O}}_i (\Gamma)$. We find that quadruplets of the form: $\{ \widetilde{\mathcal{O}}_i (\Gamma), \widetilde{\mathcal{O}}_j (\Gamma \gamma_{\nu_1} \gamma_{\nu_2}), \widetilde{\mathcal{O}}_k (\Gamma \gamma_{\nu_1}), \widetilde{\mathcal{O}}_l (\Gamma \gamma_{\nu_2})\}$ have the same symmetry properties, where $i,j,k,l$ are all different and their values depend on which Dirac matrix $\Gamma$ is employed. Switching back to the original basis, we conclude that when the fermion action breaks chiral symmetry (e.g., Wilson/clover fermions, twisted-mass fermions), asymmetric staple operators will mix in groups of 4, as follows\footnote{In principle, the symmetry properties of $\widetilde{\mathcal{O}}_i (\Gamma)$ reduce the possible mixing among the $8 \times 16$ operators in the original basis, to multiplets of $8 \times 4$ operators with Gamma structures given by $(\mathcal{O}_{\Gamma}, \mathcal{O}_{ \Gamma \gamma_{\nu_1} \gamma_{\nu_2}}, \mathcal{O}_{\Gamma \gamma_{\nu_1}}, \mathcal{O}_{\Gamma \gamma_{\nu_2}})$. The mixing sets are further reduced to multiplets of 4 operators by excluding mixing among staple operators which involve different paths but the same shape of the Wilson line~\cite{Dotsenko:1979wb,Brandt:1981kf,Dorn:1986dt}.}:
\begin{equation}
  (\mathcal{O}_{\Gamma}, \mathcal{O}_{ \Gamma \gamma_{\nu_1} \gamma_{\nu_2}}, \mathcal{O}_{\Gamma \gamma_{\nu_1}}, \mathcal{O}_{\Gamma \gamma_{\nu_2}}).  
\end{equation}
When chiral fermions (e.g., massless overlap/domain-wall/continuum fermions) are employed, the mixing pattern is minimized to $(\mathcal{O}_{\Gamma}, \mathcal{O}_{\Gamma \gamma_{\nu_1} \gamma_{\nu_2}})$. 

In the specific case of symmetric staple operators ($y' = y$), the basis of independent operators is reduced: By employing appropriate translations, $(\mathcal{O}_\Gamma^{++})^\dagger = \mathcal{O}_\Gamma^{-+}$, $(\mathcal{O}_\Gamma^{-+})^\dagger = \mathcal{O}_\Gamma^{++}$, $(\mathcal{O}_\Gamma^{+-})^\dagger = \mathcal{O}_\Gamma^{--}$, $(\mathcal{O}_\Gamma^{--})^\dagger = \mathcal{O}_\Gamma^{+-}$, and thus, the independent operators are only 4 for each $\Gamma$. Therefore, the mixing pattern takes the form: $(\mathcal{O}_\Gamma, \mathcal{O}_{[\Gamma, \gamma_{\nu_1} \gamma_{\nu_2}]/2}, \mathcal{O}_{[\Gamma, \gamma_{\nu_1}]/2}, \mathcal{O}_{[\Gamma, \gamma_{\nu_2}]/2})$ for nonchiral fermions and $(\mathcal{O}_\Gamma, \mathcal{O}_{[\Gamma, \gamma_{\nu_1} \gamma_{\nu_2}]/2})$ for chiral fermions. Depending on the Dirac matrix $\Gamma$, one or three out of the three commutators $[\Gamma, \gamma_{\nu_1}]$, $[\Gamma, \gamma_{\nu_2}]$, $[\Gamma, \gamma_{\nu_1} \gamma_{\nu_2}]$ appearing in the mixing pattern of the symmmetric staple operators with nonchiral fermions will be zero, thus leading to a mixing triplet or singlet (multiplicatively renormalizable operator), respectively. In particular, one commutator vanishes when $\Gamma= \gamma_5$, $\gamma_{\mu}$ ($\mu$ can be any direction), $\gamma_5 \gamma_{\nu_1}$, $\gamma_5 \gamma_{\nu_2}$, $\sigma_{\nu_1 \nu_2}$, $\sigma_{\nu_1 \mu}$ ($\mu \neq \nu_1, \nu_2$), $\sigma_{\nu_2 \mu}$ ($\mu \neq \nu_1, \nu_2$), and three commutators vanish when $\Gamma= \openone$, $\gamma_5 \gamma_{\mu}$ (for $\mu \neq \nu_1, \nu_2$), $\sigma_{\mu \rho}$ (for $\mu, \rho \neq \nu_1, \nu_2$). Similarly, when $[\Gamma, \gamma_{\nu_1} \gamma_{\nu_2}] \neq 0$, the mixing pattern of the symmetric staple operators with chiral fermions gives a mixing pair; otherwise it leads to a multiplicatively renormalizable operator. The sets of mixing operators in the case of symmetric staple are given explicitly below: 
\begin{itemize}
    \item [(a)] For non-chiral fermions, the mixing sets are: triplets: $(\sigma_{\nu_1 \nu_2}, \gamma_{\nu_1}, \gamma_{\nu_2})$, $(\gamma_5, \gamma_5 \gamma_{\nu_1}, \gamma_5 \gamma_{\nu_2})$, $(\gamma_{\nu_3}, \sigma_{\nu_3 \nu_1}, \sigma_{\nu_3 \nu_2})$, $(\gamma_{\nu_4}, \sigma_{\nu_4 \nu_1}, \sigma_{\nu_4 \nu_2})$, and singlets: $\openone, \gamma_5 \gamma_{\nu_3}, \gamma_5 \gamma_{\nu_4}, \sigma_{\nu_4 \nu_3}$, 
\item [(b)] For chiral fermions, the mixing sets are: doublets: $(\gamma_{\nu_1}, \gamma_{\nu_2})$, $(\gamma_5 \gamma_{\nu_1}, \gamma_5 \gamma_{\nu_2})$, $(\sigma_{\nu_3 \nu_1}, \sigma_{\nu_3 \nu_2})$, $(\sigma_{\nu_4 \nu_1}, \sigma_{\nu_4 \nu_2})$, and singlets:  $\openone, \gamma_5, \gamma_{\nu_3}, \gamma_{\nu_4}, \gamma_5 \gamma_{\nu_3}, \gamma_5\gamma_{\nu_4}, \sigma_{\nu_1 \nu_2}, \sigma_{\nu_4 \nu_3}$, 
\end{itemize}
where $(\nu_1, \nu_2, \nu_3, \nu_4)$ correspond to different orthogonal directions in the 4D Euclidean space. In contrast to the symmetric case, symmetry arguments alone are not sufficient to establish multiplicative renormalization for any asymmetric operator. 

We note that all-order perturbative studies in the continuum show that a multiplicative renormalization can address all the divergences of the nonlocal Wilson-line operators~\cite{Dotsenko:1979wb,Brandt:1981kf,Dorn:1986dt}. Thus, the mixing which is allowed by symmetries is absent in minimal subtraction schemes. However, it can occur as finite mixing in nonminimal schemes. On the lattice, the mixing is present when employing $\MSbar$ or any nonperturbative intermediate scheme; while $\MSbar$ is a minimal continuum scheme, on the lattice, additional finite regularization-dependent terms contribute to the renormalization functions in order to be able to match the continuum $\MSbar$-renormalized Green's functions. 
  \\

\begin{table}[thb]
  \centering
  \begin{tabular}{|c|c|c|c|c|}
  \hline
 & \ $\widetilde{\mathcal{O}}_1 (\Gamma) \ [{ \widetilde{\mathcal{O}}_5 (\Gamma)}]$ & \ $\widetilde{\mathcal{O}}_2 (\Gamma) \ [{ \widetilde{\mathcal{O}}_6 (\Gamma)}]$ & \ $\widetilde{\mathcal{O}}_3 (\Gamma) \ [{ \widetilde{\mathcal{O}}_7 (\Gamma)}]$ & \ $\widetilde{\mathcal{O}}_4 (\Gamma) \ [{ \widetilde{\mathcal{O}}_8 (\Gamma)}]$ \\
\hline
\hline
$\mathcal{P}_{\nu_1}$ & \ $+d_{\nu_1} (\Gamma)$ & \ $-d_{\nu_1} (\Gamma)$ & \ $+d_{\nu_1} (\Gamma)$ & \ $-d_{\nu_1} (\Gamma)$ \\
\hline
$\mathcal{P}_{\nu_2}$ & \ $+d_{\nu_2} (\Gamma)$ & \ $+d_{\nu_2} (\Gamma)$ & \ $-d_{\nu_2} (\Gamma)$ & \ $-d_{\nu_2} (\Gamma)$  \\
\hline
$\mathcal{P}_{\nu_3}$ & \ $+d_{\nu_3} (\Gamma)$ & \ $-d_{\nu_3} (\Gamma)$ & \ $-d_{\nu_3} (\Gamma)$ & \ $+d_{\nu_3} (\Gamma)$  \\
\hline
$\mathcal{P}_{\nu_4}$ & \ $+d_{\nu_4} (\Gamma)$ & \ $-d_{\nu_4} (\Gamma)$ & \ $-d_{\nu_4} (\Gamma)$ & \ $+d_{\nu_4} (\Gamma)$ \\
\hline
$\mathcal{T}_{\nu_1}$ & \ $+d_{\nu_1} (\Gamma) d_5 (\Gamma)$ & \ $+d_{\nu_1} (\Gamma) d_5 (\Gamma)$ & \ $-d_{\nu_1} (\Gamma) d_5 (\Gamma)$ & \ $-d_{\nu_1} (\Gamma) d_5 (\Gamma)$  \\
\hline
$\mathcal{T}_{\nu_2}$ & \ $+d_{\nu_2} (\Gamma) d_5 (\Gamma)$ & \ $-d_{\nu_2} (\Gamma) d_5 (\Gamma)$ & \ $+d_{\nu_2} (\Gamma) d_5 (\Gamma)$ & \ $-d_{\nu_2} (\Gamma) d_5 (\Gamma)$ \\
\hline
$\mathcal{T}_{\nu_3}$ & \ $+d_{\nu_3} (\Gamma) d_5 (\Gamma)$ & \ $+d_{\nu_3} (\Gamma) d_5 (\Gamma)$ & \ $+d_{\nu_3} (\Gamma) d_5 (\Gamma)$ & \ $+d_{\nu_3} (\Gamma) d_5 (\Gamma)$ \\
\hline
$\mathcal{T}_{\nu_4}$ & \ $+d_{\nu_4} (\Gamma) d_5 (\Gamma)$ & \ $+d_{\nu_4} (\Gamma) d_5 (\Gamma)$ & \ $+d_{\nu_4} (\Gamma) d_5 (\Gamma)$ & \ $+d_{\nu_4} (\Gamma) d_5 (\Gamma)$ \\
\hline
$C$ & \ $(+1)[(-1)] d_{\rm h.c.} (\Gamma) d_5 (\Gamma) d_4 (\Gamma)$ & \ $(+1)[(-1)] d_{\rm h.c.} (\Gamma) d_5 (\Gamma) d_4 (\Gamma)$ & \ $(+1)[(-1)] d_{\rm h.c.} (\Gamma) d_5 (\Gamma) d_4 (\Gamma)$ & \ $(+1)[(-1)] d_{\rm h.c.} (\Gamma) d_5 (\Gamma) d_4 (\Gamma)$ \\
\hline
  \end{tabular}
  \caption{Symmetry transformations of $\widetilde{\mathcal{O}}_i (\Gamma)$ [defined in Eqs. (\ref{Otilde1} -- \ref{Otilde8})] under $\mathcal{C}, \mathcal{P}, \mathcal{T}$. The operators are odd or even under these transformations: $\widetilde{\mathcal{O}}_i (\Gamma) \rightarrow \pm \widetilde{\mathcal{O}}_i (\Gamma)$. The relative sign for each transformation and for each operator is given in the table. The sign is affected by the commutation and anti-commutation relations of $\Gamma$ with $\gamma_\mu$ and $\gamma_5$, as well as the hermiticity or anti-hermiticity of $\Gamma$: $d_\mu (\Gamma) = \pm 1$ when $\Gamma \gamma_\mu = \pm \gamma_\mu \Gamma$, $d_5 (\Gamma) = \pm 1$ when $\Gamma \gamma_5 = \pm \gamma_5 \Gamma$, and $d_{\rm h.c.} (\Gamma) = \pm 1$ when $\Gamma^\dagger = \pm \Gamma$. Operators $\widetilde{\mathcal{O}}_5 - \widetilde{\mathcal{O}}_8$ (given in square brackets) share the same signs as operators $\widetilde{\mathcal{O}}_1 - \widetilde{\mathcal{O}}_4$, respectively, for all symmetry transformations, except under $\mathcal{C}$ (see last row), where the signs are opposite: plus (minus) signs correspond to operators $\widetilde{\mathcal{O}}_1 - \widetilde{\mathcal{O}}_4$ ($\widetilde{\mathcal{O}}_5 - \widetilde{\mathcal{O}}_8$).}
  \label{tab:CPT}
\end{table}

In our work, we focus on the wider mixing pattern arising in the case of asymmetric staple operators when a chirality-breaking fermion action is employed. Our goal is to construct a common prescription for renormalizing the staple operators in any regularization (regularization independent). To this end, we consider the following quadruplets of the sixteen independent bilinear operators $\mathcal{O}_\Gamma$ (with $\Gamma \in S_i$)\footnote{There is freedom in choosing the signs in front of each operator, leading to different conventions.}: 
\begin{eqnarray}
 S_1 &\equiv& \ \{ \openone \ \ , \ \sigma_{\nu_1 \nu_2}, \ \gamma_{\nu_1} \ \ \ , \ \gamma_{\nu_2} \ \ \ \ \}, \label{S1} \\
 S_2 &\equiv& \ \{ \gamma_5 \ , \ \sigma_{\nu_4 \nu_3}, \ \gamma_5 \gamma_{\nu_1}, \ \gamma_5 \gamma_{\nu_2} \ \}, \\
 S_3 &\equiv& \ \{ \gamma_{\nu_3}, \ \gamma_5 \gamma_{\nu_4}, \ \sigma_{\nu_3 \nu_1}, \ \sigma_{\nu_3 \nu_2}\ \}, \\
S_4 &\equiv& \ \{ \gamma_{\nu_4}, \ \gamma_5 \gamma_{\nu_3}, \ \sigma_{\nu_4 \nu_1}, \ \sigma_{\nu_4 \nu_2}\ \}. \label{S4}
\end{eqnarray}

In the present study, we do not consider possible mixing (on the lattice) with higher dimensional operators multiplied by the appropriate power of the lattice spacing. In the case of local operators, such mixing is present only for finite values of the lattice spacing, and it vanishes when taking the continuum limit. However, in the case of nonlocal operators, where power divergences $a^{-n}, \ n \in \mathbb{Z}^+$ are present, $\mathcal{O} (a)$ effects in the bare Green's functions can contribute to the renormalized Green's functions at two loops\footnote{$\mathcal{O} (a \ g^2)$ terms from the bare Green's function multiplied by $\mathcal{O} (1/a \ g^2)$ terms from the renormalization function lead to $\mathcal{O} (a^0 \ g^4)$ contributions to the renormalized Green's function.}. 
Alternatively, one can suppress these unwanted effects in two ways: (1) by removing power divergences from the Green's function through an appropriate ratio with another Green's function which has the same power divergences, e.g., a closed Wilson loop, (2) by subtracting artifacts from the bare Green's functions calculated in lattice perturbation theory. Our group has successfully applied this method to the renormalization of local quark bilinear operators~\cite{Constantinou:2009tr,Constantinou:2013ada,Alexandrou:2015sea}, and more recently to the renormalization of nonlocal straight Wilson-line operators for quasi-PDFs~\cite{Constantinou:2022aij}. We plan to study one-loop discretization effects for the staple operators to all orders in the lattice spacing in future work. 

\subsection{Green's functions of staple-shaped operators with external fermions}

As is standard practice, one-particle-irreducible (1-PI) two-point amputated Green's functions of the operators under study with external elementary fields, e.g., fermion fields, can be used for the extraction of renormalization functions \footnote{The removal of the standard momentum-space delta functions which appear in Green's functions such as Eq. \eqref{LambdaGamma} and Eq. \eqref{Zq} is understood.}:
\begin{equation}
\Lambda_\Gamma (q,z,y,y') =  \sum_x {\langle \psi (q)| \mathcal{O}_\Gamma (x,z,y,y') | \bar{\psi} (q) \rangle}_{\rm amp.}. \label{LambdaGamma}
\end{equation}
A summation over the position of the staple-shaped operator is taken, which is allowed by translational symmetry, in order to simplify the calculations. Such Green's functions using local operators are easily calculated in continuum perturbation theory to very high order. However, due to the nonlocal nature of the staple-shaped operators, additional scales (staple lengths) appear in Green's functions, which make the computation more complex even at the one-loop level. The corresponding calculation on the lattice is even more demanding since the procedure for isolating divergences from the Feynman integrals, as well as the procedure for taking the continuum limit $a \rightarrow 0$ (where $a$ is the lattice spacing) are more complicated (see, e.g., Refs.~\cite{Constantinou:2017sej,Constantinou:2019vyb}).

\begin{figure}[ht]
    \centering
    \includegraphics[width=0.75\textwidth]{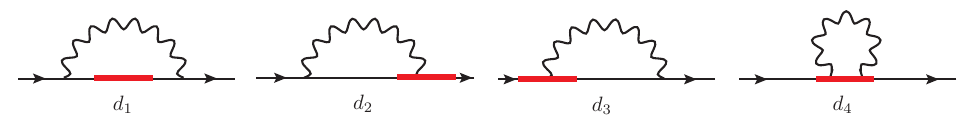}
    \caption{One-loop Feynman diagrams contributing to the Green’s functions of the asymmetric staple-shaped operator with external fermions. The straight (wavy) lines represent fermions (gluons). The operator insertion is denoted by a filled rectangle.}
    \label{fig:one-loop_diagrams}
  \end{figure}

\begin{figure}[ht]
    \centering
    \includegraphics[width=0.65\textwidth]{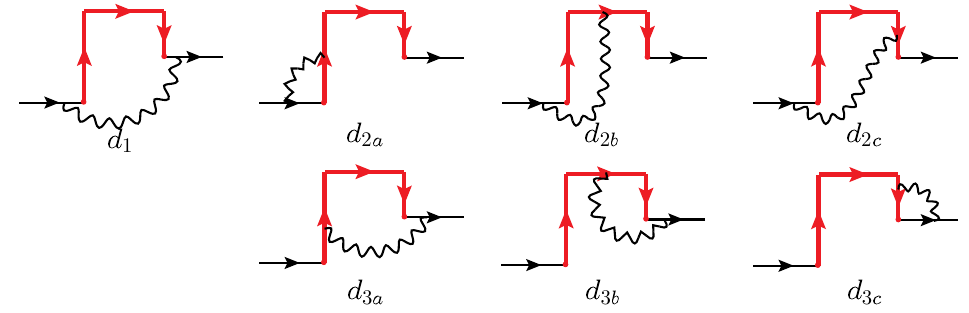}
    \includegraphics[width=0.65\textwidth]{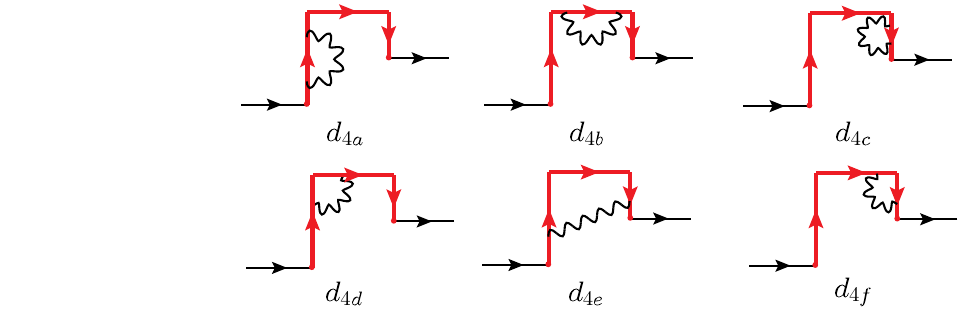}
    \caption{One-loop subdiagrams contributing to the Green’s functions of the asymmetric staple-shaped operator with external fermions. The straight (wavy) lines represent fermions (gluons). The operator insertion is denoted by an asymmetric staple-shaped line.}
    \label{fig:one-loop_subdiagrams}
\end{figure}

There are four one-loop Feynman diagrams contributing to $\Lambda_\Gamma (q,z,y,y')$, shown in Fig. \eqref{fig:one-loop_diagrams}. These
diagrams will appear in both continuum and lattice regularizations since all vertices are present in both regularizations. To this perturbative order, zero ($d_1$), one ($d_2 - d_3$) or two ($d_4$) gluons stem from the staple-shaped Wilson line. Due to the shape of the staple, the diagrams are further divided into thirteen subdiagrams, shown in Fig. \eqref{fig:one-loop_subdiagrams}, depending on the side of the staple from which gluons emanate.  
  
The transformation properties of $\Lambda_\Gamma (q,z,y,y')$ under $\mathcal{C}, \mathcal{P}, \mathcal{T}$ are given below ($\vec{q}$ is the momentum 3-vector which is perpendicular to the $\mu$ direction appearing in each transformation):
\begin{eqnarray}
 \Lambda_\Gamma (q,z,y,y') &\xrightarrow{\mathcal{P}_\mu}& \gamma_\mu \ \Lambda_{\gamma_{\mu} \Gamma \gamma_{\mu}} \left((-\vec{q},q_\mu),(-1)^{\delta_{\mu \nu_1} + 1} \ z, (-1)^{\delta_{\mu \nu_2} + 1} \ y, (-1)^{\delta_{\mu \nu_2} + 1} \ y'\right) \ \gamma_\mu, \\
 \Lambda_\Gamma (q,z,y,y') &\xrightarrow{\mathcal{T}_\mu}&
 \gamma_\mu \gamma_5 \ \Lambda_{\gamma_5 \gamma_{\mu} \Gamma \gamma_{\mu} \gamma_5} \left((\vec{q},-q_\mu),(-1)^{\delta_{\mu \nu_1}} \ z, (-1)^{\delta_{\mu \nu_2}} \ y, (-1)^{\delta_{\mu \nu_2}} \ y'\right) \ \gamma_5 \gamma_\mu, \\
\Lambda_\Gamma (q,z,y,y') &\xrightarrow{\mathcal{C}}&  C^T \ \gamma_4 \ \Lambda^\dagger_{\gamma_4 (C \Gamma C^{-1})^{\ast} \gamma_4} (-q,z,y,y') \ \gamma_4 \ {(C^{-1})}^T. \quad
  \end{eqnarray}
  By combining $\mathcal{P}_\mu$, $\mathcal{T}_\mu$ and $\mathcal{C}$, $\Lambda_\Gamma (q,z,y,y')$ transforms to:
  \begin{equation}
      \Lambda_\Gamma (q,z,y,y') \xrightarrow{\mathcal{P}_\mu \cdot \mathcal{T}_\mu \cdot \mathcal{C}} \gamma_5 \ C^T \ \gamma_4 \ \Lambda^\dagger_{\gamma_5 \gamma_4 (C \Gamma C^{-1})^{\ast} \gamma_4 \gamma_5} (q,-z,-y,-y') \ \gamma_4 \ {(C^{-1})}^T \gamma_5.
      \label{Lambda_symmetry}
  \end{equation}

\subsection{Renormalization conditions}
\label{Renormalization conditions}

We formulate below different versions of appropriate regularization-independent (RI$'$) prescriptions which address all possible divergences and mixing of the staple-shaped operators. In contrast to our previous study regarding the renormalization of symmetric staple operators, we construct renormalization matrices that address the mixing as observed by studying symmetries and not by studying one-loop lattice perturbation theory; thus, the renormalization matrices will be $4 \times 4$. Several of the nondiagonal elements of these matrices will vanish at one-loop. In this way, the renormalization prescription will be more appropriate for nonperturbative calculations addressing possible mixing that can be seen in higher loops. We first give our conventions regarding the renormalization of the operators under study, as well as of the relevant elementary fields and parameters that enter our perturbative calculations:
\begin{equation}
    \mathcal{O}^R_\Gamma (x,z,y,y') = Z^{R,X}_{\Gamma \Gamma'} \mathcal{O}^X_{\Gamma'} (x,z,y,y'),  \qquad \psi^R (x) = {(Z^{R,X}_\psi)}^{1/2} \psi^X (x), \qquad g^R = \mu^{(4-d)/2} Z^{R,X}_g g^X, \label{Z_GG:def}
\end{equation}
where $\psi^X (\psi^R)$ is the bare (renormalized) fermion field, $g^X (g^R)$ is the bare (renormalized) coupling constant. $X$ denotes dimensional (DR) or lattice (LR) regularization, $R$ denotes renormalization schemes of RI$'$ or $\MSbar$, $\mu$ is related to the $\overline{\rm MS}$ renormalization scale $\bar{\mu}$ [ $\bar{\mu} \equiv \mu {(4 \pi / e^{\gamma_E})}^{1/2}$, $\gamma_E$ is Euler’s constant] and $d$ is the number of Euclidean spacetime dimensions (in DR: $d \equiv 4 - 2 \varepsilon$, in LR: $d = 4$)\footnote{Superscripts $X$ and $R$ are omitted when they are clear from the context.}. A sum over $\Gamma'$ matrices, which belong to the mixing set of $\Gamma$ (cf. Eqs. \eqref{S1} -- \eqref{S4}), is implicit in Eq. \eqref{Z_GG:def}\footnote{From now on, sums over repeated $\Gamma$ matrices are understood.}. Given the above conventions, the renormalized Green's function $\Lambda_\Gamma^R (q,z,y,y')$ under study is defined through:
\begin{equation}
  \Lambda_\Gamma^R (q,z,y,y') = {(Z^{R,X}_\psi)}^{-1} Z^{R,X}_{\Gamma \Gamma'} \Lambda_{\Gamma'}^X (q,z,y,y'). \label{LambdaR}
\end{equation}

$\bullet$ {\bf ${\rm RI}'$ conditions:} \\
We first employ the typical RI$'$ scheme as defined for the renormalization of local fermion bilinear operators~\cite{Martinelli:1994ty} by extending the renormalization conditions consistently with the mixing and the definition in Eq. \eqref{LambdaR}:  
\begin{equation}
  \frac{1}{4 N_c} {(Z^{{\rm RI}',X}_\psi)}^{-1} Z^{{\rm RI}',X}_{\Gamma \Gamma'} {\rm Tr}[\Lambda^X_{\Gamma'} (q,z,y,y') \mathcal{P}_{\Gamma''}] \Big|_{q = \bar{q}} = \delta_{\Gamma \Gamma''}, \qquad \Gamma, \Gamma'' \in S_i,
  \label{RIcond1}
  \end{equation}  
where 
\begin{equation}
Z_\psi^{{\rm RI}',X} = \frac{1}{4 N_c} \ {\rm Tr}\left[{\langle \psi^X (q) \bar{\psi}^X (q)\rangle}^{-1} \cdot \frac{i \slashed{q}}{q^2}\right] \Big|_{q = \bar{q}}, \label{Zq}
\end{equation}
$\langle \psi^X (q) \bar{\psi}^X (q)\rangle$ is the fermion propagator, $\bar{q}$ is the RI$'$ renormalization 4-vector scale and $N_c$ is the number of colors. Note that the traces appearing in Eqs. (\ref{RIcond1} -- \ref{Zq}) regard both Dirac and color indices. Also, Eq. \eqref{RIcond1} corresponds to $16 \times 4 = 64$ conditions, which determine all the elements of the $4 \times 4$ renormalization matrices for the 4 mixing sets $S_i$.

We use two different choices of projectors in Eq. \eqref{RIcond1}:
\begin{eqnarray}
\mathcal{P}_{\Gamma}^{[1]} &=& \quad \ e^{-i {\sf q} \cdot {\sf r}} \ \Gamma^\dagger, \\
  \mathcal{P}_{\Gamma}^{[2]} &=& 
  \begin{cases}
  e^{-i {\sf q} \cdot {\sf r}} \left(\openone - \frac{\slashed{q}_T \ \slashed{q}_L}{q_T^2} \right) \Gamma^\dagger, & \qquad \Gamma \in S_1, S_2 \\
  e^{-i {\sf q} \cdot {\sf r}} \left(\openone - \frac{(\slashed{q}_T - \slashed{q}_{\nu_3}) (\slashed{q}_L + \slashed{q}_{\nu_3})}{q_T^2 - q_{\nu_3}^2} \right)\Gamma^\dagger, & \qquad \Gamma \in \{\gamma_{\nu_3}, \gamma_5 \gamma_{\nu_3}, \sigma_{\nu_3 \nu_1}, \sigma_{\nu_3 \nu_2}\} \\
  e^{-i {\sf q} \cdot {\sf r}} \left(\openone - \frac{(\slashed{q}_T - \slashed{q}_{\nu_4}) (\slashed{q}_L + \slashed{q}_{\nu_4})}{q_T^2 - q_{\nu_4}^2} \right)\Gamma^\dagger, & \qquad \Gamma \in \{\gamma_{\nu_4}, \gamma_5 \gamma_{\nu_4}, \sigma_{\nu_4 \nu_1}, \sigma_{\nu_4 \nu_2}\}
  \end{cases},
\end{eqnarray}
 where $\vec{\sf r} \equiv z \ \hat{\nu}_1 + (y-y') \ \hat{\nu}_2$, $\vec{q}_L \equiv q_{\nu_1} \hat{\nu}_1 + q_{\nu_2} \hat{\nu}_2$ and $\vec{q}_T \equiv \vec{q} - \vec{q}_L = q_{\nu_3} \hat{\nu}_3 + q_{\nu_4} \hat{\nu}_4$. Use of these two choices of projectors amounts to the implementation of two different ${\rm RI}'$ prescriptions, which we will denote as ${\rm RI}'_1$, ${\rm RI}'_2$ from now on. Compared to the first choice of projectors, the second one can further remove finite contributions of some Dirac structures, allowed by Lorentz symmetry, from the elements of the renormalization matrices. Similar projectors have also been studied in the renormalization of local operators, leading to reduced contributions from hadronic contamination in the nonperturbative data, especially for small values of the renormalization scale $\bar{q}$.  

 The RI$'$ scheme can address the power\footnote{Some power divergences may remain beyond one loop~\cite{Zhang:2022xuw}.} and logarithmic divergences, as well as the mixing between different Dirac structures, in the same way as $\MSbar$ does. However, in contrast to the $\MSbar$ scheme, RI$'$ can also treat the pinch-pole singularity when $y \rightarrow \infty$. This means that this infinite limit can be taken only in the RI$'$-renormalized Green's functions and not in the $\MSbar$-renormalized Green's functions of the staple operators. This is true because both schemes are defined for finite values of $y$, where pinch-pole singularities are not present. Then, terms that diverge with $y$ (in the $y \rightarrow \infty$ limit) do not contribute to the $\MSbar$ renormalization function, but they do so in RI$'$. Thus, when multiplying the bare Green's functions with their renormalization functions, only in RI$'$ these terms are eliminated. However, since RI$'$ is just an intermediate scheme entering the procedure of renormalizing the operators in the reference scheme $\MSbar$,  nonperturbatively, we cannot benefit from this additional feature of RI$'$; after conversion to the $\MSbar$ scheme, the pinch-pole singularity comes back. Thus, the standard $\MSbar$ prescription is not appropriate for renormalizing staple operators in the infinite limit $y \rightarrow \infty$.  \\

$\bullet$ {\bf Alternative prescriptions:}

Alternative prescriptions which are also applicable in the limit $y \rightarrow \infty$ are described below. The main feature in these prescriptions stems from the fact that the same pinch-pole singularity arises in a closed Wilson loop. Also, this object shows power and cusp divergences as the staple-shaped operators. As proposed in Ref.~\cite{Ji:2021uvr}, we can redefine the standard staple-shaped operator $\mathcal{O}_\Gamma$ by dividing it with the square root of the vacuum expectation value of a rectangular $z \times (y + y')$ Wilson loop $L (z, y+y')$: 
\begin{equation}
    \overline{\mathcal{O}}_\Gamma (x,z,y,y') \equiv \frac{\mathcal{O}_\Gamma (x,z,y,y')}{\langle L (z, y+y') \rangle^{1/2}},
\end{equation}
where 
\begin{equation}
L (z, y+y') \equiv \frac{1}{N_c} {\rm Tr} [ \mathcal{P} \{\mathcal{U} (x, (y+y') \hat{\nu}_2) \ \mathcal{U} (x + (y+y') \hat{\nu}_2, z \hat{\nu}_1) \ \mathcal{U}^\dagger (x + z \hat{\nu}_1, (y+y') \hat{\nu}_2) \ \mathcal{U}^\dagger (x, z \hat{\nu}_1) \}]. 
\end{equation}
The dimensions of the Wilson loop are chosen in such a way as to cancel the power\footnote{As in the original RI$'$ scheme, some power divergences may remain beyond one loop. However, by dividing with the Wilson loop, we expect that the residual power divergences are suppressed.}, cusp and pinch-pole divergences of $\mathcal{O}_\Gamma$. Both the standard $\MSbar$ prescription and the RI$'$ prescription (Eq. \ref{RIcond1}) can now be applied to the operator $\overline{\mathcal{O}}_\Gamma$. In particular, the condition for RI$'$ (we will call it RI$'$-bar; the related renormalization functions are denoted by $\overline{Z}^{R,X}_{\Gamma \Gamma'}$) now reads:
\begin{equation}
  \frac{1}{4 N_c} {(Z^{{\rm RI}',X}_\psi)}^{-1} \overline{Z}^{{\rm RI}',X}_{\Gamma \Gamma'} {\rm Tr}[\overline{\Lambda}^X_{\Gamma'} (q,z,y,y') \mathcal{P}_{\Gamma''}] \Bigg|_{q = \bar{q}} = \delta_{\Gamma \Gamma''},
  \end{equation}
where 
\begin{equation}
\overline{\Lambda}^X_\Gamma (q,z,y,y') =  \sum_x {\langle \psi^X (q)| \overline{\mathcal{O}}^X_\Gamma (x,z,y,y') | \bar{\psi}^X (q) \rangle}_{\rm amp.} = \frac{\Lambda^X_{\Gamma} (q,z,y,y')}{\langle L^X (z, y+y') \rangle^{1/2}},  \label{Lambdabar}
\end{equation}
\begin{equation}
    \overline{Z}^{{\rm RI}',X}_{\Gamma \Gamma'} = Z^{{\rm RI}',X}_{\Gamma \Gamma'} \ \langle L^X (z, y+y') \rangle^{1/2}, \qquad \overline{\mathcal{O}}^R_\Gamma (x,z,y,y') = \overline{Z}^{R,X}_{\Gamma \Gamma'} \overline{\mathcal{O}}^X_{\Gamma'} (x,z,y,y').
\end{equation}
$\overline{Z}^{R,X}_{\Gamma \Gamma'}$ addresses the remaining end-point divergences, as well as the mixing. 

\begin{figure}
    \centering    
    \includegraphics[width=0.6\textwidth]{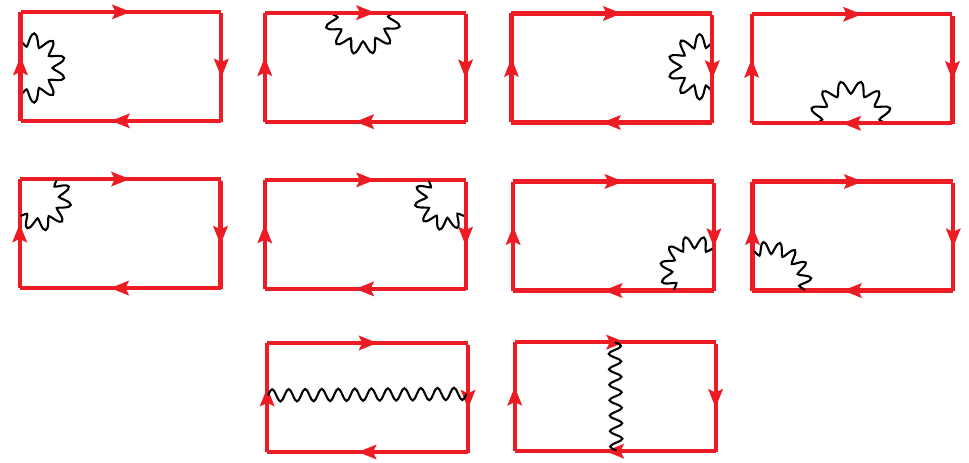}
    \caption{One-loop diagrams contributing to the vacuum expectation value of the rectangular Wilson loop. The wavy lines represent gluons.}
    \label{fig:one-loop_Wilson_loop}
\end{figure}

In order to extract $\overline{Z}^{R,X}_{\Gamma \Gamma'}$, we need to calculate the one-loop Feynman diagrams of Fig.~\ref{fig:one-loop_Wilson_loop} contributing to $\langle L^X(z, y{+}y') \rangle$. Older studies of Wilson loops can be found, e.g., in Refs.~\cite{Ebert:2019tvc,Martinelli:1998vt} in both continuum and lattice using Wilson gluons. Here, we extend these calculations to the case of Symanzik-improved gluons. For completeness, we have also repeated the continuum calculation.

Since the end-point divergences do not depend on the dimensions of the staple\footnote{On the contrary, power divergences depend on the total length of the staple line and thus, renormalization functions in ordinary RI$'$ must be computed separately for each choice of the dimensions of the staple.}, a nonperturbative determination of $\overline{Z}^{R,X}_{\Gamma \Gamma'}$ is expected to exhibit a much milder dependence on the staple lengths $z,y, y'$ which lie in the renormalization window: $a \ll z \ll \Lambda_{\rm QCD}^{-1}$, $a \ll y \ll \Lambda_{\rm QCD}^{-1}$, $a \ll y' \ll \Lambda_{\rm QCD}^{-1}$. In this way it becomes more acceptable to renormalize the modified operators $\overline{\mathcal{O}}_{\Gamma}$ defined at large values of the lengths $z,y, y'$, using renormalization functions $\overline{Z}^{RI',X}_{\Gamma \Gamma'}$ defined at smaller values of $z,y, y'$ within the perturbative region. To distinguish the lengths appearing in the bare operators from the reference lengths appearing in the renormalization functions, we will call the latter as $\bar{z}, \bar{y}, \bar{y}'$. If one is interested in taking the limit $y \rightarrow \infty$, then there are only two relevant lengths: $\bar{z}$ and $\bar{y}-\bar{y}'$.

Another option that we do not consider in this work is the short-distance ratio (SDR) scheme, described in Ref.~\cite{Zhang:2022xuw}. In this scheme, ratios of hadron matrix elements of the modified operators at different external momenta are considered. All the good features of RI$'$-bar are also valid in the SDR scheme. However, we stress that SDR is valid when operator mixing is absent or negligible. 

\section{Calculation in dimensional regularization}
\label{ResultsDR}

\subsection{Green's functions:}
\label{GFsDR}

We provide below our one-loop results for the bare Green's functions $\Lambda^{\rm DR}_\Gamma (q,z,y,y')$ in dimensional regularization (DR) up to $\mathcal{O} (\varepsilon^0)$. For the calculation of the one-loop momentum integrals, we make use of the formulae given in Appendix \ref{IntegrationMethod}. The results depend on several Dirac structures allowed by the residual rotational symmetry, and they are expressed in terms of integrals over Feynman parameters and/or over $\zeta$-variables running over the sides of the staples: $F_i \equiv F_i(q,r)$, $G_i \equiv G_i(q,y,z)$, $\bar{G}_i \equiv \bar{G}_i(q,y',z)$, $H_i \equiv H_i(q,y,z)$, $\bar{H}_i \equiv \bar{H}_i(q,y',z)$, $I_i \equiv I_i(q,y-y',z)$. These integrals are listed explicitly in Appendix \ref{IntegralList}. The following notation is employed: $\vec{\sf r} \equiv z \ \hat{\nu}_1 + (y-y') \ \hat{\nu}_2$, $\mu = \nu_3, \nu_4$.
\begin{eqnarray}
    \Lambda^{\rm DR}_{\bmb{\openone}} (q,z,y,y') \ \ &=& \ \ \Sigma_{1,1} \ \bmb{ \openone} \qquad \ \ + \ \ \Sigma_{1,2} \ \bmb{\sigma_{\nu_1 \nu_2}} \qquad \quad \ \ + \ \ \Sigma_{1,3} \ \bmb{\gamma_{\nu_1} \slashed{q}} \quad \ \, + \ \ \Sigma_{1,4} \ \bmb{\gamma_{\nu_2} \slashed{q}}, \label{Lambda_S} \\ 
   \Lambda^{\rm DR}_{\bmb{\gamma_{\nu_1}}} (q,z,y,y') \ \ &=& \ \ \Sigma_{2,1} \ \bmb{\gamma_{\nu_1}} \quad \ \ \, + \ \ \Sigma_{2,2} \ \bmb{\gamma_{\nu_2}} \qquad \qquad \ \ + \ \ \Sigma_{2,3} \ \bmb{\openone \slashed{q}} \qquad \ + \ \ \Sigma_{2,4} \ \bmb{\sigma_{\nu_1 \nu_2} \slashed{q}}, \\ 
   \Lambda^{\rm DR}_{\bmb{\gamma_{\nu_2}}} (q,z,y,y') \ \ &=& \ \ \Sigma_{3,1} \ \bmb{\gamma_{\nu_2}} \quad \ \ \, + \ \ \Sigma_{3,2} \ \bmb{\gamma_{\nu_1}} \qquad \qquad \ \ + \ \ \Sigma_{3,3} \ \bmb{\openone \slashed{q}} \qquad \ + \ \ \Sigma_{3,4} \ \bmb{\sigma_{\nu_1 \nu_2} \slashed{q}}, \\
   \Lambda^{\rm DR}_{\bmb{\gamma_{\mu}}} (q,z,y,y') \ \ &=& \ \ \Sigma_{4,1} \ \bmb{\gamma_{\mu}} \qquad \, + \ \ \Sigma_{4,2} \ \bmb{\varepsilon_{\nu_1 \nu_2 \mu \rho} \gamma_5 \gamma_{\rho}} \ \ + \ \ \Sigma_{4,3} \ \bmb{q_\mu \gamma_{\nu_1}} \quad + \ \ \Sigma_{4,4} \ \bmb{q_\mu \gamma_{\nu_2}} \nonumber \\
   &+& \ \ \Sigma_{4,5} \ \bmb{\sigma_{\mu \nu_1} \slashed{q}} \ \ + \ \ \Sigma_{4,6} \ \bmb{\sigma_{\mu \nu_2} \slashed{q}} \qquad \quad \ \, + \ \ \Sigma_{4,7} \ \bmb{q_\mu \openone \slashed{q}}, \\
      \Lambda^{\rm DR}_{\bmb{\sigma_{\nu_1 \nu_2}}} (q,z,y,y') \ \ &=&  \ \ \Sigma_{5,1} \ \bmb{\sigma_{\nu_1 \nu_2}} \ \ \ + \ \,  \Sigma_{5,2} \ \bmb{\openone} \qquad \qquad \quad \ \ + \ \ \Sigma_{5,3} \ \bmb{\gamma_{\nu_1} \slashed{q}} \quad \ \, + \ \ \Sigma_{5,4} \ \bmb{\gamma_{\nu_2} \slashed{q}}, \\
      \Lambda^{\rm DR}_{\bmb{\sigma_{\mu \nu_1}}} (q,z,y,y') \ \ &=& \ \ \Sigma_{6,1} \ \bmb{\sigma_{\mu \nu_1}} \ \ \ \ + \ \, \Sigma_{6,2} \ \bmb{\sigma_{\mu \nu_2}} \qquad \quad \ \ \ \, + \ \ \Sigma_{6,3} \ \bmb{q_\mu \openone} \quad \ \ + \ \ \Sigma_{6,4} \ \bmb{q_\mu \sigma_{\nu_1 \nu_2}} \nonumber \\
   &+& \ \ \Sigma_{6,5} \ \bmb{\gamma_{\mu} \slashed{q}} \ \ \ \ \ + \ \ \Sigma_{6,6} \ \bmb{\varepsilon_{\nu_1 \nu_2 \mu \rho} \gamma_5 \gamma_{\rho} \slashed{q}} \, + \ \ \Sigma_{6,7} \ \bmb{q_\mu \gamma_{\nu_1} \slashed{q}} \ + \ \ \Sigma_{6,8} \ \bmb{q_\mu \gamma_{\nu_2} \slashed{q}}, \\
   \Lambda^{\rm DR}_{\bmb{\sigma_{\mu \nu_2}}} (q,z,y,y') \ \ &=& \ \ \Sigma_{7,1} \ \bmb{\sigma_{\mu \nu_2}} \ \ \ \ + \ \, \Sigma_{7,2} \ \bmb{\sigma_{\mu \nu_1}} \qquad \quad \ \ \, \, \, \, + \ \ \Sigma_{7,3} \ \bmb{q_\mu \openone} \quad \ \ + \ \ \Sigma_{7,4} \ \bmb{q_\mu \sigma_{\nu_1 \nu_2}} \nonumber \\
   &+& \ \ \Sigma_{7,5} \ \bmb{\gamma_{\mu} \slashed{q}} \ \ \ \ \ + \ \ \Sigma_{7,6} \ \bmb{\varepsilon_{\nu_1 \nu_2 \mu \rho} \gamma_5 \gamma_{\rho} \slashed{q}} \, + \ \ \Sigma_{7,7} \ \bmb{q_\mu \gamma_{\nu_1} \slashed{q}} \ + \ \ \Sigma_{7,8} \ \bmb{q_\mu \gamma_{\nu_2} \slashed{q}}, \\
   \nonumber \\
   \Lambda^{\rm DR}_{\bmb{\gamma_5}} (q,z,y,y') \ \ &=& \ \ \bmb{\gamma_5} \  \Lambda^{\rm DR}_{\bmb{\openone}} (q,z,y,y'), \qquad \Lambda^{\rm DR}_{\bmb{\gamma_5 \gamma_{\nu_1}}} (q,z,y,y') \ \ = \ \ \bmb{\gamma_5} \  \Lambda^{\rm DR}_{\bmb{\gamma_{\nu_1}}} (q,z,y,y'), \\
   \Lambda^{\rm DR}_{\bmb{\gamma_5 \gamma_{\nu_2}}} (q,z,y,y') \ \ &=& \ \ \bmb{\gamma_5} \  \Lambda^{\rm DR}_{\bmb{\gamma_{\nu_2}}} (q,z,y,y'), \quad \ \ \, \Lambda^{\rm DR}_{\bmb{\gamma_5 \gamma_{\mu}}} (q,z,y,y') \ \ \ = \ \ \bmb{\gamma_5} \  \Lambda^{\rm DR}_{\bmb{\gamma_{\mu}}} (q,z,y,y'), \\
   \Lambda^{\rm DR}_{\bmb{\sigma_{\nu_4 \nu_3}}} (q,z,y,y') \ \ &=& \ \ \bmb{\gamma_5} \  \Lambda^{\rm DR}_{\bmb{\sigma_{\nu_1 \nu_2}}} (q,z,y,y'), \label{Lambda_Tnu4nu3} 
\end{eqnarray}
where $\Sigma_{i,j} \equiv \Sigma_{i,j} (\bar{\mu}^2, q_{\nu_1}, q_{\nu_2}, q^2, z, y, y')$:
\begin{equation}
\Sigma_{i,j} = e^{i {\sf q} \cdot {\sf r}} \left( \delta_{i1} + \frac{g^2 C_F}{16 \pi^2} \left\{\delta_{i1} \left[\frac{(8-\beta)}{\varepsilon} + (1-\beta) \ (1 + \ln \left(\frac{\bar{\mu}^2}{q^2}\right)) + \bmp{s_0}\right] + \bmp{s_{i,j}}\right\} + \mathcal{O} (g^4) \right),
\label{Sigma_ij}
\end{equation}
and $s_0 \equiv s_0(\bar{\mu}^2,q^2,z,y,y')$, $s_{i,j} \equiv s_{i,j}(q_{\nu_1},q_{\nu_2},q^2,z,y,y')$: 
\begin{eqnarray}
\bmp{s_0} &=& 2 (6 + \beta) \gamma_E + 4 \left[\frac{y}{z} \tan^{-1} \left(\frac{y}{z}\right) + \frac{y'}{z} \tan^{-1} \left(\frac{y'}{z}\right) - \frac{y-y'}{z} \tan^{-1} \left(\frac{y-y'}{z}\right)\right] \nonumber \\
&& + (1 - \beta) \ln \left(\frac{\bar{\mu}^2}{q^2}\right) + (2+\beta) \ln \left( \frac{\bar{\mu}^2 r^2}{4}\right) + 4 \ln \left( \frac{\bar{\mu}^2 z^2}{4}\right) + 2 \left[\ln \left(1 + \frac{z^2}{y^2}\right) + \ln \left(1 + \frac{z^2}{y'^2}\right)\right], \label{s0} \\
\nonumber \\
\bmp{s_{1,1}} &=& 15 + 4 F_1 +\beta \ \left(-1 + 2 F_1 - 2 F_4\right) + 2 \beta \ \left(F_1 - F_2\right) \ i {\sf q \cdot r} -2 \ \left(2 \bar{G}_1 + G_2 + \bar{G}_2\right) \ i q_{\nu_1} z \nonumber \\
&& -2 \ \left(H_2 - H_4\right) \ i q_{\nu_2} y + 2 \ \left(2 \bar{H}_1 + \bar{H}_2 - 2 \bar{H}_3 - \bar{H}_4\right) \ i q_{\nu_2} y' - 4 \ \left(I_1 + I_2\right) \ i q_{\nu_2} (y-y'), \\  
\bmp{s_{1,2}} &=& \bmp{s_{4,2}} = 2 \left(G_3 + H_5\right) q^2 y z + 2 \left(\bar{G}_3 + \bar{H}_5\right) q^2 y' z, \\
\bmp{s_{1,3}} &=& \bmp{s_{4,5}} = -2 \ \left(G_1 - \bar{G}_1\right) i z, \\
\bmp{s_{1,4}} &=& \bmp{s_{2,4}} = \bmp{s_{4,6}} = \bmp{s_{6,6}} = -2 \left(H_1 - H_3\right) i y -2 \left(\bar{H}_1 - \bar{H}_3\right) i y', \\
\bmp{s_{2,1}} &=& 15 - \left(2 F_1 - F_4\right) + \beta \ \left(-1 + 2 F_1 - F_4\right) + \left(2 F_1 - 4 F_2 - F_4\right) \ \frac{(y-y')^2 - z^2}{r^2} - \beta F_3 \ q^2 z^2 \nonumber \\
&& + \left(3 \beta \ F_1 - 4 \beta \ F_2 + 4 F_3 \ \frac{z^2}{r^2}\right) \ i {\sf q \cdot r} + \left[-4 F_3 + \frac{1}{2} \beta \ F_4 - 2 \left(2 G_1 + 2 \bar{G}_1 + G_2 + \bar{G}_2\right)\right] \ i q_{\nu_1} z \nonumber \\
&& + \beta F_3 \ ({\sf q \cdot r}) \ q_{\nu_1} z -2 \left(H_2 - H_4\right) \ i q_{\nu_2} y + 2 \left(2 \bar{H}_1 + \bar{H}_2 - 2 \bar{H}_3 - \bar{H}_4\right) \ i q_{\nu_2} y' \nonumber \\
&& - \left[\beta \left(F_1 - 2 F_2\right) + 4 \left(I_1 + I_2\right)\right] \ i q_{\nu_2} \left(y-y'\right), \\
\bmp{s_{2,2}} &=& -2 \ \left(G_3 + H_5\right) \ q^2 y z + 2 \ \left(\bar{G}_3 + \bar{H}_5\right) \ q^2 y' z + \left(-\beta F_3 + 4 I_3\right) \ q^2 (y-y') z \nonumber \\
&& + 2 \ \left(-2 F_1 + 4 F_2 + F_4 + 2 F_3 \ i {\sf q \cdot r}\right) \ \frac{(y-y') z}{r^2} -4 \ \left(H_1 - H_3\right) \ i q_{\nu_1} y \nonumber \\
&& + \left[\beta \ \left(F_1 - 2 F_2 + \frac{F_4}{2}\right) - 4 \ \left(F_3 + I_1\right)\right] \ i q_{\nu_1} (y-y') + \beta F_3 \ ({\sf q \cdot r}) \ q_{\nu_1} (y - y'), \\
\bmp{s_{2,3}} &=& - \beta F_3 \ \left(y - y'\right) \ \left[q_{\nu_1} \left(y-y'\right) - q_{\nu_2} z\right] + \left[-\beta \left(F_1 - 2 F_2 + \frac{F_4}{2}\right) + 2 \left(-2 F_3 + G_1 + \bar{G}_1\right)\right] \ i z \nonumber \\
&& + 2 \ \left[-2 F_1 + 4 F_2 + \left(-1 + \beta\right) F_4 + 2 F_3 \ i {\sf q \cdot r}\right] \ \frac{q_{\nu_1}}{q^2},
\end{eqnarray}
\begin{eqnarray}
\bmp{s_{3,1}} &=& 15 - \left(2 F_1 - F_4\right) + \beta \ \left(-1 + 2 F_1 - F_4\right) - \left(2 F_1 - 4 F_2 - F_4\right) \ \frac{(y-y')^2 - z^2}{r^2} - \beta F_3 \ q^2 (y-y')^2 \nonumber \\
&& + \left[2 \beta \ \left(F_1 - F_2\right) + 4 F_3 \ \frac{z^2}{(y-y')^2}\right] \ i {\sf q \cdot r} - 2 \left(2 \bar{G}_1 + G_2 + \bar{G}_2\right) \ i q_{\nu_1} z \nonumber \\
&& -2 \left(2 H_1 + H_2 -2 H_3 - H_4\right) \ i q_{\nu_2} y + 2 \left(2 \bar{H}_1 + \bar{H}_2 - 2 \bar{H}_3 - \bar{H}_4\right) \ i q_{\nu_2} y' \nonumber \\
&& - \left[\beta \left(F_1 - 2 F_2 + \frac{F_4}{2}\right) - 4 \left(F_3 + 2 I_1 + I_2\right)\right] \ i q_{\nu_2} \left(y-y'\right) + \beta F_3 \ \left(y-y'\right) {\sf q \cdot r}, \\
\bmp{s_{3,2}} &=& 2 \ \left(G_3 + H_5\right) \ q^2 y z - 2 \ \left(\bar{G}_3 + \bar{H}_5\right) \ q^2 y' z - \left(\beta F_3 + 4 I_3\right) \ q^2 (y-y') z \nonumber \\
&& + 2 \ \left(-2 F_1 + 4 F_2 + F_4 + 2 F_3 \ i {\sf q \cdot r}\right) \ \frac{(y-y') z}{r^2} \nonumber \\
&& + \left[\beta \ \left(F_1 - 2 F_2 + \frac{F_4}{2}\right) - 4 \ \left(F_3 + G_1\right)\right] \ i q_{\nu_2} z + \beta F_3 \ ({\sf q \cdot r}) \ q_{\nu_2} z, \\
\bmp{s_{3,3}} &=& \beta F_3 \ z \ \left[q_{\nu_1} \left(y-y'\right) - q_{\nu_2} z\right] + 2 \ \left(H_1 - H_3\right) \ i y - 2 \ \left(\bar{H}_1 - \bar{H}_3\right) \ i y' \nonumber \\
&& + \left[-\beta \left(F_1 - 2 F_2 + \frac{F_4}{2}\right) + 4 \left(-F_3 + I_1\right)\right] \ i (y-y') \nonumber \\
&& + 2 \ \left[-2 F_1 + 4 F_2 + \left(-1 + \beta\right) F_4 + 2 F_3 \ i {\sf q \cdot r}\right] \ \frac{q_{\nu_2}}{q^2}, \\
\bmp{s_{3,4}} &=& \bmp{s_{7,6}} = 2 \ \left(G_1 - \bar{G}_1 \right) \ i z, \\
\bmp{s_{4,1}} &=& 15 - 4 F_2 +\beta \ \left(-1 + 2 F_1 - F_4\right) + 2 \beta \ \left(F_1 - F_2\right) \ i {\sf q \cdot r} -2 \ \left(2 \bar{G}_1 + G_2 + \bar{G}_2\right) \ i q_{\nu_1} z \nonumber \\
&& -2 \ \left(H_2 - H_4\right) \ i q_{\nu_2} y + 2 \ \left(2 \bar{H}_1 + \bar{H}_2 - 2 \bar{H}_3 - \bar{H}_4\right) \ i q_{\nu_2} y' - 4 \ \left(I_1 + I_2\right) \ i q_{\nu_2} (y-y'),  \\
\bmp{s_{4,3}} &=& \left[ \beta \ \left(F_1 - 2 F_2 + \frac{F_4}{2}\right) - 4 \left(F_3 + G_1\right)\right] \ i z + \beta F_3 z \ ({\sf q \cdot r}), \\
\bmp{s_{4,4}} &=& -4 \ \left(H_1 - H_3\right) \ i y + \left[ \beta \ \left(F_1 - 2 F_2 + \frac{F_4}{2} \right) - 4 \left(F_3 + I_1 \right) \right] \ i (y-y') + \beta F_3 \ \left(y - y'\right) \ ({\sf q \cdot r}), \\
\bmp{s_{4,7}} &=& 2 \left[-2 F_1 + 4 F_2 - \left(1 - \beta\right) F_4 \right] \ \frac{1}{q^2} - \beta F_3 r^2 + 4 F_3 \frac{i {\sf q \cdot r}}{q^2}, \\
\bmp{s_{5,1}} &=& 15 - 2 \beta \ \left(1 - F_1\right) - 4 F_1 - \beta F_3 q^2 r^2 + \beta \ \left(3 F_1 - 4 F_2 + \frac{F_4}{2}\right) \ i {\sf q \cdot r} + \beta F_3 \ ({\sf q \cdot r})^2 \nonumber \\
&& - 2 \ \left( 2 G_1 + 2 \bar{G}_1 + G_2 + \bar{G}_2 \right) \ i q_{\nu_1} z - 2 \ \left(2 H_1 +H_2 - 2 H_3 - H_4 \right) \ i q_{\nu_2} y \nonumber \\
&& + 2 \ \left(2 \bar{H}_1 + \bar{H}_2 - 2 \bar{H}_3 - \bar{H}_4 \right) \ i q_{\nu_2} y' - 4 \ \left(2 I_1 + I_2\right) \ i q_{\nu_2} (y-y'), \\
\bmp{s_{5,2}} &=& \beta \left( F_1 - 2 F_2 + \frac{F_4}{2} \right) \ i \left[ q_{\nu_1} \ \left(y - y'\right) - q_{\nu_2} z \right] + 2 \ \left(G_3 + H_5\right) \ q^2 y z - 2 \left(\bar{G}_3 + \bar{H}_5 \right) \ q^2 y' z \nonumber \\
&& - \beta F_3 \left[ q_{\nu_1} \ \left(y - y'\right) - q_{\nu_2} z \right] \ ({\sf q \cdot r}) - 4 \ \left( H_1 - H_3 \right) \ i q_{\nu_1} y - 4 I_1 \ i q_{\nu_1} (y - y') + 4 G_1 q_{\nu_2} i z, \\
\bmp{s_{5,3}} &=& - \beta F_3 \left[ q_{\nu_1} \ \left(y - y'\right) - q_{\nu_2} z \right] z + 2 \ \left(H_1 - H_3 \right) \ i y -2 \ \left(\bar{H}_1 - \bar{H}_3 \right) \ i y' \nonumber \\
&& + \left[ -\beta \left(F_1 - 2 F_2 + \frac{F_4}{2}\right) + 4 I_1 \right] \ i (y-y'), \\
\bmp{s_{5,4}} &=& - \beta F_3 \left[ q_{\nu_1} \ \left(y - y'\right) - q_{\nu_2} z \right] \ \left(y-y'\right) + \left[ \beta \left( F_1 - 2 F_2 + \frac{F_4}{2} \right) -2 \ \left( G_1 + \bar{G}_1 \right) \right] \ i z, \\
\bmp{s_{6,1}} &=& 15 + \beta \left(-1 + 2 F_1\right) - 4 F_1 - \beta F_3 q^2 z^2 + \beta (3 F_1 - 4 F_2) i {\sf q \cdot r} \nonumber \\
&& + \left[ \beta \frac{F_4}{2} - 2 \left(2 G_1 + 2 \bar{G}_1 + G_2 + \bar{G}_2 \right)\right] \ i q_{\nu_1} z + \beta F_3 ({\sf q \cdot r}) \ q_{\nu_1} z -2 \left(H_2 - H_4 \right) \ i q_{\nu_2} y \nonumber \\
&& + 2 \left( 2 \bar{H}_1 + \bar{H}_2 - 2 \bar{H}_3 - \bar{H}_4 \right) \ i q_{\nu_2} y' + \left[ - \beta \left(F_1 - 2 F_2 \right) -4 \left(I_1 + I_2 \right)\right] \ i q_{\nu_2} (y-y'), \\
\bmp{s_{6,2}} &=& -2 \left(G_3 + H_5 \right) q^2 y z + 2 \left(\bar{G}_3 + \bar{H}_5\right) q^2 y' z + \left(-\beta F_3 + 4 I_3 \right) q^2 (y-y') z - 4 \left(H_1 - H_3 \right) \ i q_{\nu_1} y \nonumber \\
&& + \left[ \beta \left(F_1 - 2 F_2 + \frac{F_4}{2} \right) - 4 I_1 \right] \ i q_{\nu_1} (y-y') + \beta F_3 \ ({\sf q \cdot r}) \ q_{\nu_1} (y-y'), \\
\bmp{s_{6,3}} &=& \bmp{s_{7,4}} = \left[ \beta \left( F_1 -2 F_2 + \frac{F_4}{2} \right) - 4 G_1 \right] \ i z + \beta F_3 z \ ({\sf q \cdot r}),
\end{eqnarray}
\begin{eqnarray}
\bmp{s_{6,4}} &=& -\bmp{s_{7,3}} = 4 \left(H_1 - H_3 \right) \ i y + \left[ -\beta \left( F_1 - 2 F_2 + \frac{F_4}{2} \right) + 4 I_1 \right] \ i (y-y') - \beta F_3 (y-y') \ ({\sf q \cdot r}), \\
\bmp{s_{6,5}} &=& - \beta F_3 (y-y') \left[ q_{\nu_1} \ \left(y - y'\right) - q_{\nu_2} z \right] + \left[ -\beta \left(F_1 - 2 F_2 + \frac{F_4}{2} \right) + 2 \left( G_1 + \bar{G}_1 \right)\right] \ i z, \\
\bmp{s_{6,7}} &=& -\bmp{s_{7,8}} = \beta F_3 \left[ \left( y - y' \right)^2 - z^2 \right], \\
\bmp{s_{6,8}} &=& \bmp{s_{7,7}} = -2 \beta F_3 (y-y') z, \\
\bmp{s_{7,1}} &=& 15 + \beta \left(-1 + 2 F_1 \right) - 4 F_1 - \beta F_3 q^2 (y-y')^2 + 2 \beta \left(F_1 - F_2\right) \ i {\sf q \cdot r} - 2 \left(2 \bar{G}_1 + G_2 + \bar{G}_2 \right) \ i q_{\nu_1} z \nonumber \\
&& -2 \left( 2 H_1 + H_2 - 2 H_3 - H_4 \right) \ i q_{\nu_2} y + 2 \left( 2 \bar{H}_1 + \bar{H}_2 - 2 \bar{H}_3 - \bar{H}_4 \right) \ i q_{\nu_2} y' \nonumber \\
&& + \left[ \beta \left( F_1 - 2 F_2 + \frac{F_4}{2} \right) - 4 \left(2 I_1 + I_2 \right) \right] \ i q_{\nu_2} (y-y') + \beta F_3 \ ({\sf q \cdot r}) \ q_{\nu_2} (y-y'), \\
\bmp{s_{7,2}} &=& 2 \left(G_3 + H_5 \right) q^2 y z -2 \left( \bar{G}_3 + \bar{H}_5 \right) q^2 y' z - \left( \beta F_3 + 4 I_3 \right) q^2 (y-y') z \nonumber \\
&& + \left[ \beta \left(F_1 - 2 F_2 + \frac{F_4}{2} \right) - 4 G_1 \right] \ i q_{\nu_2} z + \beta F_3 \ ({\sf q \cdot r}) \ q_{\nu_2} z, \\
\bmp{s_{7,5}} &=& 2 \left(H_1 - H_3\right) i y - 2 \left( \bar{H}_1 - \bar{H}_3 \right) \ i y' + \left[ -\beta \left( F_1 - 2 F_2 + \frac{F_4}{2}\right) + 4 I_1 \right] \ i (y-y') \nonumber \\
&& + \beta F_3 z \left[ q_{\nu_1} \left(y - y'\right) - q_{\nu_2} z \right]. \label{s75}
\end{eqnarray}
$\beta$ is the gauge-fixing parameter [$\beta = 0 \ (1)$ in Feynman (Landau) gauge], and $\varepsilon_{\mu \nu \rho \tau}$ is the Levi-Civita tensor ($\varepsilon_{1234} = +1$).  
The pole term $\mathcal{O}(1/\varepsilon)$ (Eq. \eqref{Sigma_ij}) comes from the sum of cusp ($d_{4d}$, $d_{4f}$), end-point ($d_{2a}$, $d_{3c}$) and contact singularities ($d_{4a}$, $d_{4b}$, $d_{4c}$)\footnote{For a per-diagram determination of the pole terms, see Ref.~\cite{Constantinou:2019vyb}.}. Linear divergences are not present in DR. Also, the term $[y/z \tan^{-1} (y/z) + y'/z \tan^{-1} (y'/z)]$ appearing in $s_0$ (Eq. \eqref{s0}) gives rise to a pinch-pole singularity (linear divergence) in the limit $y \rightarrow \infty$ for fixed values of $(y-y')$. This term stems from the sub-diagram $d_{4e}$. Our results agree with previous studies in Refs.~\cite{Constantinou:2019vyb,Ebert:2019tvc} which consider specific cases of the staple operators. More specifically, when $y' \rightarrow y$, in which the asymmetric staple gauge link becomes symmetric, the above expressions reproduce the results of Ref.~\cite{Constantinou:2019vyb}, where a slightly different basis of Feynman-parameter integrals has been employed, which is related to our basis (when $y' = y$) through linear combinations and/or integration by parts. Other vanishing limits of the lengths for one or more staple segments, which classically result to simpler shapes, e.g., straight-line gauge link of length $y-y'$ when ($z \rightarrow 0$), or, straight-line gauge link of length $z$ when ($y=y'$ and $y \rightarrow 0$), or, single point when ($y=y'$ and $y \rightarrow 0$ and $z \rightarrow 0$), are singular because cusp and endpoints do not vanish smoothly in these limits giving rise to linear or logarithmic divergences (see Eqs. (\ref{vanishingz} -- \ref{vanishingy})). Thus, in these cases a complete consistency check is not applicable. However, by excluding the divergent parts of our results ($s_0$), the aforementioned limits can be taken in order to reproduce the finite parts of the corresponding Green's functions with simpler shapes of Wilson line~\cite{Constantinou:2017sej}. In particular, some of the form factors $\Sigma_{i,j}$ are zero resulting to fewer Lorentz structures for each operator, as expected by symmetries. The basis of Feynman-parameter integrals is now reduced due to vanishing terms, or identical integrals.

The different Dirac structures appearing in Eqs. (\ref{Lambda_S} -- \ref{Lambda_Tnu4nu3}) can be classified into two categories: (1) the structures corresponding to the tree-level Green's functions of the staple operators that belong to the same mixing set according to symmetries (structures multiplied by $\Sigma_{i,1}$ or $\Sigma_{i,2}$), and (2) all the remaining structures. The former structures are multiplied by regularization-dependent form factors ($\Sigma_{i,j}$), while the form factors of the latter structures do not depend on regularization. This is confirmed by comparing our one-loop results in both continuum and lattice regularizations (Eq. \eqref{general_WL}). In particular, the bare Green's functions in the continuum consist of two instead of four structures of the first category because chiral symmetry is preserved (see discussion in Sec. \ref{Sym}). On the other hand, the lattice bare Green's functions at one loop also contain a pair of the four first-category structures, which is different from that obtained in DR (see Sec. \ref{ResultsLR}). Since the lattice regularization that we employ breaks chiral symmetry, we expect that the missing two structures of the first category will appear in higher loops.  

By taking into account the classification of the Dirac structures in the Green's functions, we conclude that a renormalization prescription that considers a wider mixing pattern, i.e., mixing among all 16 independent operators of different $\Gamma$ matrices, is not an optimal choice for renormalizing the staple operators. In this case, several form factors $\Sigma_{i,j}$ will contribute to the $16 \times 16$ mixing matrix, including pure dependence on the regularization-independent form factors $\Sigma_{i,j>2}$ in some nondiagonal elements. Given that we are interested in matching bare lattice Green’s functions to the $\MSbar$ scheme (through an intermediate nonperturbative scheme), in which the regularization-independent form factors cannot contribute, it is more economic to consider an intermediate renormalization prescription that includes only the minimum number of operators necessary for disentangling the mixing occurring in the $\MSbar$ scheme on the lattice. In our study, we consider such a minimal set of operators as dictated by symmetries.

The one-loop renormalized Green's functions in the $\MSbar$ scheme can be obtained by removing the $\mathcal{O}(1/\varepsilon)$ term in Eq. \eqref{Sigma_ij}. Also, the renormalized Green's functions in any variant of the RI$'$ scheme can be obtained by imposing the corresponding renormalization conditions to the above results. The extraction of the conversion functions among these schemes (RI$'$ variants and $\MSbar$) is then straightforward; they are given in Eqs. (\ref{Conversion}, \ref{Cmatrix} -- \ref{Clast}) for the schemes under study. 

Specific combinations of the form factors $\Sigma_{i,j}$ contribute in the renormalization conditions of Eq. \eqref{RIcond1} for each choice of projectors $\mathcal{P}_{\Gamma}^{[1]}$ and $\mathcal{P}_{\Gamma}^{[2]}$. The latter choice has the advantage that in most cases, only $\Sigma_{i,1}$ and $\Sigma_{i,2}$ survive, which multiply the relevant structures for the study of mixing. However, in the case of projectors $\mathcal{P}_{\gamma_{\mu}}^{[2]}$ and $\mathcal{P}_{\gamma_5 \gamma_{\mu}}^{[2]}$, for $\mu = \nu_3, \nu_4$, some additional form factors can survive when taking the trace with the amputated Green's functions $\Lambda_{\gamma_5 \gamma_{\rho}}$ and $\Lambda_{ \gamma_{\rho}}$, respectively (where $\rho \neq (\nu_1, \nu_2, \mu)$). For instance, Eq. \eqref{RIcond1} for the case $\Gamma = \Gamma'' = \gamma_{\nu_3}$ takes the following form (in terms of $\Sigma_{i,j}$), when employing $\mathcal{P}_{\gamma_{\nu_3}}^{[1]}$:
\begin{equation}
    {(Z^{{\rm RI}',{\rm DR}}_\psi)}^{-1} [Z^{{\rm RI}'_1,{\rm DR}}_{\gamma_{\nu_3}, \gamma_{\nu_3}} (\Sigma_{4,1} + \Sigma_{4,5} \ q_{\nu_1} + \Sigma_{4,6} \ q_{\nu_2} + \Sigma_{4,7} \ q_{\nu_3}^2) - Z^{{\rm RI}'_1,{\rm DR}}_{\gamma_{\nu_3}, \gamma_5 \gamma_{\nu_4}} (\Sigma_{4,2} + \Sigma_{4,5} \ q_{\nu_2} - \Sigma_{4,6} \ q_{\nu_1})] \Big|_{q = \bar{q}} = 1,
    \label{RC1example}
\end{equation}
while employing $\mathcal{P}_{\gamma_{\nu_3}}^{[2]}$ gives: 
\begin{equation}
    {(Z^{{\rm RI}',{\rm DR}}_\psi)}^{-1} \left[Z^{{\rm RI}'_2,{\rm DR}}_{\gamma_{\nu_3}, \gamma_{\nu_3}} \ \Sigma_{4,1} - Z^{{\rm RI}'_2,{\rm DR}}_{\gamma_{\nu_3}, \gamma_5 \gamma_{\nu_4}} (\Sigma_{4,2} + \Sigma_{4,3} \ q_{\nu_2} - \Sigma_{4,4} \ q_{\nu_1})\right] \Big|_{q = \bar{q}} = 1.
    \label{RC2example}
\end{equation}
We see that the second choice of the projector can give a simpler combination of $\Sigma_{i,j}$, at least for the term multiplying $Z^{{\rm DR},{\rm RI}'}_{\gamma_{\nu_3}, \gamma_{\nu_3}}$. But here, the main advantage of using the second choice of projectors is that the renormalization matrices are independent of the individual components of the momentum scale, which are orthogonal to the plane in which the staple lies. Eq. \eqref{RC1example} depends explicitly on $\bar{q}_{\nu_3}$, while Eq. \eqref{RC2example} only depends on $\bar{q}_{\nu_3}$ through $\bar{q}_T^2 \equiv \bar{q}_{\nu_3}^2 + \bar{q}_{\nu_4}^2$. Indeed, the same dependence on transverse components of $\bar q$ regards all renormalization functions resulting from the choice of projectors $\mathcal{P}_{\Gamma}^{[2]}$. Consequently, the nonperturbative determination of renormalization functions in this prescription will only depend on ${\bar q}_T^2$, rather than on its individual components. This gives us the possibility of increasing statistics in the nonperturbative calculations of renormalization functions on the lattice by averaging over different $\bar{q}_{\nu_3}$ and $\bar{q}_{\nu_4}$ components which have the same $\bar{q}_T^2$, according to the residual 2D symmetry. Furthermore, we expect that by eliminating contributions coming from the form factors $\Sigma_{i,j} \ (j>2)$, hadronic contaminations, which may be present in the nonperturbative calculation of the renormalization functions, will be reduced, as it happens in the case of local quark bilinear operators.  

We also provide the one-loop result for the bare Green’s function of the Wilson loop $\langle L^{\rm DR}(z,y+y') \rangle$:
\begin{eqnarray}
    \langle L^{\rm DR}(z,y+y') \rangle &=& 1 + 8 \ \frac{g^2}{16 \pi^2} C_F \Bigg[2 + \frac{1}{\varepsilon} + 2 \gamma_E + \frac{y+y'}{z} \tan^{-1} (\frac{y+y'}{z}) + \frac{z}{y+y'} \tan^{-1} (\frac{z}{y+y'}) \nonumber \\
    && \qquad \qquad \qquad \quad + \ln (\frac{\bar{\mu}^2 z^2}{4}) - \ln (1 + \frac{z^2}{(y+y')^2}) \Bigg] + \mathcal{O} (g^4).
    \label{WLbare}
\end{eqnarray}

In the limit $y \rightarrow \infty$ (keeping $(y-y')$ fixed) the divergent term $(y+y')/z \tan^{-1} ((y+y')/z)$ in Eq. \eqref{WLbare} cancels the divergent term $[y/z \tan^{-1} (y/z) + y'/z \tan^{-1} (y'/z)]$ of $s_0$ (Eq. \eqref{s0}) when calculating $\overline{\Lambda}_{\Gamma} (q,z,y,y')$ (Eq. \eqref{Lambdabar}) for the RI$'$-bar schemes.

\subsection{Renormalization functions}

The renormalization functions of $\mathcal{O}_\Gamma$ and $\overline{\mathcal{O}}_\Gamma$ in the $\MSbar$ scheme have been determined by imposing that the $\MSbar$-renormalized Green's functions of the two operators are equal to the finite parts (exclude pole terms) of the corresponding bare Green's functions:
\begin{eqnarray}
    Z^{{\overline{\rm MS}},{\rm DR}}_{\Gamma \Gamma'} &=& \delta_{\Gamma \Gamma'} \left[ 1 - \frac{(g^{\overline{\rm MS}})^2}{16 \pi^2} C_F \frac{7}{\varepsilon} + \mathcal{O} \left((g^{\overline{\rm MS}})^4\right)\right], 
    \label{ZGMSbar}    \\
    \overline{Z}^{{\overline{\rm MS}},{\rm DR}}_{\Gamma \Gamma'} &=& \delta_{\Gamma \Gamma'} \left[ 1 - \frac{(g^{\overline{\rm MS}})^2}{16 \pi^2} C_F \frac{3}{\varepsilon} + \mathcal{O} \left((g^{\overline{\rm MS}})^4\right)\right].
    \label{ZGbarMSbar}
\end{eqnarray}
As expected, the renormalization functions in Eqs. (\ref{ZGMSbar} -- \ref{ZGbarMSbar}) are diagonal (there is no mixing) and independent of the Dirac matrices $\Gamma$, $\Gamma'$, and the lengths of the staple. In addition, the result coincides with that obtained for the symmetric staple operators in our previous work~\cite{Constantinou:2019vyb}; thus, the asymmetry in the shape of the staple does not affect the renormalization in $\MSbar$. The divergent term $7/\varepsilon$ of Eq. \eqref{ZGMSbar} comes from the sum of the pole terms in the bare Green's function \eqref{Sigma_ij} and in the renormalization factor of the external quark fields (see Appendix \ref{ap.C}). The divergent term $3/\varepsilon$ of Eq. \eqref{ZGbarMSbar} includes the additional $\mathcal{O} (1/\varepsilon)$ terms coming from the vacuum expectation value of the Wilson loop (Eq. \ref{WLbare}). In the latter case the resulting $3/\varepsilon$ term coincides with that found in the renormalization of a straight Wilson-line operator.  

In the RI$'$-type schemes (RI$'_1$, RI$'_2$, RI$'_1$-bar, RI$'_2$-bar), the renormalization functions are (in general) nondiagonal matrices due to the operator mixing. Their expressions are given below in terms of the conversion matrices that connect the RI$'$-type schemes with the $\MSbar$ scheme, as defined in the next subsection (Eq. \eqref{Conversion}): 

\begin{eqnarray}
    Z^{{\rm RI}'_i,{\rm DR}}_{\Gamma \Gamma'} &=& \delta_{\Gamma \Gamma'} + Z^{\MSbar,{\rm DR}}_{\Gamma \Gamma'} - C^{\overline{\rm MS}, {\rm RI}'_i}_{\Gamma \Gamma'} + \mathcal{O} ((g^{\overline{\rm MS}})^4), \\
    \overline{Z}^{{\rm RI}'_i,{\rm DR}}_{\Gamma \Gamma'} &=& \delta_{\Gamma \Gamma'} + \overline{Z}^{\MSbar,{\rm DR}}_{\Gamma \Gamma'} - \overline{C}^{\overline{\rm MS}, {\rm RI}'_i}_{\Gamma \Gamma'} + \mathcal{O} ((g^{\overline{\rm MS}})^4).
\end{eqnarray}  

\subsection{Conversion matrices}

By using our results for the bare Green's functions, we extract the conversion matrices that match the renormalization matrices of the staple operators from RI$'$-type (RI$'_1$, RI$'_2$, RI$'_1$-bar, RI$'_2$-bar) schemes to $\MSbar$:
\begin{equation}
Z^{\overline{\rm MS},X}_{\Gamma \Gamma'} = C^{\overline{\rm MS}, {\rm RI}'_i}_{\Gamma\Gamma''} Z^{{\rm RI}'_i,X}_{\Gamma'' \Gamma'}, \qquad \overline{Z}^{\overline{\rm MS},X}_{\Gamma \Gamma'} = \overline{C}^{\overline{\rm MS}, {\rm RI}'_i}_{\Gamma\Gamma''} \overline{Z}^{{\rm RI}'_i,X}_{\Gamma'' \Gamma'}.
\label{Conversion}
\end{equation}
The conversion matrices are regularization (X) independent, and thus, our results in DR are also applicable to the lattice. They can be used in lattice simulations in order to translate the renormalized hadron matrix elements of the staple operators, such as the quasi-beam function~\cite{Alexandrou:2023ucc} and the soft function~\cite{Li:2021wvl}, from the intermediate RI$'$-type schemes to $\MSbar$. According to the mixing pattern, we need to calculate a total of 4 (one for each mixing set $S_i$) conversion matrices of dimensions $4 \times 4$ for each RI$'$-type scheme. We found that the matrices are block diagonal in 2 sub-matrices of dimensions $2 \times 2$ at one-loop level\footnote{The form of Eqs. \eqref{Cform} and \eqref{Cgamma5} is valid for all four RI$'$-type schemes studied in this work (RI$'_1$, RI$'_2$, RI$'_1$-bar, RI$'_2$-bar).}; these blocks have the same generic structure for all $\Gamma$:
\begin{equation}
C^{\overline{\rm MS}, {\rm RI}'} = \begin{pmatrix}
    C^{\overline{\rm MS}, {\rm RI}'}_{\Gamma, \Gamma} & C^{\overline{\rm MS}, {\rm RI}'}_{\Gamma, \Gamma \gamma_{\nu_1} \gamma_{\nu_2}} & 0 & 0 \\
    C^{\overline{\rm MS}, {\rm RI}'}_{\Gamma \gamma_{\nu_1} \gamma_{\nu_2}, \Gamma} & C^{\overline{\rm MS}, {\rm RI}'}_{\Gamma \gamma_{\nu_1}\gamma_{\nu_2}, \Gamma \gamma_{\nu_1} \gamma_{\nu_2}} & 0 & 0 \\
    0 & 0 & C^{\overline{\rm MS}, {\rm RI}'}_{\Gamma \gamma_{\nu_1},\Gamma \gamma_{\nu_1}} & C^{\overline{\rm MS}, {\rm RI}'}_{\Gamma \gamma_{\nu_1}, \Gamma \gamma_{\nu_2}} \\
    0 & 0 & C^{\overline{\rm MS}, {\rm RI}'}_{\Gamma \gamma_{\nu_2}, \Gamma \gamma_{\nu_1}} & C^{\overline{\rm MS}, {\rm RI}'}_{\Gamma \gamma_{\nu_2}, \Gamma \gamma_{\nu_2}} 
\end{pmatrix} + \ \mathcal{O} ((g^{\overline{\rm MS}})^4). 
\label{Cform}
\end{equation}
Introducing a $\gamma_5$ matrix in both operator and projector leaves the conversion matrix unaffected to one loop\footnote{Different definitions of $\gamma_5$ in d dimensions do not affect the one-loop result.}:
\begin{equation}
C_{\gamma_5 \Gamma, \gamma_5 \Gamma'}^{\overline{\rm MS}, {\rm RI}'} = C_{\Gamma, \Gamma'}^{\overline{\rm MS}, {\rm RI}'} + \ \mathcal{O} ((g^{\overline{\rm MS}})^4).
\label{Cgamma5}
\end{equation}
Thus, the conversion matrix for the mixing set $S_2$ is identical to the conversion matrix for the mixing set $S_1$. Also the conversion matrices for the mixing sets $S_3$ and $S_4$ are related through the interchange $\nu_3 \leftrightarrows \nu_4$\footnote{For the second choice of projectors $\mathcal{P}_\Gamma^{[2]}$, the interchange $\nu_3 \leftrightarrows \nu_4$ is not needed.}. 

We provide below all nonzero elements of the conversion matrices for each mixing set for the RI$'_1$ and RI$'_2$ schemes; they are expressed in terms of the coefficients $s_0$ and $s_{i,j}$ given in Eqs. (\ref{s0} -- \ref{s75}) by setting $q = \bar{q}$. Here we use the notation $\varepsilon_{\rm LC} \equiv \varepsilon_{\nu_1 \nu_2 \nu_3 \nu_4}$, and $\mu = \nu_3, \nu_4$.
\begin{equation}
  C^{\overline{\rm MS}, {\rm RI}'_i}_{\Gamma \Gamma'} = \delta_{\Gamma \Gamma'} + \frac{{(g^{\overline{\rm MS}})}^2}{16 \pi^2} C_F \ \left(\delta_{\Gamma \Gamma'} \ \bmp{s_0} + \bmr{c^{[i]}_{\Gamma, \Gamma'}} \right) + \ \mathcal{O} ((g^{\overline{\rm MS}})^4), \label{Cmatrix} 
\end{equation}
where for the RI$'_1$ scheme ($i=1$):
\begin{eqnarray}
&&\bmr{c^{[1]}_{\openone, \openone}} \qquad \quad \, = \ \bmr{c^{[1]}_{\gamma_5, \gamma_5}} \qquad \ \  = \ \bmp{s_{1,1}} + \bmp{s_{1,3}} \ \bar{q}_{\nu_1} + \bmp{s_{1,4}} \ \bar{q}_{\nu_2}, \\
&&\bmr{c^{[1]}_{\openone, \sigma_{\nu_1 \nu_2}}} \quad \, \ = \ \varepsilon_{\rm LC} \ \bmr{c^{[1]}_{\gamma_5, \sigma_{\nu_4 \nu_3}}} = \ \bmp{s_{1,2}} + \bmp{s_{1,3}} \ \bar{q}_{\nu_2} - \bmp{s_{1,4}} \ \bar{q}_{\nu_1}, \\
&&\bmr{c^{[1]}_{\sigma_{\nu_1 \nu_2}, \openone}} \quad \ \, \, = \ \varepsilon_{\rm LC} \ \bmr{c^{[1]}_{\sigma_{\nu_4 \nu_3}, \gamma_5}} = \ \bmp{s_{5,2}} + \bmp{s_{5,3}} \ \bar{q}_{\nu_1} + \bmp{s_{5,4}} \ \bar{q}_{\nu_2}, \\
&&\bmr{c^{[1]}_{\sigma_{\nu_1 \nu_2}, \sigma_{\nu_1 \nu_2}}} = \ \bmr{c^{[1]}_{\sigma_{\nu_4 \nu_3}, \sigma_{\nu_4 \nu_3}}} = \ \bmp{s_{5,1}} + \bmp{s_{5,3}} \ \bar{q}_{\nu_2} - \bmp{s_{5,4}} \ \bar{q}_{\nu_1}, \\
&&\bmr{c^{[1]}_{\gamma_{\nu_1}, \gamma_{\nu_1}}} \quad \ \ \, = \ \bmr{c^{[1]}_{\gamma_5 \gamma_{\nu_1}, \gamma_5 \gamma_{\nu_1}}} = \ \bmp{s_{2,1}} + \bmp{s_{2,3}} \ \bar{q}_{\nu_1} + \bmp{s_{2,4}} \ \bar{q}_{\nu_2}, \\
&&\bmr{c^{[1]}_{\gamma_{\nu_1}, \gamma_{\nu_2}}} \quad \ \ \, = \ \bmr{c^{[1]}_{\gamma_5 \gamma_{\nu_1}, \gamma_5 \gamma_{\nu_2}}} = \ \bmp{s_{2,2}} + \bmp{s_{2,3}} \ \bar{q}_{\nu_2} - \bmp{s_{2,4}} \ \bar{q}_{\nu_1}, \\
&&\bmr{c^{[1]}_{\gamma_{\nu_2}, \gamma_{\nu_1}}} \quad \ \ \, = \ \bmr{c^{[1]}_{\gamma_5 \gamma_{\nu_2}, \gamma_5 \gamma_{\nu_1}}} = \ \bmp{s_{3,2}} + \bmp{s_{3,3}} \ \bar{q}_{\nu_1} + \bmp{s_{3,4}} \ \bar{q}_{\nu_2}, \\
&&\bmr{c^{[1]}_{\gamma_{\nu_2}, \gamma_{\nu_2}}} \quad \ \ \, = \ \bmr{c^{[1]}_{\gamma_5 \gamma_{\nu_2}, \gamma_5 \gamma_{\nu_2}}} = \ \bmp{s_{3,1}} + \bmp{s_{3,3}} \ \bar{q}_{\nu_2} - \bmp{s_{3,4}} \ \bar{q}_{\nu_1}, \\
&&\bmr{c^{[1]}_{\gamma_{\mu}, \gamma_{\mu}}} \qquad \ \, = \ \bmr{c^{[1]}_{\gamma_5 \gamma_{\mu}, \gamma_5 \gamma_{\mu}}} \ \ = \ \bmp{s_{4,1}} + \bmp{s_{4,5}} \ \bar{q}_{\nu_1} + \bmp{s_{4,6}} \ \bar{q}_{\nu_2} + \bmp{s_{4,7}} \ \bar{q}_{\mu}^2, \\
&&\bmr{c^{[1]}_{\gamma_{\nu_3}, \gamma_5 \gamma_{\nu_4}}} \ \, \, \, \, = \ \bmr{c^{[1]}_{\gamma_5 \gamma_{\nu_3}, \gamma_{\nu_4}}} \ \ \, = - \bmr{c^{[1]}_{\gamma_{\nu_4}, \gamma_5 \gamma_{\nu_3}}} = - \bmr{c^{[1]}_{\gamma_5 \gamma_{\nu_4}, \gamma_{\nu_3}}} = \varepsilon_{\rm LC} (\bmp{s_{4,2}} + \bmp{s_{4,5}} \ \bar{q}_{\nu_2} - \bmp{s_{4,6}} \ \bar{q}_{\nu_1}), \\
&&\bmr{c^{[1]}_{\sigma_{\mu \nu_1}, \sigma_{\mu \nu_1}}} \ \ \, = \ \bmp{s_{6,1}} + \bmp{s_{6,5}} \ \bar{q}_{\nu_1} + \bmp{s_{6,6}} \ \bar{q}_{\nu_2} - \bmp{s_{6,7}} \ \bar{q}_{\mu}^2, \\
&&\bmr{c^{[1]}_{\sigma_{\mu \nu_1}, \sigma_{\mu \nu_2}}}  \ \ \, = \ \bmp{s_{6,2}} + \bmp{s_{6,5}} \ \bar{q}_{\nu_2} - \bmp{s_{6,6}} \ \bar{q}_{\nu_1} - \bmp{s_{6,8}} \ \bar{q}_{\mu}^2, \\
&&\bmr{c^{[1]}_{\sigma_{\mu \nu_2}, \sigma_{\mu \nu_1}}}  \ \ \, = \ \bmp{s_{7,2}} + \bmp{s_{7,5}} \ \bar{q}_{\nu_1} + \bmp{s_{7,6}} \ \bar{q}_{\nu_2} - \bmp{s_{7,7}} \ \bar{q}_{\mu}^2, \\
&&\bmr{c^{[1]}_{\sigma_{\mu \nu_2}, \sigma_{\mu \nu_2}}}  \ \ \, = \ \bmp{s_{7,1}} + \bmp{s_{7,5}} \ \bar{q}_{\nu_2} - \bmp{s_{7,6}} \ \bar{q}_{\nu_1} - \bmp{s_{7,8}} \ \bar{q}_{\mu}^2,
\end{eqnarray}
and for the RI$'_2$ scheme ($i=2$):
\begin{eqnarray}
&&\bmr{c^{[2]}_{\openone, \openone}} \qquad \quad \, = \ \bmr{c^{[2]}_{\gamma_5, \gamma_5}} \qquad \ \  = \ \bmp{s_{1,1}}, \\
&&\bmr{c^{[2]}_{\openone, \sigma_{\nu_1 \nu_2}}} \quad \, \ = \ \varepsilon_{\rm LC} \ \bmr{c^{[2]}_{\gamma_5, \sigma_{\nu_4 \nu_3}}} = \ \bmp{s_{1,2}}, \\
&&\bmr{c^{[2]}_{\sigma_{\nu_1 \nu_2}, \openone}} \quad \ \, \, = \ \varepsilon_{\rm LC} \ \bmr{c^{[2]}_{\sigma_{\nu_4 \nu_3}, \gamma_5}} = \ \bmp{s_{5,2}}, \\
&&\bmr{c^{[2]}_{\sigma_{\nu_1 \nu_2}, \sigma_{\nu_1 \nu_2}}} = \ \bmr{c^{[2]}_{\sigma_{\nu_4 \nu_3}, \sigma_{\nu_4 \nu_3}}} = \ \bmp{s_{5,1}}, \\
&&\bmr{c^{[2]}_{\gamma_{\nu_1}, \gamma_{\nu_1}}} \quad \ \ \, = \ \bmr{c^{[2]}_{\gamma_5 \gamma_{\nu_1}, \gamma_5 \gamma_{\nu_1}}} = \ \bmp{s_{2,1}}, \\
&&\bmr{c^{[2]}_{\gamma_{\nu_1}, \gamma_{\nu_2}}} \quad \ \ \, = \ \bmr{c^{[2]}_{\gamma_5 \gamma_{\nu_1}, \gamma_5 \gamma_{\nu_2}}} = \ \bmp{s_{2,2}}, \\
&&\bmr{c^{[2]}_{\gamma_{\nu_2}, \gamma_{\nu_1}}} \quad \ \ \, = \ \bmr{c^{[2]}_{\gamma_5 \gamma_{\nu_2}, \gamma_5 \gamma_{\nu_1}}} = \ \bmp{s_{3,2}}, \\
&&\bmr{c^{[2]}_{\gamma_{\nu_2}, \gamma_{\nu_2}}} \quad \ \ \, = \ \bmr{c^{[2]}_{\gamma_5 \gamma_{\nu_2}, \gamma_5 \gamma_{\nu_2}}} = \ \bmp{s_{3,1}}, \\
&&\bmr{c^{[2]}_{\gamma_{\mu}, \gamma_{\mu}}} \qquad \ \, = \ \bmr{c^{[2]}_{\gamma_5 \gamma_{\mu}, \gamma_5 \gamma_{\mu}}} \ \ = \ \bmp{s_{4,1}}, \\
&&\bmr{c^{[2]}_{\gamma_{\nu_3}, \gamma_5 \gamma_{\nu_4}}} \ \, \, \, \, = \ \bmr{c^{[2]}_{\gamma_5 \gamma_{\nu_3}, \gamma_{\nu_4}}} \ \ \, = - \bmr{c^{[2]}_{\gamma_{\nu_4}, \gamma_5 \gamma_{\nu_3}}} = - \bmr{c^{[2]}_{\gamma_5 \gamma_{\nu_4}, \gamma_{\nu_3}}} = \varepsilon_{\rm LC} (\bmp{s_{4,2}} + \bmp{s_{4,3}} \ \bar{q}_{\nu_2} - \bmp{s_{4,4}} \ \bar{q}_{\nu_1}), \\
&&\bmr{c^{[2]}_{\sigma_{\mu \nu_1}, \sigma_{\mu \nu_1}}} \ \ \, = \ \bmp{s_{6,1}}, \\
&&\bmr{c^{[2]}_{\sigma_{\mu \nu_1}, \sigma_{\mu \nu_2}}}  \ \ \, = \ \bmp{s_{6,2}}, \\
&&\bmr{c^{[2]}_{\sigma_{\mu \nu_2}, \sigma_{\mu \nu_1}}}  \ \ \, = \ \bmp{s_{7,2}}, \\
&&\bmr{c^{[2]}_{\sigma_{\mu \nu_2}, \sigma_{\mu \nu_2}}}  \ \ \, = \ \bmp{s_{7,1}}. \label{Clast}
\end{eqnarray}

The conversion matrices for the RI$'_1$-bar and RI$'_2$-bar schemes are related to those of RI$'_1$ and RI$'_2$ schemes, respectively, through:
\begin{eqnarray}
\overline{C}^{\overline{\rm MS}, {\rm RI}'_i}_{\Gamma \Gamma'} &=& C^{\overline{\rm MS}, {\rm RI}'_i}_{\Gamma \Gamma'} / \langle L^{\overline{\rm MS}} (z, y+y') \rangle^{1/2} \nonumber \\
&=& C^{\overline{\rm MS}, {\rm RI}'_i}_{\Gamma \Gamma'} - 4 \ \delta_{\Gamma \Gamma'} \frac{(g^{\overline{\rm MS}})^2}{16 \pi^2} C_F \Bigg[2 + 2 \gamma_E + \frac{y+y'}{z} \tan^{-1} (\frac{y+y'}{z}) + \frac{z}{y+y'} \tan^{-1} (\frac{z}{y+y'}) \nonumber \\
    && \qquad \qquad \qquad \qquad \qquad \qquad \ + \ln (\frac{\bar{\mu}^2 z^2}{4}) - \ln (1 + \frac{z^2}{(y+y')^2}) \Bigg] + \mathcal{O} ((g^{\overline{\rm MS}})^4).
    \label{Cbarmatrix}
\end{eqnarray}
As observed, the conversion matrices in the RI$'_i$ and RI$'_i$-bar schemes exhibit differences in their diagonal elements while remaining unchanged in the nondiagonal elements up to one loop.

There is a nontrivial dependence of the conversion matrices (as well as of the bare Green's functions) on the staple lengths ($z, y, y'$) and momentum scale $\bar{q}$ leading to singular limits at vanishing or infinite values of these parameters due to the appearance of contact terms beyond tree level. In particular, the limit $z \rightarrow 0$, despite the fact that classically it results in a straight Wilson line of length $(y-y')$, gives rise to a linear $\sim 1/|z|$ and a logarithmic $\sim \ln (z^2)$ divergence in the conversion matrices of $\mathcal{O}_\Gamma$ coming from the cusp and pinch-pole divergent terms of the original operator: 
\begin{equation}
    C^{\overline{\rm MS}, {\rm RI}'_i}_{\Gamma \Gamma'} \xrightarrow[]{z \rightarrow 0} \ 4 \ \delta_{\Gamma \Gamma'} \ \frac{{(g^\MSbar)}^2}{16 \pi^2} \ C_F \ \left[ \frac{\pi}{2} \ \frac{|y| + |y'| - |y-y'|}{|z|} + \ln (\bar{\mu}^2 z^2) \right] + \mathcal{O} (z^0) + \mathcal{O} ((g^{\overline{\rm MS}})^4).
    \label{vanishingz}
\end{equation}

Setting $y' = y$, and considering the limit $y \rightarrow 0$, where the staple becomes a straight line of length $z$, cusp points do not vanish smoothly giving logarithmic divergences $\sim \ln (y^2)$:
\begin{equation}
    C^{\overline{\rm MS}, {\rm RI}'_i}_{\Gamma \Gamma'} \xrightarrow[]{y' = y, \ y \rightarrow 0} \ -4 \ \delta_{\Gamma \Gamma'} \ \frac{{(g^\MSbar)}^2}{16 \pi^2} \ C_F \ \ln (\bar{\mu}^2 y^2) + \mathcal{O} (y^0) + \mathcal{O} ((g^{\overline{\rm MS}})^4).
    \label{vanishingy}
\end{equation}
As discussed in Sec. \ref{GFsDR}, the limit $y \rightarrow \infty$, when $(y-y')$ is fixed, results in a pinch-pole linear singularity ($\sim |y|$):  
\begin{equation}
    C^{\overline{\rm MS}, {\rm RI}'_i}_{\Gamma \Gamma'} \xrightarrow[]{y \rightarrow \infty, \ (y-y') {\rm \ fixed}} \ -8 \ \delta_{\Gamma \Gamma'} \ \frac{{(g^\MSbar)}^2}{16 \pi^2} \ C_F \ \frac{\pi}{2} \ \frac{|y|}{|z|} + \mathcal{O} (y^0) + \mathcal{O} ((g^{\overline{\rm MS}})^4),
\end{equation}
which is eliminated in $\overline{C}^{\overline{\rm MS}, {\rm RI}'_i}_{\Gamma \Gamma'}$. In contrast to the aforementioned limits, the limit $(y-y') \rightarrow 0$, in which the asymmetric staple Wilson line becomes symmetric, is not singular, and thus, our present results can reproduce our previous results in Ref.~\cite{Constantinou:2019vyb} for the case of symmetric staples. Furthermore, the limit $\bar{q}^2 \rightarrow 0$ is nonsmooth for some of the operators $\mathcal{O}_\Gamma$ giving a logarithmic divergence ($\sim \ln (\bar{q}^2)$), as follows:  
\begin{equation}
    C^{\overline{\rm MS}, {\rm RI}'_i}_{\Gamma \Gamma'} \xrightarrow[]{\bar{q}^2 \rightarrow 0} \ \delta_{\Gamma \Gamma'} \ d_\Gamma \ \frac{{(g^\MSbar)}^2}{16 \pi^2} \ C_F \ \ln (\frac{\bar{q}^2}{\bar{\mu}^2}) + \mathcal{O} ((\bar{q}^2)^0) + \mathcal{O} ((g^{\overline{\rm MS}})^4),
\end{equation}
where $d_\Gamma = -3$ for ($\Gamma = \openone, \gamma_5$), 1 for ($\Gamma = \sigma_{\nu_i \nu_j}$), and 0 for ($\Gamma = \gamma_{\nu_i}, \gamma_5 \gamma_{\nu_i}$). Note that the limit is taken by rescaling simultaneously all components of the 4-vector scale $\bar{q}$. Moreover, the limit $\bar{q}^2 \rightarrow \infty$ also gives a logarithmic divergence, which is independent of the operator $\mathcal{O}_\Gamma$:
\begin{equation}
    C^{\overline{\rm MS}, {\rm RI}'_i}_{\Gamma \Gamma'} \xrightarrow[]{\bar{q}^2 \rightarrow \infty} \ 7 \ \delta_{\Gamma \Gamma'} \ \frac{{(g^\MSbar)}^2}{16 \pi^2} \ C_F \ \ln (\frac{\bar{q}^2}{\bar{\mu}^2}) + \mathcal{O} ((\bar{q}^2)^0) + \mathcal{O} ((g^{\overline{\rm MS}})^4).
\end{equation}

Another property of the conversion matrices comes from the combination of $\mathcal{P}_\mu$, $\mathcal{T}_\mu, \mathcal{C}$ symmetries: one can prove (see Eq. \eqref{Lambda_symmetry}) that the real (imaginary) parts of the conversion matrices are even (odd) under ($z \rightarrow -z$, $y \rightarrow -y$, $y' \rightarrow -y'$). This is confirmed by our one-loop computation. 

\begin{figure}
    \centering  
    \includegraphics[width=\textwidth]{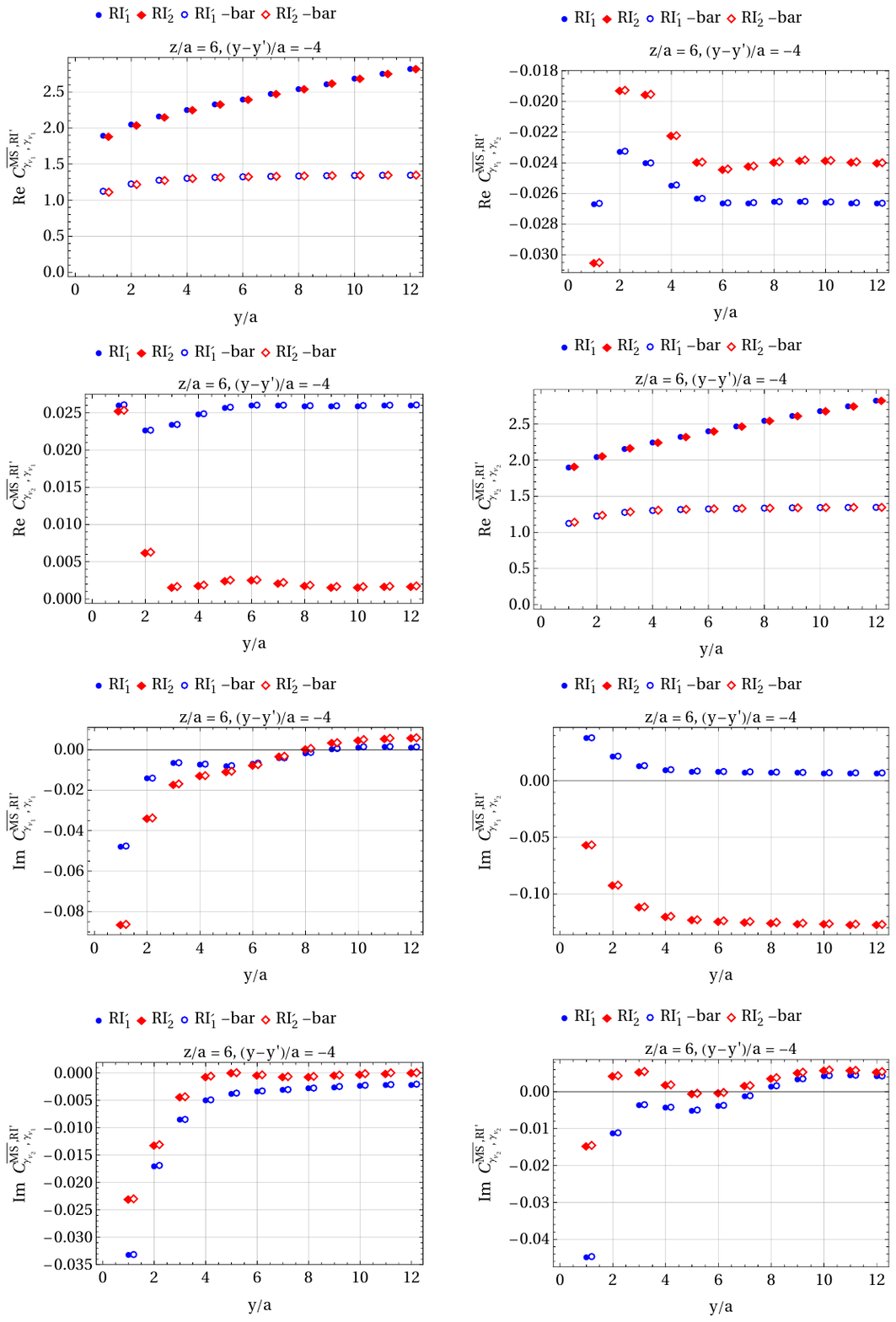} 
    \caption{Real and imaginary parts of the conversion matrix elements $C^{\overline{\rm MS}, {\rm RI}'}_{\Gamma \Gamma'}$, for $\Gamma, \Gamma' = \gamma_{\nu_1}, \gamma_{\nu_2}$ and for the four RI$'$-type schemes: RI$'_1$, RI$'_2$, RI$'_1$-bar, RI$'_2$-bar, as functions of $y/a$ [$\bar{\mu} = 2$ GeV, $\beta = 1$ (Landau gauge), $a \bar{q} = (\frac{2 \pi}{L} n_1, \frac{2 \pi}{L} n_2, \frac{2 \pi}{L} n_3, \frac{2 \pi}{T} (n_4 + \frac{1}{2}))$, $n_1=n_2=n_3=4$, $n_4=5$, $L=32$, $T=64$ and $a=0.09$ fm].}
    \label{fig:plot1}
\end{figure}
\begin{figure}
   \centering  
    \includegraphics[width=\textwidth]{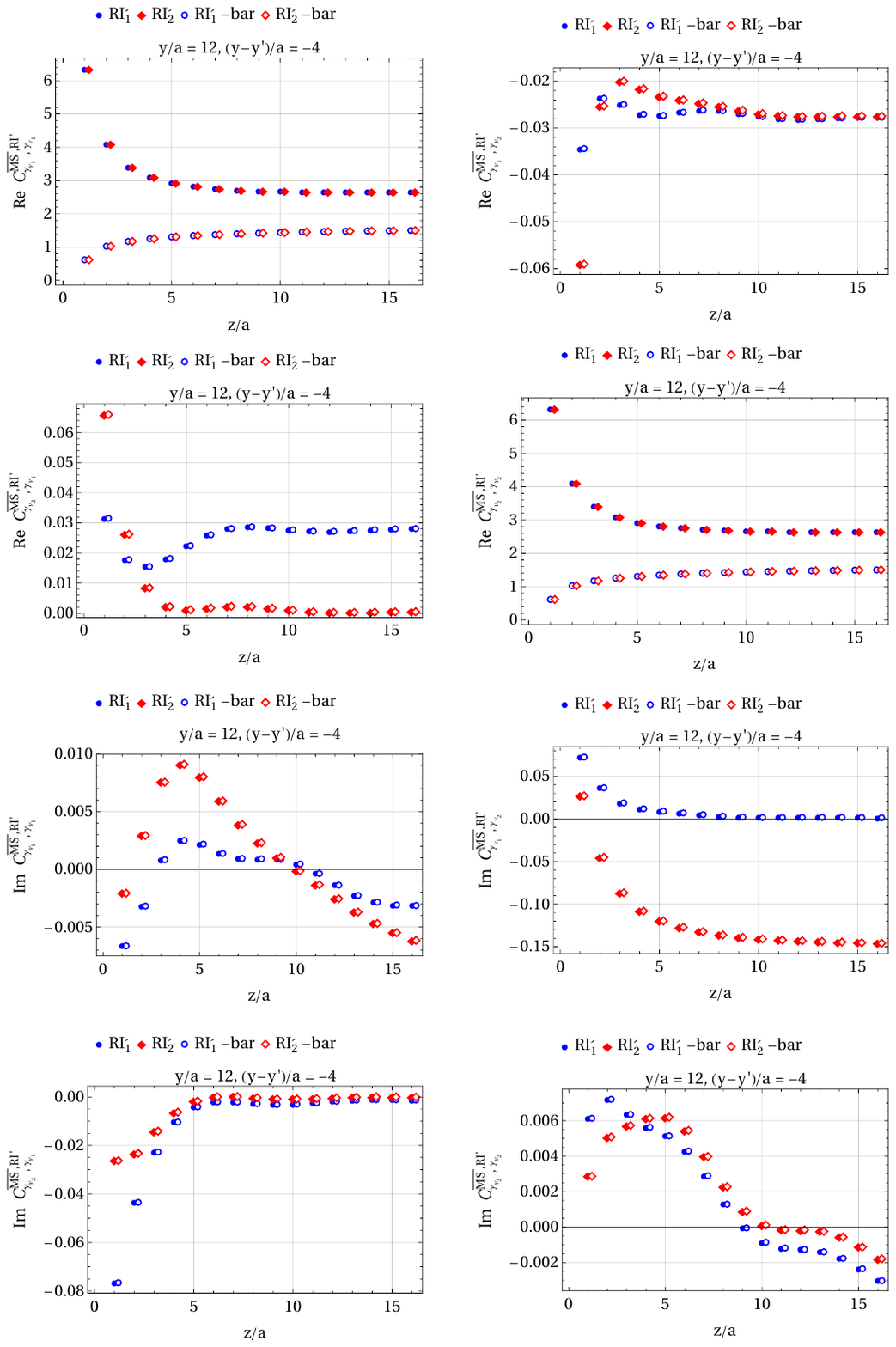}
    \caption{Real and imaginary parts of the conversion matrix elements $C^{\overline{\rm MS}, {\rm RI}'}_{\Gamma \Gamma'}$, for $\Gamma, \Gamma' = \gamma_{\nu_1}, \gamma_{\nu_2}$ and for the four RI$'$-type schemes: RI$'_1$, RI$'_2$, RI$'_1$-bar, RI$'_2$-bar, as functions of $z/a$ [$\bar{\mu} = 2$ GeV, $\beta = 1$ (Landau gauge), $a \bar{q} = (\frac{2 \pi}{L} n_1, \frac{2 \pi}{L} n_2, \frac{2 \pi}{L} n_3, \frac{2 \pi}{T} (n_4 + \frac{1}{2}))$, $n_1=n_2=n_3=4$, $n_4=5$, $L=32$, $T=64$ and $a=0.09$ fm].}
    \label{fig:plot2}
\end{figure}
\begin{figure}
   \centering  
    \includegraphics[width=\textwidth]{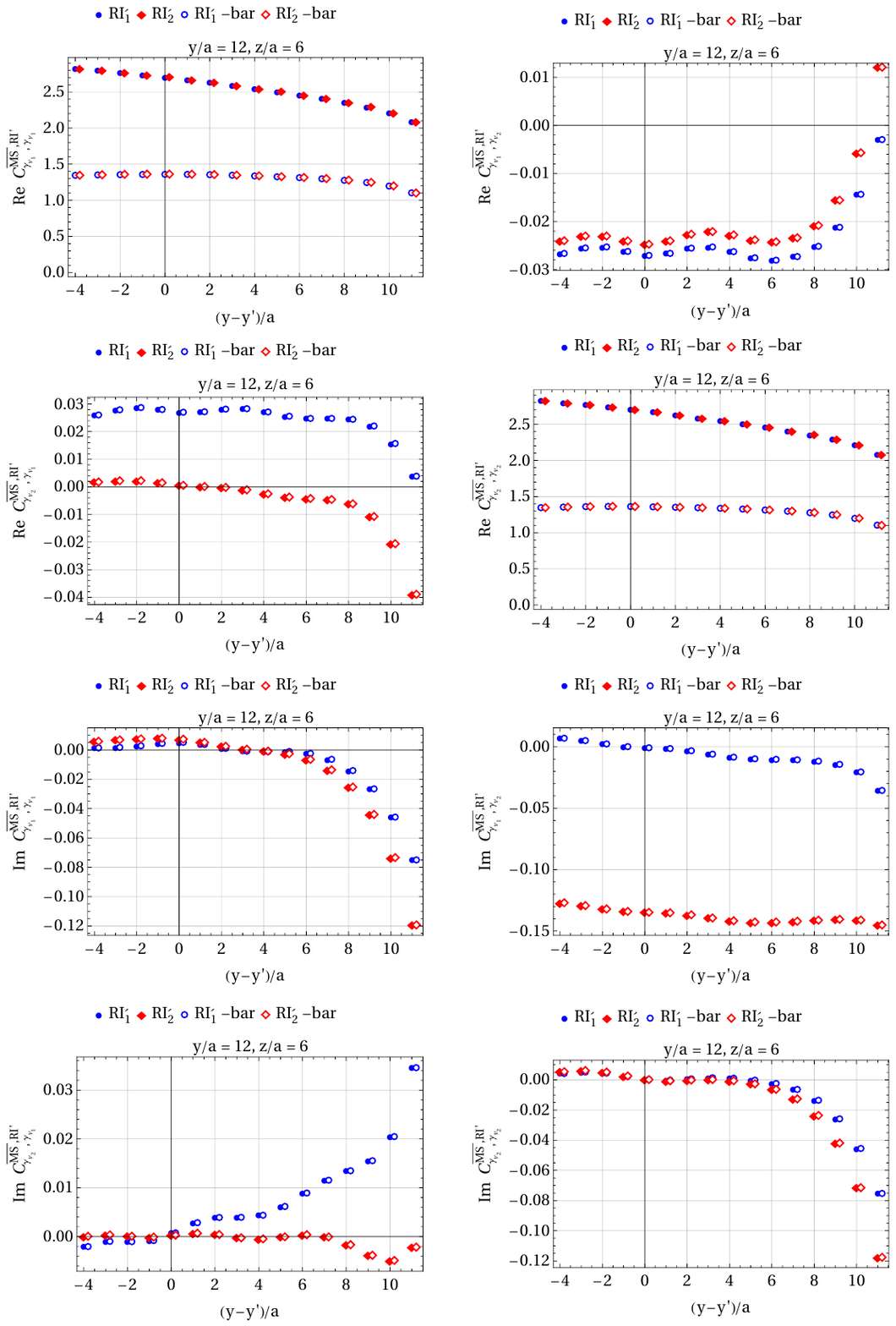}
    \caption{Real and imaginary parts of the conversion matrix elements $C^{\overline{\rm MS}, {\rm RI}'}_{\Gamma \Gamma'}$, for $\Gamma, \Gamma' = \gamma_{\nu_1}, \gamma_{\nu_2}$ and for the four RI$'$-type schemes: RI$'_1$, RI$'_2$, RI$'_1$-bar, RI$'_2$-bar, as functions of $(y-y')/a$ [$\bar{\mu} = 2$ GeV, $\beta = 1$ (Landau gauge), $a \bar{q} = (\frac{2 \pi}{L} n_1, \frac{2 \pi}{L} n_2, \frac{2 \pi}{L} n_3, \frac{2 \pi}{T} (n_4 + \frac{1}{2}))$, $n_1=n_2=n_3=4$, $n_4=5$, $L=32$, $T=64$ and $a=0.09$ fm].}
    \label{fig:plot3}
\end{figure}

We illustrate our results for the conversion matrices in the plots of Figs. (\ref{fig:plot1} -- \ref{fig:plot3}) by employing certain values of the free parameters used in lattice simulations: $\bar{\mu} = 2$ GeV, $\beta = 1$ (Landau gauge), $a \bar{q} = (\frac{2 \pi}{L} n_1, \frac{2 \pi}{L} n_2, \frac{2 \pi}{L} n_3, \frac{2 \pi}{T} (n_4 + \frac{1}{2}))$, where $a$ is the lattice spacing, $L^3 \times T$ is the lattice volume and $n_i \in \mathbb{Z}$. Following simulations by ETMC, we choose isotropic momentum scales in the spatial directions ($n_1 = n_2 = n_3$) and a nonzero twist of $1/2$ in the temporal component; the latter choice is compatible with the antiperiodic boundary conditions applied on the fermion fields in the temporal direction. As an example, we apply $n_1=n_2=n_3=4$, $n_4=5$, $L=32$, $T=64$ and $a=0.09$ fm. We note that specific choices of $n_i$ can lead to a vanishing imaginary part. In particular, by setting to zero the two momentum components parallel to the staple segments, the one-loop expression for the conversion matrices becomes purely real. However, such a choice of momentum gives rise to unwanted Lorentz noninvariant contributions in the nonperturbative calculations. Thus, in our example, we follow the common practice of employing democratic momenta with reduced Lorentz noninvariant contributions at the cost of introducing an imaginary part in the RI$'$-type renormalization matrices and the conversion matrices. In Figs. (\ref{fig:plot1} -- \ref{fig:plot3}), we examine the dependence of some representative conversion matrix elements on the Wilson-line scales $y$, $z$, and $y-y'$ (in lattice units), respectively. In particular, we plot the real and imaginary parts of the $2 \times 2$ block $C^{\overline{\rm MS}, {\rm RI}'}_{\Gamma \Gamma'}$, $(\Gamma, \Gamma' = \gamma_{\nu_1}, \gamma_{\nu_2})$, for the four renormalization prescriptions: RI$'_1$, RI$'_2$, RI$'_1$-bar, RI$'_2$-bar.

As shown in Fig.~\ref{fig:plot1}, the real part of the diagonal elements (first plot in the left column and second plot in the right column) has an almost linear (flat) dependence on $y/a$ in the RI$'_i$ (RI$'_i$-bar) schemes due to the presence (absence) of the pinch-pole singularity. As expected, the convergent behavior of RI$'_i$-bar appears for large values of $y/a$ ($\gtrsim 6$). There are no significant differences in the real diagonal elements between schemes with index 1 and 2. The real part of the nondiagonal elements (first plot in the second column and second plot in the first column of Fig.~\ref{fig:plot1}) also converges for large values of $y/a$ ($\gtrsim 8$) for all RI$'$-type schemes. Note that the nondiagonal elements are identical between RI$'_i$ and RI$'_i$-bar schemes (for the same $i$) at one loop. Differences between schemes with index 1 and 2 are now visible: the real nondiagonal elements for the RI$'_2$ scheme have smaller absolute values compared to RI$'_1$, which are closer to zero. The relative size of the real nondiagonal elements compared to the diagonal ones is $\sim 20 \%$. The imaginary part of the conversion matrix elements (the four plots in the last two rows of Fig.~\ref{fig:plot1}) gives a much milder contribution compared to the real one ($\lesssim 10 \%$). The imaginary (diagonal and nondiagonal) elements for RI$'_i$-bar coincide with the corresponding elements for RI$'_i$ (for the same $i$) at one loop. A plateau is observed at large values of $y/a$; a more stable behavior is seen for the nondiagonal elements. As in the case of real nondiagonal elements, noticeable distinctions can also be spotted in the imaginary parts between schemes with index 1 and 2. However, there is no consistent pattern regarding which scheme yields smaller contributions, as it varies for each element.

Similar conclusions are extracted from Fig.~\ref{fig:plot2} by considering the dependence of the conversion matrix elements on $z/a$. Here, we observe a convergent dependence for $z/a \gtrsim 6$ for all conversion matrix elements, except for the imaginary diagonal parts (the four plots in the last two rows of Fig.~\ref{fig:plot2}). Now, the real diagonal parts have flat behavior in both RI$'_i$ and RI$'_i$-bar schemes since the pinch-pole singularity does not arise for large values of $z/a$.

Examining the dependence of the conversion matrix elements on $(y-y')/a$ in Fig.~\ref{fig:plot3}, we conclude that a more flat behavior is observed for smaller values of $(y-y')/a$, while for larger values the elements decrease (in most cases) rapidly. An almost linear dependence on $(y-y')/a$ with a negative slope is obtained in the real diagonal parts of the RI$'_i$ schemes (see the first plot in the first column and the second plot in the second column of Fig.~\ref{fig:plot3}), coming from the pinch-pole divergent term.

In summary, the conversion matrices exhibit significant contributions from the real diagonal components, whereas the imaginary diagonal and nondiagonal elements make comparatively milder yet perceptible contributions. Safer conclusions by comparing the different types of renormalization schemes can be obtained when combining the conversion matrices with the nonperturbative data. 

\section{Calculation in the lattice regularization}
\label{ResultsLR}

\subsection{Green's functions}

In the lattice calculation, we employ the Wilson/clover fermion action (see \cite{Sheikholeslami:1985ij}) and a family of gluon Symanzik improved actions \cite{Horsley:2004mx} of the form:
\begin{equation}
\hspace{-1cm}
S_G=\frac{2}{g^2} \Bigl[ c_0 \sum_{\rm plaq.} {\rm Re\,Tr\,}\{1-U_{\rm plaq.}\} \,+\, c_1 \sum_{\rm rect.} {\rm Re \, Tr\,}\{1- U_{\rm rect.}\} \Bigr]\,,
\label{Symanzik}
\end{equation}
where $U_{\rm plaq.}$ and $U_{\rm rect.}$ are the standard 4-link ``plaquette'', and 6-link ``$2{\times}1$ rectangle'' Wilson loops. We selected three of the most common choices of the Symanzik coefficients $c_i$, called Wilson, Tree-Level Symanzik, and Iwasaki gluon actions, as shown in Table \ref{tab1}.
\begin{table}[!ht]
\centering
\begin{tabular}{|l|l|l|}
\hline
\ \textbf{Gluon action} \ & \ $\boldsymbol{c_0}$ \ & \ $\boldsymbol{c_1}$ \ \\
\hline
\ Wilson \ & \ 1 \ & \ \ 0 \ \\
\ Tree-Level Symanzik \ & \ 5/3 \ & \ -1/12 \ \\
\ Iwasaki \ & \  3.648 \ & \ -0.331$\,$ \ \\
\hline
\end{tabular}
\caption{Selected sets of values for the Symanzik coefficients. In all cases, they satisfy $c_0 + 8 c_1 = 1$.}
\label{tab1}
\end{table} 

Since we consider mass-independent renormalization schemes, we set the quark mass equal to zero. Consequently, the results from this study are also applicable to the twisted mass fermions \cite{Shindler:2007vp} in the chiral limit. One should, however, keep in mind that, in going from the twisted basis to the physical basis, operator identifications are modified (e.g., the scalar density, under ``maximal twist'', turns into a pseudoscalar density, etc.).

The results for the one-loop lattice bare Green’s functions of the asymmetric staple operators $\Lambda_\Gamma^{\rm LR}$ are presented below in terms of the $\MSbar$-renormalized Green’s functions, derived by the corresponding calculation in DR. The methodology for calculating the one-loop momentum integrals on the lattice is described in Refs.~\cite{Constantinou:2017sej,Constantinou:2019vyb}. Here, we cite results for a general Wilson-line lattice operator with $n$ cusps [$(n+1)$ segments] and no self-intersections, as calculated in our previous study considering symmetric staples~\cite{Constantinou:2019vyb}. We confirm that the formula constructed in the latter publication gives the correct result for the difference between the bare lattice Green’s functions and the $\MSbar$-renormalized Green’s functions, also in the case of asymmetric staple operators studied in this work. Thus, for a general Wilson-line lattice operator we have:
\begin{eqnarray}
\delta \Lambda_\Gamma &\equiv& \Lambda_\Gamma^{\rm LR} (q,r,n,\{\ell_1, \ldots, \ell_{n+1}\},\hat{\nu}_i,\hat{\nu}_f) - \Lambda_\Gamma^{\MSbar} (q,r,n,\{\ell_1, \ldots, \ell_{n+1}\},\hat{\nu}_i,\hat{\nu}_f) \nonumber \\
    &=& - \frac{g^2\,C_F}{16\,\pi^2}\, e^{i\,{\sf q} \cdot {\sf r}}\, \Bigg\{ 2 \ \Gamma \left[ \boldsymbol{\alpha_1} + 16 \pi^2 P_2 \, \beta + (1 - \beta) \log (a^2 \bar{\mu}^2)\right] + \frac{1}{2} (\Gamma \hat{\slashed{\nu}}_i + \hat{\slashed{\nu}}_{\hspace{-0.4mm}f} \Gamma) (\boldsymbol{\alpha_2} + \boldsymbol{\alpha_3} c_{SW}) \nonumber \\
    && \qquad \qquad \qquad \ \ + \Gamma \Big[ (n + 1) \boldsymbol{\alpha_4} + n \boldsymbol{\alpha_5} - 16 \pi^2 P_2 \, \beta + \left( 2 (n + 1) + \beta \right) \log (a^2 \bar{\mu}^2) + \boldsymbol{\alpha_6} \frac{l}{a} \Big] \Bigg\} + \mathcal{O} (g^4), \qquad
    \label{general_WL}
\end{eqnarray}
where $r$ is the 4-vector that connects the two endpoints of the Wilson line, $\ell_j$ is the length of the $j^{\rm th}$ straight-line segment and $l \equiv \sum_{j=1}^{n+1} \ell_j$ is the total length of the Wilson line, $c_{SW}$ is the clover coefficient in the fermion action, and $\hat{\nu}_{_i} \ (\hat{\nu}_{_{\hspace{-0.5mm}f}})$ is the direction of the Wilson line in the initial (final) endpoint. $P_2 = 0.02401318111946489(1)$ \cite{Luscher:1995np} and $\alpha_i$ are numerical constants which depend on the gluon action and the Wilson parameter of the fermion action $\mathcalboondox{r}$; their values are given in Table \ref{tab:stapleLR} for the Wilson, Tree-level Symanzik and Iwasaki gluon actions and for $\mathcalboondox{r} = 1$. The first two terms in the curly brackets of Eq. \eqref{general_WL} come from the sum of Feynman diagrams $d_2$ and $d_3$, while the last term comes from diagram $d_4$. The expression for the case under study (asymmetric staple operator with two cusps) can be extracted by setting $n=2$, $l = |z| + |y| + |y'|$, $\hat{\nu}_i = {\rm sgn} (y) \ \hat{\nu}_2$, and $\hat{\nu}_f = -{\rm sgn} (y') \ \hat{\nu}_2$. The corresponding $\MSbar$-renormalized Green's function $\Lambda_\Gamma^{\MSbar}$ can be read from Eqs. (\ref{Lambda_S} -- \ref{s75}) by removing $1/\varepsilon$ terms.

\begin{table}[thb]
  \centering
  \begin{tabular}{|l|l|l|l|l|l|l|}
  \hline
\ \textbf{Gluon action} & \ \qquad \ $\boldsymbol{\alpha_1}$ & \ \qquad \ $\boldsymbol{\alpha_2}$ & \ \qquad \ $\boldsymbol{\alpha_3}$ & \ \quad \ \ $\boldsymbol{\alpha_4}$ & \ \quad \ \ \ $\boldsymbol{\alpha_5}$ & \ \qquad \ $\boldsymbol{\alpha_6}$\\
\hline
\hline
\ Wilson & \ -4.464066(5) \ & \ \ \ 14.44991(1) \ & \ \ -8.284666(8) \ & \ -4.52575(1) \ & \ \ \qquad 0 \ & \ \ 19.95484(2) \ \\
\ Tree-Level Symanzik \ & \ -4.341269(5) \ & \ \ \ 12.75582(1) \ & \ \ -7.673556(8) \ & \ -3.93028(1) \ & \ -0.809890(1) \ & \ \ 17.29374(2) \ \\ 
\ Iwasaki & \ -4.163735(5) \ & \ \ \ \phantom{0}9.93653(1) \ & \ \ -6.527638(6) \ & \ -1.90532(1) \ & \ -2.101083(2) \ & \ \ 12.97809(1) \ \\
\hline
  \end{tabular}
  \caption{Numerical values of the coefficients $\alpha_1 - \alpha_6$ appearing in the difference $\delta \Lambda_\Gamma$ of Eq. \eqref{general_WL} for $\mathcalboondox{r} = 1$. A systematic error is quoted coming from the numerical integration over loop momenta.}
  \label{tab:stapleLR}
\end{table}

Conclusions from this calculation are summarized below:
\begin{enumerate}
\item {\bf Linear divergence:} On the lattice, there is a linear divergence, which depends on the length of the Wilson line and the gluon action that is employed. This divergence comes from diagrams contributing to the Wilson-line self-energy. At one loop, the contributing diagram is only $d_4$.
\item {\bf Logarithmic divergences:}
In both regularizations (DR and lattice), there are end-point logarithmic divergences, coming from diagrams $d_2$ and $d_3$, and cusp and contact logarithmic divergences, coming from diagram $d_4$. The coefficients in front of these divergences are regularization-independent.
\item {\bf Operator mixing:} Diagrams $d_2$ and $d_3$ give rise (upon summation) to the Dirac structure $\Gamma' = (\Gamma \hat{\slashed{\nu}}_i + \hat{\slashed{\nu}}_{\hspace{-0.4mm}f} \Gamma)/2$, which differs from the tree-level structure $\Gamma$ of the operator. This indicates that operator mixing is present: In order to remove this additional structure, we need to renormalize the Wilson-line operators $\mathcal{O}_\Gamma$ and $\mathcal{O}_{\Gamma'}$ as a doublet by introducing a $2\times 2$ mixing renormalization matrix. However, as concluded by symmetries, the employment of $4\times4$ mixing matrices for renormalizing quadruplets of asymmetric staple-shaped operators is expected to be required at higher loops. The mixing contributions at one loop depend solely on the direction of the Wilson line entering the endpoints, regardless of the shape of the Wilson line. Also, the coefficient $\alpha_2 + \alpha_3 \ c_{\rm SW}$ in front of the structure $\Gamma'$ depends on $\mathcalboondox{r}$ and $c_{SW}$; in particular, $\alpha_2$ vanishes when $\mathcalboondox{r}=0$. Thus, the one-loop mixing contributions originate from the chirality-breaking parts of the fermion action. As concluded by symmetries, these specific contributions are expected to be absent when a chiral-fermion action is employed.
\item {\bf Finite contributions:} Diagram $d_1$ is identical in DR and lattice regularization (up to discretization effects $\mathcal{O}(a)$) giving no contribution to the difference $\delta \Lambda_\Gamma$. Finite contributions of diagrams $d_2$ and $d_3$ stem only from the endpoints of the Wilson line. Any parts of a segment that do not include the endpoints give finite contributions which differ between DR and lattice regularization only by discretization effects. In diagram $d_4$, the finite contributions come from the cusps and the straight-line segments of the Wilson line; they depend on both the number of cusps ($n$) and the number of segments ($n+1$).
\end{enumerate}
Hence, the exact shape of the Wilson line does not affect the one-loop continuum and lattice Green's functions of the Wilson-line operators differently; the only additional contributions on the lattice depend on the total length, the number of cusps, and the direction of the Wilson line entering the endpoints.

In order to investigate RI$'$-bar schemes on the lattice, we have also calculated the one-loop bare Green's function of the Wilson loop $\langle L^{\rm LR}(z,y+y') \rangle$, given below in terms of the corresponding $\MSbar$-renormalized Green's function:
\begin{eqnarray}
    \langle L^{\rm LR}(z,y+y') \rangle &=& \langle L^{\overline{\rm MS}}(z,y+y') \rangle - \frac{g^2}{16 \pi^2} C_F \left[ 4 (\boldsymbol{\alpha_4} + \boldsymbol{\alpha_5}) + 8 \ln (a^2 \bar{\mu}^2) + 2 \boldsymbol{\alpha_6} \ \frac{| z | + | y | + | y' |}{a} \right] + \mathcal{O} (g^4). \qquad
    \label{WLlat}
\end{eqnarray}  
$\langle L^{\overline{\rm MS}}(z,y+y') \rangle$ can be read from Eq. \eqref{WLbare} by removing $1/\varepsilon$ terms.  
The result agrees with~\cite{Martinelli:1998vt} in the case of Wilson gluon action. As expected, the linearly divergent term ($1/a$) of Eq. \eqref{WLlat} cancels the linearly divergent term of Eq. \eqref{general_WL} when calculating $\overline{\Lambda}_{\Gamma}^{\rm LR} (q,z,y,y')$ (Eq. \eqref{Lambdabar}).

\subsection{Renormalization matrices}

The lattice renormalization matrices of the Wilson-line operators $\mathcal{O}_\Gamma$ in the $\MSbar$ scheme ($Z^{\MSbar,{\rm LR}}_{\Gamma \Gamma'}$) can be extracted from Eq. \eqref{general_WL} by imposing that the terms in the curly bracket are canceled when renormalizing both operator and external fermion fields in the lattice Green's function.
\begin{eqnarray}
    Z^{\MSbar,{\rm LR}}_{\Gamma \Gamma'} &=& \delta_{\Gamma \Gamma'} \Big[1 - \frac{(g^\MSbar)^2 C_F}{16 \pi^2} \Big(1 - 2 \alpha_1 - (n+1) \alpha_4 - n \alpha_5 - \alpha_6 \frac{l}{a} + e_1^{\psi} + e_2^{\psi} c_{\rm SW} + e_3^{\psi} c_{\rm SW}^2 \nonumber \\
    && \qquad - (2n + 3) \ln (a^2 \bar{\mu}^2) \Big)\Big] + \delta_{\Gamma', (\Gamma \hat{\slashed{\nu}}_i + \hat{\slashed{\nu}}_{\hspace{-0.4mm}f} \Gamma)/2} \frac{(g^\MSbar)^2 C_F}{16 \pi^2} \left(\alpha_2 + \alpha_3 c_{\rm SW}\right) + \mathcal{O} ((g^{\overline{\rm MS}})^4),
    \label{ZLRMS}
\end{eqnarray}
where $e^{\psi}_i$ comes from the renormalization factor of the external fermion fields (Eq. \eqref{ZpsiLRMSbar}). In the case under study ($n=2$, $l=|z| + |y| + |y'|$, $\hat{\nu}_i = {\rm sgn} (y) \ \hat{\nu}_2$, $\hat{\nu}_f = -{\rm sgn} (y') \ \hat{\nu}_2$), the one-loop renormalization matrices take the following form, where mixing in quadruplets $(\mathcal{O}_{\Gamma}, \mathcal{O}_{ \Gamma \gamma_{\nu_1} \gamma_{\nu_2}}, \mathcal{O}_{\Gamma \gamma_{\nu_1}}, \mathcal{O}_{\Gamma \gamma_{\nu_2}})$ is employed:
\begin{equation}
Z^{\MSbar,{\rm LR}} = \begin{pmatrix}
    Z^{\MSbar,{\rm LR}}_{\Gamma, \Gamma} & 0 & 0 & Z^{ \MSbar,{\rm LR}}_{\Gamma, \Gamma \gamma_{\nu_2}} \\
    0 & Z^{\MSbar, {\rm LR}}_{\Gamma \gamma_{\nu_1}\gamma_{\nu_2}, \Gamma \gamma_{\nu_1} \gamma_{\nu_2}} & Z^{\MSbar, {\rm LR}}_{\Gamma \gamma_{\nu_1} \gamma_{\nu_2}, \Gamma \gamma_{\nu_1}} &  0 \\
    0 & Z^{\MSbar, {\rm LR}}_{\Gamma \gamma_{\nu_1}, \Gamma \gamma_{\nu_1} \gamma_{\nu_2}} & Z^{\MSbar, {\rm LR}}_{\Gamma \gamma_{\nu_1}, \Gamma \gamma_{\nu_1}} & 0 \\
    Z^{\MSbar, {\rm LR}}_{\Gamma \gamma_{\nu_2}, \Gamma} & 0 & 0 & Z^{\MSbar, {\rm LR}}_{\Gamma \gamma_{\nu_2}, \Gamma \gamma_{\nu_2}} 
\end{pmatrix} + \ \mathcal{O} ((g^{\overline{\rm MS}})^4).
\label{ZLRMSbar}
\end{equation}
The elements of the above matrix can be read from Eq. \eqref{ZLRMS} by setting $(\Gamma, \Gamma')$ equal to pairs of $\{\Gamma, \Gamma \gamma_{\nu_1} \gamma_{\nu_2}, \Gamma \gamma_{\nu_1}, \Gamma \gamma_{\nu_2}\}$. Note that only two nondiagonal elements survive in each renormalization matrix depending on whether $(\tilde{\Gamma} \hat{\slashed{\nu}}_i + \hat{\slashed{\nu}}_{\hspace{-0.4mm}f} \tilde{\Gamma})/2 = {\rm sgn}(y) \ [\tilde{\Gamma}, \gamma_{\nu_2}]/2$ vanishes or not for each $\tilde{\Gamma} \in \{\Gamma, \Gamma \gamma_{\nu_1} \gamma_{\nu_2}, \Gamma \gamma_{\nu_1}, \Gamma \gamma_{\nu_2}\}$. Thus, the nondiagonal elements $Z^{ \MSbar, {\rm LR}}_{\Gamma \Gamma'}$ which are nonvanishing are those corresponding to $(\Gamma, \Gamma') = (\gamma_5, \gamma_5 \gamma_{\nu_2})$, $(\gamma_5 \gamma_{\nu_2}, \gamma_5)$, $(\gamma_{\nu_1}, \sigma_{\nu_1 \nu_2})$, $(\sigma_{\nu_1 \nu_2}, \gamma_{\nu_1})$, $(\gamma_{\nu_3}, \sigma_{\nu_3 \nu_2})$, $(\sigma_{\nu_3 \nu_2}, \gamma_{\nu_3})$, $(\gamma_{\nu_4}, \sigma_{\nu_4 \nu_2})$, $(\sigma_{\nu_4 \nu_2}, \gamma_{\nu_4})$.

The lattice renormalization matrices for the modified asymmetric staple-shaped Wilson-line operators $\overline{\mathcal{O}}_\Gamma$ in the $\MSbar$ scheme have the same form as \eqref{ZLRMSbar}. Their elements are given by:
\begin{eqnarray}
    \overline{Z}^{\MSbar, {\rm LR}}_{\Gamma \Gamma'} &=& \delta_{\Gamma \Gamma'} \Big[1 - \frac{(g^\MSbar)^2 C_F}{16 \pi^2} \Big(1 - 2 \alpha_1 - \alpha_4 + e_1^{\psi} + e_2^{\psi} c_{\rm SW} + e_3^{\psi} c_{\rm SW}^2 - 3 \ln (a^2 \bar{\mu}^2) \Big)\Big] \nonumber \\
&& + \delta_{\Gamma', [\Gamma, \gamma_{\nu_2}]/2} \ {\rm sgn}(y) \ \frac{(g^\MSbar)^2 C_F}{16 \pi^2} \left(\alpha_2 + \alpha_3 c_{\rm SW}\right) + \mathcal{O} ((g^{\overline{\rm MS}})^4).
    \label{ZbarLRMS}
\end{eqnarray}    
The linear divergence is now absent from the renormalization matrices. The nondiagonal elements are identical to \eqref{ZLRMSbar}. 

In the RI$'$-type schemes, the lattice renormalization matrices can be obtained from Eq. \eqref{Conversion} by combining the corresponding renormalization matrices in the $\MSbar$ scheme (Eqs. \eqref{ZLRMS}, \eqref{ZbarLRMS}), and the conversion matrices calculated in Eqs. (\ref{Cmatrix} -- \ref{Cbarmatrix}), as follows:
\begin{eqnarray}
    Z^{{\rm RI}'_i,{\rm LR}}_{\Gamma \Gamma'} &=& \delta_{\Gamma \Gamma'} + Z^{\MSbar,{\rm LR}}_{\Gamma \Gamma'} - C^{\overline{\rm MS}, {\rm RI}'_i}_{\Gamma \Gamma'} + \mathcal{O} ((g^{\overline{\rm MS}})^4), \\
    \overline{Z}^{{\rm RI}'_i,{\rm LR}}_{\Gamma \Gamma'} &=& \delta_{\Gamma \Gamma'} + \overline{Z}^{\MSbar,{\rm LR}}_{\Gamma \Gamma'} - \overline{C}^{\overline{\rm MS}, {\rm RI}'_i}_{\Gamma \Gamma'} + \mathcal{O} ((g^{\overline{\rm MS}})^4).
\end{eqnarray}  
They take the following matrix form\footnote{The form of Eq. \eqref{ZLRRI} is valid for all four RI$'$-type schemes studied in this work (RI$'_1$, RI$'_2$, RI$'_1$-bar, RI$'_2$-bar).}:
\begin{equation}
Z^{{\rm RI}',{\rm LR}} = \begin{pmatrix}
    Z^{{\rm RI}',{\rm LR}}_{\Gamma, \Gamma} & Z^{{\rm RI}',{\rm LR}}_{\Gamma, \Gamma \gamma_{\nu_1} \gamma_{\nu_2}} & 0 & Z^{{\rm RI}',{\rm LR}}_{\Gamma, \Gamma \gamma_{\nu_2}} \\
    Z^{{\rm RI}',{\rm LR}}_{\Gamma \gamma_{\nu_1}\gamma_{\nu_2}, \Gamma} & Z^{{\rm RI}',{\rm LR}}_{\Gamma \gamma_{\nu_1}\gamma_{\nu_2}, \Gamma \gamma_{\nu_1} \gamma_{\nu_2}} & Z^{{\rm RI}',{\rm LR}}_{\Gamma \gamma_{\nu_1} \gamma_{\nu_2}, \Gamma \gamma_{\nu_1}} &  0 \\
    0 & Z^{{\rm RI}',{\rm LR}}_{\Gamma \gamma_{\nu_1}, \Gamma \gamma_{\nu_1} \gamma_{\nu_2}} & Z^{{\rm RI}',{\rm LR}}_{\Gamma \gamma_{\nu_1}, \Gamma \gamma_{\nu_1}} & Z^{{\rm RI}',{\rm LR}}_{\Gamma \gamma_{\nu_1}, \Gamma \gamma_{\nu_2}} \\
    Z^{{\rm RI}',{\rm LR}}_{\Gamma \gamma_{\nu_2}, \Gamma} & 0 & Z^{{\rm RI}',{\rm LR}}_{\Gamma \gamma_{\nu_2}, \Gamma \gamma_{\nu_1}} & Z^{{\rm RI}',{\rm LR}}_{\Gamma \gamma_{\nu_2}, \Gamma \gamma_{\nu_2}} 
\end{pmatrix} + \ \mathcal{O} ((g^{\overline{\rm MS}})^4).
\label{ZLRRI}
\end{equation}
As in the $\MSbar$ scheme, there are further zero elements in \eqref{ZLRRI} depending on the commutation relations of $\Gamma$, $\Gamma \gamma_{\nu_1}$, $\Gamma \gamma_{\nu_2}$, and $\Gamma \gamma_{\nu_1} \gamma_{\nu_2}$ with $\gamma_{\nu_2}$.

\section{Summary and Future plans}
\label{Conclusions}

In this work, we present an extensive and comprehensive study of the renormalization of nonlocal quark bilinear operators featuring an asymmetric staple-shaped Wilson line. This project is motivated by the increased interest in studying TMDPDFs from lattice QCD using novel approaches, such as large momentum effective theory and short-distance factorization, which require matrix elements of the operators under study. More details can be found in the TMD Handbook~\cite{Boussarie:2023izj}.

The analysis is based on a one-loop perturbative calculation of Green's functions of such operators in both lattice and continuum (dimensional) regularizations. Based on our results, we identify the mixing pattern of these operators and propose renormalization prescriptions applicable to perturbative and nonperturbative data. More precisely, we discuss RI$'$-type conditions by using different projectors that effectively address power and logarithmic divergences, as well as the finite mixing among operators with different Dirac structures.  We have systematically analyzed the mixing patterns of these operators, leveraging symmetry arguments for both chiral and nonchiral fermions. We also introduce a variant of the RI$'$ prescription, which removes the pinch-pole singularities inherent in staple operators of infinite length by incorporating rectangular Wilson loops. This strategy also eliminates residual power divergences. 
Another novel aspect of this work is the extraction of the conversion factors to the $\overline{\rm MS}$ scheme using the results in dimensional regularization.
Our calculations are performed for arbitrary values of the renormalization momentum scale and the spatial dimensions of the staple. This ensures their applicability across a broad spectrum of nonperturbative investigations that may use the results of this work.

Potential future work includes an ambitious extension of this calculation to two-loop perturbation theory. This direction has the potential to offer valuable insights and improvements to ongoing nonperturbative investigations. By performing higher-order loop calculations, we aim to refine the renormalization procedure and eliminate sources of systematic uncertainties. Additionally, we plan to calculate one-loop lattice discretization effects across a range of staple lengths and momentum scales, which has significant potential for enhancing nonperturbative estimates, particularly at short distances. Other possible directions include calculations with different lattice formulations, which might be interesting for the TMD community. 

\begin{acknowledgements}
  G.S. and H.P. acknowledge financial support from the European Regional Development Fund and the Republic of Cyprus through the Research and Innovation Foundation under contract number EXCELLENCE/0421/0025.
M.C. acknowledges financial support from the U.S. Department of Energy, Office of Nuclear Physics, Early Career Award under Grant No. DE-SC0020405.
M.C. is grateful for the hospitality of the University of Cyprus during the completion of this manuscript under the project 3D-nucleon, ID number EXCELLENCE/0421/0043, co-financed by the European Regional Development Fund and the Republic of Cyprus through the Research and Innovation Foundation.  
\end{acknowledgements}

\appendix

\numberwithin{equation}{section}

\section{One-loop Feynman integrals with one external momentum in the presence of phase factors}
\label{IntegrationMethod}

In this appendix, we collect some useful formulae for the calculation of one-loop Feynman integrals with one external momentum in the presence of phase factors. We briefly describe the procedure that we follow for the derivation of these formulae in dimensional regularization. For supplementary material, we refer to Refs.~\cite{Constantinou:2017sej, Spanoudes:2018zya}.  

We consider the following d-dimensional tensor Feynman integral with one external momentum $q$, a phase factor $e^{i p \cdot \xi}$, where $\xi$ is a nonzero 4-vector and an arbitrary number ($n$) of momentum-loop components $p_{\mu_i}$ in the numerator:  
\begin{equation}
I^{d,\{\mu_1, \ldots, \mu_n\}}_{\alpha, \beta} (\xi,q) \equiv \int \frac{d^d p}{(2 \pi)^d} \frac{e^{i p \cdot \xi} \ p_{\mu_1} \ldots p_{\mu_n}}{(p^2)^{\alpha} \ ((-p + q)^2)^{\beta}}.
\label{tensor_integral}
\end{equation}
In our calculation, $\xi$ is a 2-vector lying in the staple's plane. In the case of $\xi = 0$, the phase vanishes, and the integral is simplified to the standard-form one-loop Feynman integral calculated in Ref.~\cite{Chetyrkin:1981qh}.

The tensor integral of Eq. \eqref{tensor_integral} can be written in terms of derivatives w.r.t. $\xi$ of the scalar integral $I_{\alpha, \beta}^d (\xi, q)$:
\begin{eqnarray}
  I^{d,\{\mu_1, \ldots, \mu_n\}}_{\alpha, \beta} (\xi,q) &=& (-i)^n \frac{d}{d \xi_{\mu_1}} \ldots \frac{d}{d \xi_{\mu_n}} I^d_{\alpha, \beta} (\xi,q), \\
I^d_{\alpha, \beta} (\xi,q) &\equiv& \int \frac{d^d p}{(2 \pi)^d} \frac{e^{i p \cdot \xi}}{(p^2)^{\alpha} \ ((-p + q)^2)^{\beta}}.  
\end{eqnarray}
The scalar integral $I^d_{\alpha, \beta} (\xi,q)$ can be computed using Feynman or Schwinger parametrization leading to the following expressions ($s \equiv \alpha+\beta-d/2$):
\begin{eqnarray}
  I^d_{\alpha, \beta} (\xi,q \neq 0) &=& \frac{2^{1-s-d}}{\pi^{d/2} \Gamma(\alpha) \Gamma(\beta)} \left(\frac{\xi^2}{q^2}\right)^{s/2} \int_0^1 dx \ K_{s} \left(\sqrt{q^2 \ \xi^2 \ (1-x) \ x}\right) e^{i x q \cdot \xi} \ x^{-1+\beta-s/2} \ {(1-x)}^{-1+\alpha-s/2}, \qquad \ \ \ \\
  I^d_{\alpha, \beta} (\xi,q = 0) &=& \frac{4^{\alpha + \beta} \ \Gamma (-s)}{\pi^{d/2} \ \Gamma (\alpha + \beta)} (\xi^2)^{s}.
  \end{eqnarray}

\section{Feynman-parameter integrals}
\label{IntegralList}

We write down a list of Feynman-parameter integrals appearing in the one-loop Green's functions of the asymmetric staple-shaped Wilson-line operators; they are classified into four categories:
\begin{itemize}
\item {\bf Category F:} $F_i \equiv F_i (q,r)$
  \begin{eqnarray}
    F_1 (q,r) &=& \int_0^1 dx \ K_0 (\sqrt{x \ (1-x) \ q^2 r^2}) \ e^{-i x {\sf q} \cdot {\sf r}}, \\
    F_2 (q,r) &=& \int_0^1 dx \ K_0 (\sqrt{x \ (1-x) \ q^2 r^2}) \ e^{-i x {\sf q} \cdot {\sf r}} \ x, \\
    F_3 (q,r) &=& \int_0^1 dx \ K_0 (\sqrt{x \ (1-x) \ q^2 r^2}) \ e^{-i x {\sf q} \cdot {\sf r}} \ x \ (1-x), \\
    F_4 (q,r) &=& \int_0^1 dx \ K_1 (\sqrt{x \ (1-x) \ q^2 r^2}) \ e^{-i x {\sf q} \cdot {\sf r}} \ \sqrt{x \ (1-x) \ q^2 r^2},
  \end{eqnarray}  
\item {\bf Category G:} $G_i \equiv G_i (q,y,z), \ \bar{G}_i \equiv \bar{G}_i (q,y',z)$
  \begin{eqnarray}
    G_1 (q,y,z) &=& \int_0^1 dx \int_0^1 d\zeta \ K_0 (\sqrt{x \ (1-x) \ q^2 \ (y^2 + z^2 \zeta^2)}) \ e^{-i x {\sf q} \cdot (z \zeta \hat{\nu}_1 + y \hat{\nu}_2)} \ (1-x), \\
    G_2 (q,y,z) &=& \int_0^1 dx \int_0^1 d\zeta \ K_0 (\sqrt{x \ (1-x) \ q^2 \ (y^2 + z^2 \zeta^2)}) \ e^{-i x {\sf q} \cdot (z \zeta \hat{\nu}_1 + y \hat{\nu}_2)} \ x, \\
    G_3 (q,y,z) &=& \int_0^1 dx \int_0^1 d\zeta \ K_1 (\sqrt{x \ (1-x) \ q^2 \ (y^2 + z^2 \zeta^2)}) \ e^{-i x {\sf q} \cdot (z \zeta \hat{\nu}_1 + y \hat{\nu}_2)} \ \sqrt{\frac{x \ (1-x)}{q^2 \ (y^2 + z^2 \zeta^2)}}, \\
    \bar{G}_i (q,y',z) &=& G_i^\ast (q,y',-z), \quad (i=1,2,3),
  \end{eqnarray} 
  \item {\bf Category H:} $H_i \equiv H_i (q,y,z), \ \bar{H}_i \equiv \bar{H}_i (q,y',z)$
  \begin{eqnarray}
    H_1 (q, y, 0) &=& \int_0^1 dx \int_0^1 d\zeta \ K_0 (\sqrt{x \ (1-x) \ q^2 \ y^2 \zeta^2}) \ e^{-i x {\sf q} \cdot y \zeta \hat{\nu}_2} \ (1-x), \\
    H_2 (q, y, 0) &=& \int_0^1 dx \int_0^1 d\zeta \ K_0 (\sqrt{x \ (1-x) \ q^2 \ y^2 \zeta^2}) \ e^{-i x {\sf q} \cdot y \zeta \hat{\nu}_2} \ x, \\
    H_3 (q, y, z) &=& \int_0^1 dx \int_0^1 d\zeta \ K_0 (\sqrt{x \ (1-x) \ q^2 \ (z^2 + y^2 \zeta^2)}) \ e^{-i x {\sf q} \cdot (z \hat{\nu}_1 + y \zeta \hat{\nu}_2)} \ (1-x), \\
    H_4 (q, y, z) &=& \int_0^1 dx \int_0^1 d\zeta \ K_0 (\sqrt{x \ (1-x) \ q^2 \ (z^2 + y^2 \zeta^2)}) \ e^{-i x {\sf q} \cdot (z \hat{\nu}_1 + y \zeta \hat{\nu}_2)} \ x, \\
    H_5 (q, y, z) &=& \int_0^1 dx \int_0^1 d\zeta \ K_1 (\sqrt{x \ (1-x) \ q^2 \ (z^2 + y^2 \zeta^2)}) \ e^{-i x {\sf q} \cdot (z \hat{\nu}_1 + y \zeta \hat{\nu}_2)} \ \sqrt{\frac{x \ (1-x)}{q^2 \ (z^2 + y^2 \zeta^2)}}, \\
    \bar{H}_i (q,y',z) &=& H_i^\ast (q,y',z), \quad (i=1,2,3,4,5), 
  \end{eqnarray}
  \item {\bf Category I:} $I_i \equiv I_i (q,y-y',z)$
  \begin{eqnarray}
    I_1 (q, y-y', z) &=& \int_0^1 dx \int_0^1 d\zeta \ K_0 (\sqrt{x \ (1-x) \ q^2 \ (z^2 + (y-y')^2 \zeta^2)}) \ e^{-i x {\sf q} \cdot (z \hat{\nu}_1 + (y-y') \zeta \hat{\nu}_2)} \ (1-x), \\
    I_2 (q, y-y', z) &=& \int_0^1 dx \int_0^1 d\zeta \ K_0 (\sqrt{x \ (1-x) \ q^2 \ (z^2 + (y-y')^2 \zeta^2)}) \ e^{-i x {\sf q} \cdot (z \hat{\nu}_1 + (y-y') \zeta \hat{\nu}_2)} \ x, \\
    I_3 (q, y-y', z) &=& \int_0^1 dx \int_0^1 d\zeta \ K_1 (\sqrt{x \ (1-x) \ q^2 \ (z^2 + (y-y')^2 \zeta^2)}) \ e^{-i x {\sf q} \cdot (z \hat{\nu}_1 + (y-y') \zeta \hat{\nu}_2)} \cdot \nonumber \\
    && \qquad \qquad \qquad \sqrt{\frac{x \ (1-x)}{q^2 \ (z^2 + (y-y')^2 \zeta^2)}}.
  \end{eqnarray}
  \end{itemize}

  \section{Renormalization of fermion fields}
\label{ap.C}

The one-loop expressions for the renormalization factors of the fermion fields in $\MSbar$ and RI$'$ schemes, using both dimensional (DR) and lattice (LR) regularizations, are provided in this appendix. The expressions are taken from Refs. \cite{Gracey:2003yr} and \cite{Alexandrou:2012mt}, respectively. We also give the one-loop conversion factors between the two schemes. We use the convention $\psi^R (x) = {(Z^{R,X}_\psi)}^{1/2} \psi^X (x)$.
\begin{eqnarray} 
Z_\psi^{\MSbar,\text{DR}} &=& 1 - \frac{(g^\MSbar)^2 C_F}{16 \pi^2} (\beta - 1) \frac{1}{\varepsilon} + \mathcal{O} ((g^\MSbar)^4), \label{ZpsiDRMSbar}\\
Z_\psi^{\text{RI}',\text{DR}} &=& 1 - \frac{(g^\MSbar)^2 C_F}{16 \pi^2} (\beta - 1) \left( \frac{1}{\varepsilon} + 1 +\log \left(\frac{\bar{\mu}^2}{\bar{q}^2}\right) \right) + \mathcal{O} ((g^\MSbar)^4), \\
C_\psi^{\MSbar,\text{RI}'} &=& \frac{Z_\psi^{\MSbar,\text{DR}}}{Z_\psi^{\text{RI}',\text{DR}}} = \frac{Z_\psi^{\MSbar,\text{LR}}}{Z_\psi^{\text{RI}',\text{LR}}} = 1 + \frac{(g^\MSbar)^2 C_F}{16 \pi^2} (\beta - 1) \left(1 +\log \left(\frac{\bar{\mu}^2}{\bar{q}^2}\right) \right) + \mathcal{O} ((g^\MSbar)^4), \\
Z_\psi^{\text{RI}',\text{LR}} &=& 1 - \frac{(g^\MSbar)^2 C_F}{16 \pi^2} \left[(1 + 16 \pi^2 P_2) \, \beta + \boldsymbol{e_1^\psi} + \boldsymbol{e_2^\psi} c_{SW} + \boldsymbol{e_3^\psi} c_{SW}^2 + (1-\beta) \log \left( a^2 \bar{q}^2 \right) \right] + \mathcal{O} ((g^\MSbar)^4), \\
Z_\psi^{\MSbar,\text{LR}} &=& C_\psi^{\MSbar,\text{RI}'} \ Z_\psi^{\text{RI}',\text{LR}} \nonumber \\
&=& 1 - \frac{(g^\MSbar)^2 C_F}{16 \pi^2} \Big[ 1 + 16 \pi^2 P_2 \, \beta + \boldsymbol{e_1^\psi} + \boldsymbol{e_2^\psi} c_{SW} + \boldsymbol{e_3^\psi} c_{SW}^2 + (1-\beta) \log \left( a^2 \bar{\mu}^2 \right) \Big] + \mathcal{O} ((g^\MSbar)^4). \label{ZpsiLRMSbar}
\end{eqnarray}
The numerical constants $e_i^\psi$ depend on the gluon action in use; their values for Wilson, tree-level Symanzik, and Iwasaki improved gluon actions are given in Table \ref{tab:Zpsi}.
\begin{table}[thb]
  \centering
  \begin{tabular}{|l|l|l|l|}
  \hline
\ \textbf{Gluon action} & \ \qquad \ $\boldsymbol{e_1^\psi}$ & \ \qquad \ \ $\boldsymbol{e_2^\psi}$ & \ \qquad \ \ \ $\boldsymbol{e_3^\psi}$  \\
\hline
\hline
\ Wilson & \ 11.8524043(2) \ & \ -2.24886853(1) \ & \ -1.397367103(3) \ \\
\ Tree-level Symanzik \ & \ \phantom{a}8.2312629(2) \ & \ -2.01542508(3) \ & \ -1.242202721(2) \ \\ 
\ Iwasaki & \ \phantom{a}3.3245571(2) \ & \ -1.60101083(6) \ & \ -0.973206902(1) \ \\
\hline
  \end{tabular}
  \caption{Numerical values of the coefficients $e_1^\psi${-}$e_3^\psi$ appearing in the one-loop renormalization factors of fermion fields on the lattice for $\mathcalboondox{r} = 1$. A systematic error is quoted coming from the numerical integration over loop momenta.}
  \label{tab:Zpsi}
\end{table}

\nocite{apsrev41Control}
\bibliographystyle{apsrev4-1}
\bibliography{refs}

\begin{thebibliography}{137}%
\makeatletter
\providecommand \@ifxundefined [1]{%
 \@ifx{#1\undefined}
}%
\providecommand \@ifnum [1]{%
 \ifnum #1\expandafter \@firstoftwo
 \else \expandafter \@secondoftwo
 \fi
}%
\providecommand \@ifx [1]{%
 \ifx #1\expandafter \@firstoftwo
 \else \expandafter \@secondoftwo
 \fi
}%
\providecommand \natexlab [1]{#1}%
\providecommand \enquote  [1]{``#1''}%
\providecommand \bibnamefont  [1]{#1}%
\providecommand \bibfnamefont [1]{#1}%
\providecommand \citenamefont [1]{#1}%
\providecommand \href@noop [0]{\@secondoftwo}%
\providecommand \href [0]{\begingroup \@sanitize@url \@href}%
\providecommand \@href[1]{\@@startlink{#1}\@@href}%
\providecommand \@@href[1]{\endgroup#1\@@endlink}%
\providecommand \@sanitize@url [0]{\catcode `\\12\catcode `\$12\catcode `\&12\catcode `\#12\catcode `\^12\catcode `\_12\catcode `\%12\relax}%
\providecommand \@@startlink[1]{}%
\providecommand \@@endlink[0]{}%
\providecommand \url  [0]{\begingroup\@sanitize@url \@url }%
\providecommand \@url [1]{\endgroup\@href {#1}{\urlprefix }}%
\providecommand \urlprefix  [0]{URL }%
\providecommand \Eprint [0]{\href }%
\providecommand \doibase [0]{http://dx.doi.org/}%
\providecommand \selectlanguage [0]{\@gobble}%
\providecommand \bibinfo  [0]{\@secondoftwo}%
\providecommand \bibfield  [0]{\@secondoftwo}%
\providecommand \translation [1]{[#1]}%
\providecommand \BibitemOpen [0]{}%
\providecommand \bibitemStop [0]{}%
\providecommand \bibitemNoStop [0]{.\EOS\space}%
\providecommand \EOS [0]{\spacefactor3000\relax}%
\providecommand \BibitemShut  [1]{\csname bibitem#1\endcsname}%
\let\auto@bib@innerbib\@empty
\bibitem [{\citenamefont {Ji}(2013)}]{Ji:2013dva}%
  \BibitemOpen
  \bibfield  {author} {\bibinfo {author} {\bibfnamefont {Xiangdong}\ \bibnamefont {Ji}},\ }\bibfield  {title} {\enquote {\bibinfo {title} {{Parton Physics on a Euclidean Lattice}},}\ }\href {\doibase 10.1103/PhysRevLett.110.262002} {\bibfield  {journal} {\bibinfo  {journal} {Phys. Rev. Lett.}\ }\textbf {\bibinfo {volume} {110}},\ \bibinfo {pages} {262002} (\bibinfo {year} {2013})},\ \Eprint {http://arxiv.org/abs/1305.1539} {arXiv:1305.1539 [hep-ph]} \BibitemShut {NoStop}%
\bibitem [{\citenamefont {Ji}(2014)}]{Ji:2014gla}%
  \BibitemOpen
  \bibfield  {author} {\bibinfo {author} {\bibfnamefont {Xiangdong}\ \bibnamefont {Ji}},\ }\bibfield  {title} {\enquote {\bibinfo {title} {{Parton Physics from Large-Momentum Effective Field Theory}},}\ }\href {\doibase 10.1007/s11433-014-5492-3} {\bibfield  {journal} {\bibinfo  {journal} {Sci. China Phys. Mech. Astron.}\ }\textbf {\bibinfo {volume} {57}},\ \bibinfo {pages} {1407--1412} (\bibinfo {year} {2014})},\ \Eprint {http://arxiv.org/abs/1404.6680} {arXiv:1404.6680 [hep-ph]} \BibitemShut {NoStop}%
\bibitem [{\citenamefont {Lin}\ \emph {et~al.}(2015)\citenamefont {Lin}, \citenamefont {Chen}, \citenamefont {Cohen},\ and\ \citenamefont {Ji}}]{Lin:2014zya}%
  \BibitemOpen
  \bibfield  {author} {\bibinfo {author} {\bibfnamefont {Huey-Wen}\ \bibnamefont {Lin}}, \bibinfo {author} {\bibfnamefont {Jiunn-Wei}\ \bibnamefont {Chen}}, \bibinfo {author} {\bibfnamefont {Saul~D.}\ \bibnamefont {Cohen}}, \ and\ \bibinfo {author} {\bibfnamefont {Xiangdong}\ \bibnamefont {Ji}},\ }\bibfield  {title} {\enquote {\bibinfo {title} {{Flavor Structure of the Nucleon Sea from Lattice QCD}},}\ }\href {\doibase 10.1103/PhysRevD.91.054510} {\bibfield  {journal} {\bibinfo  {journal} {Phys. Rev. D}\ }\textbf {\bibinfo {volume} {91}},\ \bibinfo {pages} {054510} (\bibinfo {year} {2015})},\ \Eprint {http://arxiv.org/abs/1402.1462} {arXiv:1402.1462 [hep-ph]} \BibitemShut {NoStop}%
\bibitem [{\citenamefont {Alexandrou}\ \emph {et~al.}(2015)\citenamefont {Alexandrou}, \citenamefont {Cichy}, \citenamefont {Drach}, \citenamefont {Garcia-Ramos}, \citenamefont {Hadjiyiannakou}, \citenamefont {Jansen}, \citenamefont {Steffens},\ and\ \citenamefont {Wiese}}]{Alexandrou:2015rja}%
  \BibitemOpen
  \bibfield  {author} {\bibinfo {author} {\bibfnamefont {Constantia}\ \bibnamefont {Alexandrou}}, \bibinfo {author} {\bibfnamefont {Krzysztof}\ \bibnamefont {Cichy}}, \bibinfo {author} {\bibfnamefont {Vincent}\ \bibnamefont {Drach}}, \bibinfo {author} {\bibfnamefont {Elena}\ \bibnamefont {Garcia-Ramos}}, \bibinfo {author} {\bibfnamefont {Kyriakos}\ \bibnamefont {Hadjiyiannakou}}, \bibinfo {author} {\bibfnamefont {Karl}\ \bibnamefont {Jansen}}, \bibinfo {author} {\bibfnamefont {Fernanda}\ \bibnamefont {Steffens}}, \ and\ \bibinfo {author} {\bibfnamefont {Christian}\ \bibnamefont {Wiese}},\ }\bibfield  {title} {\enquote {\bibinfo {title} {{Lattice calculation of parton distributions}},}\ }\href {\doibase 10.1103/PhysRevD.92.014502} {\bibfield  {journal} {\bibinfo  {journal} {Phys. Rev. D}\ }\textbf {\bibinfo {volume} {92}},\ \bibinfo {pages} {014502} (\bibinfo {year} {2015})},\ \Eprint {http://arxiv.org/abs/1504.07455} {arXiv:1504.07455 [hep-lat]} \BibitemShut {NoStop}%
\bibitem [{\citenamefont {Chen}\ \emph {et~al.}(2016)\citenamefont {Chen}, \citenamefont {Cohen}, \citenamefont {Ji}, \citenamefont {Lin},\ and\ \citenamefont {Zhang}}]{Chen:2016utp}%
  \BibitemOpen
  \bibfield  {author} {\bibinfo {author} {\bibfnamefont {Jiunn-Wei}\ \bibnamefont {Chen}}, \bibinfo {author} {\bibfnamefont {Saul~D.}\ \bibnamefont {Cohen}}, \bibinfo {author} {\bibfnamefont {Xiangdong}\ \bibnamefont {Ji}}, \bibinfo {author} {\bibfnamefont {Huey-Wen}\ \bibnamefont {Lin}}, \ and\ \bibinfo {author} {\bibfnamefont {Jian-Hui}\ \bibnamefont {Zhang}},\ }\bibfield  {title} {\enquote {\bibinfo {title} {{Nucleon Helicity and Transversity Parton Distributions from Lattice QCD}},}\ }\href {\doibase 10.1016/j.nuclphysb.2016.07.033} {\bibfield  {journal} {\bibinfo  {journal} {Nucl. Phys. B}\ }\textbf {\bibinfo {volume} {911}},\ \bibinfo {pages} {246--273} (\bibinfo {year} {2016})},\ \Eprint {http://arxiv.org/abs/1603.06664} {arXiv:1603.06664 [hep-ph]} \BibitemShut {NoStop}%
\bibitem [{\citenamefont {Alexandrou}\ \emph {et~al.}(2017{\natexlab{a}})\citenamefont {Alexandrou}, \citenamefont {Cichy}, \citenamefont {Constantinou}, \citenamefont {Hadjiyiannakou}, \citenamefont {Jansen}, \citenamefont {Steffens},\ and\ \citenamefont {Wiese}}]{Alexandrou:2016jqi}%
  \BibitemOpen
  \bibfield  {author} {\bibinfo {author} {\bibfnamefont {Constantia}\ \bibnamefont {Alexandrou}}, \bibinfo {author} {\bibfnamefont {Krzysztof}\ \bibnamefont {Cichy}}, \bibinfo {author} {\bibfnamefont {Martha}\ \bibnamefont {Constantinou}}, \bibinfo {author} {\bibfnamefont {Kyriakos}\ \bibnamefont {Hadjiyiannakou}}, \bibinfo {author} {\bibfnamefont {Karl}\ \bibnamefont {Jansen}}, \bibinfo {author} {\bibfnamefont {Fernanda}\ \bibnamefont {Steffens}}, \ and\ \bibinfo {author} {\bibfnamefont {Christian}\ \bibnamefont {Wiese}},\ }\bibfield  {title} {\enquote {\bibinfo {title} {{Updated Lattice Results for Parton Distributions}},}\ }\href {\doibase 10.1103/PhysRevD.96.014513} {\bibfield  {journal} {\bibinfo  {journal} {Phys. Rev. D}\ }\textbf {\bibinfo {volume} {96}},\ \bibinfo {pages} {014513} (\bibinfo {year} {2017}{\natexlab{a}})},\ \Eprint {http://arxiv.org/abs/1610.03689} {arXiv:1610.03689 [hep-lat]} \BibitemShut {NoStop}%
\bibitem [{\citenamefont {Zhang}\ \emph {et~al.}(2017)\citenamefont {Zhang}, \citenamefont {Chen}, \citenamefont {Ji}, \citenamefont {Jin},\ and\ \citenamefont {Lin}}]{Zhang:2017bzy}%
  \BibitemOpen
  \bibfield  {author} {\bibinfo {author} {\bibfnamefont {Jian-Hui}\ \bibnamefont {Zhang}}, \bibinfo {author} {\bibfnamefont {Jiunn-Wei}\ \bibnamefont {Chen}}, \bibinfo {author} {\bibfnamefont {Xiangdong}\ \bibnamefont {Ji}}, \bibinfo {author} {\bibfnamefont {Luchang}\ \bibnamefont {Jin}}, \ and\ \bibinfo {author} {\bibfnamefont {Huey-Wen}\ \bibnamefont {Lin}},\ }\bibfield  {title} {\enquote {\bibinfo {title} {{Pion Distribution Amplitude from Lattice QCD}},}\ }\href {\doibase 10.1103/PhysRevD.95.094514} {\bibfield  {journal} {\bibinfo  {journal} {Phys. Rev. D}\ }\textbf {\bibinfo {volume} {95}},\ \bibinfo {pages} {094514} (\bibinfo {year} {2017})},\ \Eprint {http://arxiv.org/abs/1702.00008} {arXiv:1702.00008 [hep-lat]} \BibitemShut {NoStop}%
\bibitem [{\citenamefont {Alexandrou}\ \emph {et~al.}(2017{\natexlab{b}})\citenamefont {Alexandrou}, \citenamefont {Cichy}, \citenamefont {Constantinou}, \citenamefont {Hadjiyiannakou}, \citenamefont {Jansen}, \citenamefont {Panagopoulos},\ and\ \citenamefont {Steffens}}]{Alexandrou:2017huk}%
  \BibitemOpen
  \bibfield  {author} {\bibinfo {author} {\bibfnamefont {Constantia}\ \bibnamefont {Alexandrou}}, \bibinfo {author} {\bibfnamefont {Krzysztof}\ \bibnamefont {Cichy}}, \bibinfo {author} {\bibfnamefont {Martha}\ \bibnamefont {Constantinou}}, \bibinfo {author} {\bibfnamefont {Kyriakos}\ \bibnamefont {Hadjiyiannakou}}, \bibinfo {author} {\bibfnamefont {Karl}\ \bibnamefont {Jansen}}, \bibinfo {author} {\bibfnamefont {Haralambos}\ \bibnamefont {Panagopoulos}}, \ and\ \bibinfo {author} {\bibfnamefont {Fernanda}\ \bibnamefont {Steffens}},\ }\bibfield  {title} {\enquote {\bibinfo {title} {{A complete non-perturbative renormalization prescription for quasi-PDFs}},}\ }\href {\doibase 10.1016/j.nuclphysb.2017.08.012} {\bibfield  {journal} {\bibinfo  {journal} {Nucl. Phys. B}\ }\textbf {\bibinfo {volume} {923}},\ \bibinfo {pages} {394--415} (\bibinfo {year} {2017}{\natexlab{b}})},\ \Eprint {http://arxiv.org/abs/1706.00265} {arXiv:1706.00265 [hep-lat]} \BibitemShut {NoStop}%
\bibitem [{\citenamefont {Zhang}\ \emph {et~al.}(2019{\natexlab{a}})\citenamefont {Zhang}, \citenamefont {Jin}, \citenamefont {Lin}, \citenamefont {Sch\"afer}, \citenamefont {Sun}, \citenamefont {Yang}, \citenamefont {Zhang}, \citenamefont {Zhao},\ and\ \citenamefont {Chen}}]{Zhang:2017zfe}%
  \BibitemOpen
  \bibfield  {author} {\bibinfo {author} {\bibfnamefont {Jian-Hui}\ \bibnamefont {Zhang}}, \bibinfo {author} {\bibfnamefont {Luchang}\ \bibnamefont {Jin}}, \bibinfo {author} {\bibfnamefont {Huey-Wen}\ \bibnamefont {Lin}}, \bibinfo {author} {\bibfnamefont {Andreas}\ \bibnamefont {Sch\"afer}}, \bibinfo {author} {\bibfnamefont {Peng}\ \bibnamefont {Sun}}, \bibinfo {author} {\bibfnamefont {Yi-Bo}\ \bibnamefont {Yang}}, \bibinfo {author} {\bibfnamefont {Rui}\ \bibnamefont {Zhang}}, \bibinfo {author} {\bibfnamefont {Yong}\ \bibnamefont {Zhao}}, \ and\ \bibinfo {author} {\bibfnamefont {Jiunn-Wei}\ \bibnamefont {Chen}} (\bibinfo {collaboration} {LP3}),\ }\bibfield  {title} {\enquote {\bibinfo {title} {{Kaon Distribution Amplitude from Lattice QCD and the Flavor SU(3) Symmetry}},}\ }\href {\doibase 10.1016/j.nuclphysb.2018.12.020} {\bibfield  {journal} {\bibinfo  {journal} {Nucl. Phys. B}\ }\textbf {\bibinfo {volume} {939}},\ \bibinfo {pages} {429--446} (\bibinfo {year} {2019}{\natexlab{a}})},\ \Eprint
  {http://arxiv.org/abs/1712.10025} {arXiv:1712.10025 [hep-ph]} \BibitemShut {NoStop}%
\bibitem [{\citenamefont {Alexandrou}\ \emph {et~al.}(2018{\natexlab{a}})\citenamefont {Alexandrou}, \citenamefont {Cichy}, \citenamefont {Constantinou}, \citenamefont {Jansen}, \citenamefont {Scapellato},\ and\ \citenamefont {Steffens}}]{Alexandrou:2018pbm}%
  \BibitemOpen
  \bibfield  {author} {\bibinfo {author} {\bibfnamefont {Constantia}\ \bibnamefont {Alexandrou}}, \bibinfo {author} {\bibfnamefont {Krzysztof}\ \bibnamefont {Cichy}}, \bibinfo {author} {\bibfnamefont {Martha}\ \bibnamefont {Constantinou}}, \bibinfo {author} {\bibfnamefont {Karl}\ \bibnamefont {Jansen}}, \bibinfo {author} {\bibfnamefont {Aurora}\ \bibnamefont {Scapellato}}, \ and\ \bibinfo {author} {\bibfnamefont {Fernanda}\ \bibnamefont {Steffens}},\ }\bibfield  {title} {\enquote {\bibinfo {title} {{Light-Cone Parton Distribution Functions from Lattice QCD}},}\ }\href {\doibase 10.1103/PhysRevLett.121.112001} {\bibfield  {journal} {\bibinfo  {journal} {Phys. Rev. Lett.}\ }\textbf {\bibinfo {volume} {121}},\ \bibinfo {pages} {112001} (\bibinfo {year} {2018}{\natexlab{a}})},\ \Eprint {http://arxiv.org/abs/1803.02685} {arXiv:1803.02685 [hep-lat]} \BibitemShut {NoStop}%
\bibitem [{\citenamefont {Alexandrou}\ \emph {et~al.}(2018{\natexlab{b}})\citenamefont {Alexandrou}, \citenamefont {Cichy}, \citenamefont {Constantinou}, \citenamefont {Jansen}, \citenamefont {Scapellato},\ and\ \citenamefont {Steffens}}]{Alexandrou:2018eet}%
  \BibitemOpen
  \bibfield  {author} {\bibinfo {author} {\bibfnamefont {Constantia}\ \bibnamefont {Alexandrou}}, \bibinfo {author} {\bibfnamefont {Krzysztof}\ \bibnamefont {Cichy}}, \bibinfo {author} {\bibfnamefont {Martha}\ \bibnamefont {Constantinou}}, \bibinfo {author} {\bibfnamefont {Karl}\ \bibnamefont {Jansen}}, \bibinfo {author} {\bibfnamefont {Aurora}\ \bibnamefont {Scapellato}}, \ and\ \bibinfo {author} {\bibfnamefont {Fernanda}\ \bibnamefont {Steffens}},\ }\bibfield  {title} {\enquote {\bibinfo {title} {{Transversity parton distribution functions from lattice QCD}},}\ }\href {\doibase 10.1103/PhysRevD.98.091503} {\bibfield  {journal} {\bibinfo  {journal} {Phys. Rev. D}\ }\textbf {\bibinfo {volume} {98}},\ \bibinfo {pages} {091503} (\bibinfo {year} {2018}{\natexlab{b}})},\ \Eprint {http://arxiv.org/abs/1807.00232} {arXiv:1807.00232 [hep-lat]} \BibitemShut {NoStop}%
\bibitem [{\citenamefont {Liu}\ \emph {et~al.}(2020)\citenamefont {Liu} \emph {et~al.}}]{LatticeParton:2018gjr}%
  \BibitemOpen
  \bibfield  {author} {\bibinfo {author} {\bibfnamefont {Yu-Sheng}\ \bibnamefont {Liu}} \emph {et~al.} (\bibinfo {collaboration} {Lattice Parton}),\ }\bibfield  {title} {\enquote {\bibinfo {title} {{Unpolarized isovector quark distribution function from lattice QCD: A systematic analysis of renormalization and matching}},}\ }\href {\doibase 10.1103/PhysRevD.101.034020} {\bibfield  {journal} {\bibinfo  {journal} {Phys. Rev. D}\ }\textbf {\bibinfo {volume} {101}},\ \bibinfo {pages} {034020} (\bibinfo {year} {2020})},\ \Eprint {http://arxiv.org/abs/1807.06566} {arXiv:1807.06566 [hep-lat]} \BibitemShut {NoStop}%
\bibitem [{\citenamefont {Zhang}\ \emph {et~al.}(2019{\natexlab{b}})\citenamefont {Zhang}, \citenamefont {Chen}, \citenamefont {Jin}, \citenamefont {Lin}, \citenamefont {Sch\"afer},\ and\ \citenamefont {Zhao}}]{Zhang:2018nsy}%
  \BibitemOpen
  \bibfield  {author} {\bibinfo {author} {\bibfnamefont {Jian-Hui}\ \bibnamefont {Zhang}}, \bibinfo {author} {\bibfnamefont {Jiunn-Wei}\ \bibnamefont {Chen}}, \bibinfo {author} {\bibfnamefont {Luchang}\ \bibnamefont {Jin}}, \bibinfo {author} {\bibfnamefont {Huey-Wen}\ \bibnamefont {Lin}}, \bibinfo {author} {\bibfnamefont {Andreas}\ \bibnamefont {Sch\"afer}}, \ and\ \bibinfo {author} {\bibfnamefont {Yong}\ \bibnamefont {Zhao}},\ }\bibfield  {title} {\enquote {\bibinfo {title} {{First direct lattice-QCD calculation of the $x$-dependence of the pion parton distribution function}},}\ }\href {\doibase 10.1103/PhysRevD.100.034505} {\bibfield  {journal} {\bibinfo  {journal} {Phys. Rev. D}\ }\textbf {\bibinfo {volume} {100}},\ \bibinfo {pages} {034505} (\bibinfo {year} {2019}{\natexlab{b}})},\ \Eprint {http://arxiv.org/abs/1804.01483} {arXiv:1804.01483 [hep-lat]} \BibitemShut {NoStop}%
\bibitem [{\citenamefont {Alexandrou}\ \emph {et~al.}(2019)\citenamefont {Alexandrou}, \citenamefont {Cichy}, \citenamefont {Constantinou}, \citenamefont {Hadjiyiannakou}, \citenamefont {Jansen}, \citenamefont {Scapellato},\ and\ \citenamefont {Steffens}}]{Alexandrou:2019lfo}%
  \BibitemOpen
  \bibfield  {author} {\bibinfo {author} {\bibfnamefont {Constantia}\ \bibnamefont {Alexandrou}}, \bibinfo {author} {\bibfnamefont {Krzysztof}\ \bibnamefont {Cichy}}, \bibinfo {author} {\bibfnamefont {Martha}\ \bibnamefont {Constantinou}}, \bibinfo {author} {\bibfnamefont {Kyriakos}\ \bibnamefont {Hadjiyiannakou}}, \bibinfo {author} {\bibfnamefont {Karl}\ \bibnamefont {Jansen}}, \bibinfo {author} {\bibfnamefont {Aurora}\ \bibnamefont {Scapellato}}, \ and\ \bibinfo {author} {\bibfnamefont {Fernanda}\ \bibnamefont {Steffens}},\ }\bibfield  {title} {\enquote {\bibinfo {title} {{Systematic uncertainties in parton distribution functions from lattice QCD simulations at the physical point}},}\ }\href {\doibase 10.1103/PhysRevD.99.114504} {\bibfield  {journal} {\bibinfo  {journal} {Phys. Rev. D}\ }\textbf {\bibinfo {volume} {99}},\ \bibinfo {pages} {114504} (\bibinfo {year} {2019})},\ \Eprint {http://arxiv.org/abs/1902.00587} {arXiv:1902.00587 [hep-lat]} \BibitemShut {NoStop}%
\bibitem [{\citenamefont {Izubuchi}\ \emph {et~al.}(2019)\citenamefont {Izubuchi}, \citenamefont {Jin}, \citenamefont {Kallidonis}, \citenamefont {Karthik}, \citenamefont {Mukherjee}, \citenamefont {Petreczky}, \citenamefont {Shugert},\ and\ \citenamefont {Syritsyn}}]{Izubuchi:2019lyk}%
  \BibitemOpen
  \bibfield  {author} {\bibinfo {author} {\bibfnamefont {Taku}\ \bibnamefont {Izubuchi}}, \bibinfo {author} {\bibfnamefont {Luchang}\ \bibnamefont {Jin}}, \bibinfo {author} {\bibfnamefont {Christos}\ \bibnamefont {Kallidonis}}, \bibinfo {author} {\bibfnamefont {Nikhil}\ \bibnamefont {Karthik}}, \bibinfo {author} {\bibfnamefont {Swagato}\ \bibnamefont {Mukherjee}}, \bibinfo {author} {\bibfnamefont {Peter}\ \bibnamefont {Petreczky}}, \bibinfo {author} {\bibfnamefont {Charles}\ \bibnamefont {Shugert}}, \ and\ \bibinfo {author} {\bibfnamefont {Sergey}\ \bibnamefont {Syritsyn}},\ }\bibfield  {title} {\enquote {\bibinfo {title} {{Valence parton distribution function of pion from fine lattice}},}\ }\href {\doibase 10.1103/PhysRevD.100.034516} {\bibfield  {journal} {\bibinfo  {journal} {Phys. Rev. D}\ }\textbf {\bibinfo {volume} {100}},\ \bibinfo {pages} {034516} (\bibinfo {year} {2019})},\ \Eprint {http://arxiv.org/abs/1905.06349} {arXiv:1905.06349 [hep-lat]} \BibitemShut {NoStop}%
\bibitem [{\citenamefont {Cichy}\ \emph {et~al.}(2019)\citenamefont {Cichy}, \citenamefont {Del~Debbio},\ and\ \citenamefont {Giani}}]{Cichy:2019ebf}%
  \BibitemOpen
  \bibfield  {author} {\bibinfo {author} {\bibfnamefont {Krzysztof}\ \bibnamefont {Cichy}}, \bibinfo {author} {\bibfnamefont {Luigi}\ \bibnamefont {Del~Debbio}}, \ and\ \bibinfo {author} {\bibfnamefont {Tommaso}\ \bibnamefont {Giani}},\ }\bibfield  {title} {\enquote {\bibinfo {title} {{Parton distributions from lattice data: the nonsinglet case}},}\ }\href {\doibase 10.1007/JHEP10(2019)137} {\bibfield  {journal} {\bibinfo  {journal} {JHEP}\ }\textbf {\bibinfo {volume} {10}},\ \bibinfo {pages} {137} (\bibinfo {year} {2019})},\ \Eprint {http://arxiv.org/abs/1907.06037} {arXiv:1907.06037 [hep-ph]} \BibitemShut {NoStop}%
\bibitem [{\citenamefont {Chai}\ \emph {et~al.}(2020)\citenamefont {Chai} \emph {et~al.}}]{Chai:2020nxw}%
  \BibitemOpen
  \bibfield  {author} {\bibinfo {author} {\bibfnamefont {Yahui}\ \bibnamefont {Chai}} \emph {et~al.},\ }\bibfield  {title} {\enquote {\bibinfo {title} {{Parton distribution functions of $\Delta^+$ on the lattice}},}\ }\href {\doibase 10.1103/PhysRevD.102.014508} {\bibfield  {journal} {\bibinfo  {journal} {Phys. Rev. D}\ }\textbf {\bibinfo {volume} {102}},\ \bibinfo {pages} {014508} (\bibinfo {year} {2020})},\ \Eprint {http://arxiv.org/abs/2002.12044} {arXiv:2002.12044 [hep-lat]} \BibitemShut {NoStop}%
\bibitem [{\citenamefont {Zhang}\ \emph {et~al.}(2020{\natexlab{a}})\citenamefont {Zhang}, \citenamefont {Honkala}, \citenamefont {Lin},\ and\ \citenamefont {Chen}}]{Zhang:2020gaj}%
  \BibitemOpen
  \bibfield  {author} {\bibinfo {author} {\bibfnamefont {Rui}\ \bibnamefont {Zhang}}, \bibinfo {author} {\bibfnamefont {Carson}\ \bibnamefont {Honkala}}, \bibinfo {author} {\bibfnamefont {Huey-Wen}\ \bibnamefont {Lin}}, \ and\ \bibinfo {author} {\bibfnamefont {Jiunn-Wei}\ \bibnamefont {Chen}},\ }\bibfield  {title} {\enquote {\bibinfo {title} {{Pion and kaon distribution amplitudes in the continuum limit}},}\ }\href {\doibase 10.1103/PhysRevD.102.094519} {\bibfield  {journal} {\bibinfo  {journal} {Phys. Rev. D}\ }\textbf {\bibinfo {volume} {102}},\ \bibinfo {pages} {094519} (\bibinfo {year} {2020}{\natexlab{a}})},\ \Eprint {http://arxiv.org/abs/2005.13955} {arXiv:2005.13955 [hep-lat]} \BibitemShut {NoStop}%
\bibitem [{\citenamefont {Bhattacharya}\ \emph {et~al.}(2020{\natexlab{a}})\citenamefont {Bhattacharya}, \citenamefont {Cichy}, \citenamefont {Constantinou}, \citenamefont {Metz}, \citenamefont {Scapellato},\ and\ \citenamefont {Steffens}}]{Bhattacharya:2020xlt}%
  \BibitemOpen
  \bibfield  {author} {\bibinfo {author} {\bibfnamefont {Shohini}\ \bibnamefont {Bhattacharya}}, \bibinfo {author} {\bibfnamefont {Krzysztof}\ \bibnamefont {Cichy}}, \bibinfo {author} {\bibfnamefont {Martha}\ \bibnamefont {Constantinou}}, \bibinfo {author} {\bibfnamefont {Andreas}\ \bibnamefont {Metz}}, \bibinfo {author} {\bibfnamefont {Aurora}\ \bibnamefont {Scapellato}}, \ and\ \bibinfo {author} {\bibfnamefont {Fernanda}\ \bibnamefont {Steffens}},\ }\bibfield  {title} {\enquote {\bibinfo {title} {{One-loop matching for the twist-3 parton distribution $g_T (x)$}},}\ }\href {\doibase 10.1103/PhysRevD.102.034005} {\bibfield  {journal} {\bibinfo  {journal} {Phys. Rev. D}\ }\textbf {\bibinfo {volume} {102}},\ \bibinfo {pages} {034005} (\bibinfo {year} {2020}{\natexlab{a}})},\ \bibinfo {note} {[Erratum: Phys.Rev.D 108, 039901 (2023)]},\ \Eprint {http://arxiv.org/abs/2005.10939} {arXiv:2005.10939 [hep-ph]} \BibitemShut {NoStop}%
\bibitem [{\citenamefont {Bhattacharya}\ \emph {et~al.}(2020{\natexlab{b}})\citenamefont {Bhattacharya}, \citenamefont {Cichy}, \citenamefont {Constantinou}, \citenamefont {Metz}, \citenamefont {Scapellato},\ and\ \citenamefont {Steffens}}]{Bhattacharya:2020jfj}%
  \BibitemOpen
  \bibfield  {author} {\bibinfo {author} {\bibfnamefont {Shohini}\ \bibnamefont {Bhattacharya}}, \bibinfo {author} {\bibfnamefont {Krzysztof}\ \bibnamefont {Cichy}}, \bibinfo {author} {\bibfnamefont {Martha}\ \bibnamefont {Constantinou}}, \bibinfo {author} {\bibfnamefont {Andreas}\ \bibnamefont {Metz}}, \bibinfo {author} {\bibfnamefont {Aurora}\ \bibnamefont {Scapellato}}, \ and\ \bibinfo {author} {\bibfnamefont {Fernanda}\ \bibnamefont {Steffens}},\ }\bibfield  {title} {\enquote {\bibinfo {title} {{The role of zero-mode contributions in the matching for the twist-3 PDFs $e(x)$ and $h_{L}(x)$}},}\ }\href {\doibase 10.1103/PhysRevD.102.114025} {\bibfield  {journal} {\bibinfo  {journal} {Phys. Rev. D}\ }\textbf {\bibinfo {volume} {102}},\ \bibinfo {pages} {114025} (\bibinfo {year} {2020}{\natexlab{b}})},\ \Eprint {http://arxiv.org/abs/2006.12347} {arXiv:2006.12347 [hep-ph]} \BibitemShut {NoStop}%
\bibitem [{\citenamefont {Alexandrou}\ \emph {et~al.}(2020)\citenamefont {Alexandrou}, \citenamefont {Cichy}, \citenamefont {Constantinou}, \citenamefont {Hadjiyiannakou}, \citenamefont {Jansen}, \citenamefont {Scapellato},\ and\ \citenamefont {Steffens}}]{Alexandrou:2020zbe}%
  \BibitemOpen
  \bibfield  {author} {\bibinfo {author} {\bibfnamefont {Constantia}\ \bibnamefont {Alexandrou}}, \bibinfo {author} {\bibfnamefont {Krzysztof}\ \bibnamefont {Cichy}}, \bibinfo {author} {\bibfnamefont {Martha}\ \bibnamefont {Constantinou}}, \bibinfo {author} {\bibfnamefont {Kyriakos}\ \bibnamefont {Hadjiyiannakou}}, \bibinfo {author} {\bibfnamefont {Karl}\ \bibnamefont {Jansen}}, \bibinfo {author} {\bibfnamefont {Aurora}\ \bibnamefont {Scapellato}}, \ and\ \bibinfo {author} {\bibfnamefont {Fernanda}\ \bibnamefont {Steffens}},\ }\bibfield  {title} {\enquote {\bibinfo {title} {{Unpolarized and helicity generalized parton distributions of the proton within lattice QCD}},}\ }\href {\doibase 10.1103/PhysRevLett.125.262001} {\bibfield  {journal} {\bibinfo  {journal} {Phys. Rev. Lett.}\ }\textbf {\bibinfo {volume} {125}},\ \bibinfo {pages} {262001} (\bibinfo {year} {2020})},\ \Eprint {http://arxiv.org/abs/2008.10573} {arXiv:2008.10573 [hep-lat]} \BibitemShut {NoStop}%
\bibitem [{\citenamefont {Alexandrou}\ \emph {et~al.}(2021{\natexlab{a}})\citenamefont {Alexandrou}, \citenamefont {Constantinou}, \citenamefont {Hadjiyiannakou}, \citenamefont {Jansen},\ and\ \citenamefont {Manigrasso}}]{Alexandrou:2020uyt}%
  \BibitemOpen
  \bibfield  {author} {\bibinfo {author} {\bibfnamefont {Constantia}\ \bibnamefont {Alexandrou}}, \bibinfo {author} {\bibfnamefont {Martha}\ \bibnamefont {Constantinou}}, \bibinfo {author} {\bibfnamefont {Kyriakos}\ \bibnamefont {Hadjiyiannakou}}, \bibinfo {author} {\bibfnamefont {Karl}\ \bibnamefont {Jansen}}, \ and\ \bibinfo {author} {\bibfnamefont {Floriano}\ \bibnamefont {Manigrasso}},\ }\bibfield  {title} {\enquote {\bibinfo {title} {{Flavor decomposition for the proton helicity parton distribution functions}},}\ }\href {\doibase 10.1103/PhysRevLett.126.102003} {\bibfield  {journal} {\bibinfo  {journal} {Phys. Rev. Lett.}\ }\textbf {\bibinfo {volume} {126}},\ \bibinfo {pages} {102003} (\bibinfo {year} {2021}{\natexlab{a}})},\ \Eprint {http://arxiv.org/abs/2009.13061} {arXiv:2009.13061 [hep-lat]} \BibitemShut {NoStop}%
\bibitem [{\citenamefont {Alexandrou}\ \emph {et~al.}(2021{\natexlab{b}})\citenamefont {Alexandrou}, \citenamefont {Cichy}, \citenamefont {Constantinou}, \citenamefont {Green}, \citenamefont {Hadjiyiannakou}, \citenamefont {Jansen}, \citenamefont {Manigrasso}, \citenamefont {Scapellato},\ and\ \citenamefont {Steffens}}]{Alexandrou:2020qtt}%
  \BibitemOpen
  \bibfield  {author} {\bibinfo {author} {\bibfnamefont {Constantia}\ \bibnamefont {Alexandrou}}, \bibinfo {author} {\bibfnamefont {Krzysztof}\ \bibnamefont {Cichy}}, \bibinfo {author} {\bibfnamefont {Martha}\ \bibnamefont {Constantinou}}, \bibinfo {author} {\bibfnamefont {Jeremy~R.}\ \bibnamefont {Green}}, \bibinfo {author} {\bibfnamefont {Kyriakos}\ \bibnamefont {Hadjiyiannakou}}, \bibinfo {author} {\bibfnamefont {Karl}\ \bibnamefont {Jansen}}, \bibinfo {author} {\bibfnamefont {Floriano}\ \bibnamefont {Manigrasso}}, \bibinfo {author} {\bibfnamefont {Aurora}\ \bibnamefont {Scapellato}}, \ and\ \bibinfo {author} {\bibfnamefont {Fernanda}\ \bibnamefont {Steffens}},\ }\bibfield  {title} {\enquote {\bibinfo {title} {{Lattice continuum-limit study of nucleon quasi-PDFs}},}\ }\href {\doibase 10.1103/PhysRevD.103.094512} {\bibfield  {journal} {\bibinfo  {journal} {Phys. Rev. D}\ }\textbf {\bibinfo {volume} {103}},\ \bibinfo {pages} {094512} (\bibinfo {year} {2021}{\natexlab{b}})},\ \Eprint
  {http://arxiv.org/abs/2011.00964} {arXiv:2011.00964 [hep-lat]} \BibitemShut {NoStop}%
\bibitem [{\citenamefont {Lin}\ \emph {et~al.}(2021)\citenamefont {Lin}, \citenamefont {Chen}, \citenamefont {Fan}, \citenamefont {Zhang},\ and\ \citenamefont {Zhang}}]{Lin:2020ssv}%
  \BibitemOpen
  \bibfield  {author} {\bibinfo {author} {\bibfnamefont {Huey-Wen}\ \bibnamefont {Lin}}, \bibinfo {author} {\bibfnamefont {Jiunn-Wei}\ \bibnamefont {Chen}}, \bibinfo {author} {\bibfnamefont {Zhouyou}\ \bibnamefont {Fan}}, \bibinfo {author} {\bibfnamefont {Jian-Hui}\ \bibnamefont {Zhang}}, \ and\ \bibinfo {author} {\bibfnamefont {Rui}\ \bibnamefont {Zhang}},\ }\bibfield  {title} {\enquote {\bibinfo {title} {{Valence-Quark Distribution of the Kaon and Pion from Lattice QCD}},}\ }\href {\doibase 10.1103/PhysRevD.103.014516} {\bibfield  {journal} {\bibinfo  {journal} {Phys. Rev. D}\ }\textbf {\bibinfo {volume} {103}},\ \bibinfo {pages} {014516} (\bibinfo {year} {2021})},\ \Eprint {http://arxiv.org/abs/2003.14128} {arXiv:2003.14128 [hep-lat]} \BibitemShut {NoStop}%
\bibitem [{\citenamefont {Fan}\ \emph {et~al.}(2020)\citenamefont {Fan}, \citenamefont {Gao}, \citenamefont {Li}, \citenamefont {Lin}, \citenamefont {Karthik}, \citenamefont {Mukherjee}, \citenamefont {Petreczky}, \citenamefont {Syritsyn}, \citenamefont {Yang},\ and\ \citenamefont {Zhang}}]{Fan:2020nzz}%
  \BibitemOpen
  \bibfield  {author} {\bibinfo {author} {\bibfnamefont {Zhouyou}\ \bibnamefont {Fan}}, \bibinfo {author} {\bibfnamefont {Xiang}\ \bibnamefont {Gao}}, \bibinfo {author} {\bibfnamefont {Ruizi}\ \bibnamefont {Li}}, \bibinfo {author} {\bibfnamefont {Huey-Wen}\ \bibnamefont {Lin}}, \bibinfo {author} {\bibfnamefont {Nikhil}\ \bibnamefont {Karthik}}, \bibinfo {author} {\bibfnamefont {Swagato}\ \bibnamefont {Mukherjee}}, \bibinfo {author} {\bibfnamefont {Peter}\ \bibnamefont {Petreczky}}, \bibinfo {author} {\bibfnamefont {Sergey}\ \bibnamefont {Syritsyn}}, \bibinfo {author} {\bibfnamefont {Yi-Bo}\ \bibnamefont {Yang}}, \ and\ \bibinfo {author} {\bibfnamefont {Rui}\ \bibnamefont {Zhang}},\ }\bibfield  {title} {\enquote {\bibinfo {title} {{Isovector parton distribution functions of the proton on a superfine lattice}},}\ }\href {\doibase 10.1103/PhysRevD.102.074504} {\bibfield  {journal} {\bibinfo  {journal} {Phys. Rev. D}\ }\textbf {\bibinfo {volume} {102}},\ \bibinfo {pages} {074504} (\bibinfo {year} {2020})},\
  \Eprint {http://arxiv.org/abs/2005.12015} {arXiv:2005.12015 [hep-lat]} \BibitemShut {NoStop}%
\bibitem [{\citenamefont {Gao}\ \emph {et~al.}(2020)\citenamefont {Gao}, \citenamefont {Jin}, \citenamefont {Kallidonis}, \citenamefont {Karthik}, \citenamefont {Mukherjee}, \citenamefont {Petreczky}, \citenamefont {Shugert}, \citenamefont {Syritsyn},\ and\ \citenamefont {Zhao}}]{Gao:2020ito}%
  \BibitemOpen
  \bibfield  {author} {\bibinfo {author} {\bibfnamefont {Xiang}\ \bibnamefont {Gao}}, \bibinfo {author} {\bibfnamefont {Luchang}\ \bibnamefont {Jin}}, \bibinfo {author} {\bibfnamefont {Christos}\ \bibnamefont {Kallidonis}}, \bibinfo {author} {\bibfnamefont {Nikhil}\ \bibnamefont {Karthik}}, \bibinfo {author} {\bibfnamefont {Swagato}\ \bibnamefont {Mukherjee}}, \bibinfo {author} {\bibfnamefont {Peter}\ \bibnamefont {Petreczky}}, \bibinfo {author} {\bibfnamefont {Charles}\ \bibnamefont {Shugert}}, \bibinfo {author} {\bibfnamefont {Sergey}\ \bibnamefont {Syritsyn}}, \ and\ \bibinfo {author} {\bibfnamefont {Yong}\ \bibnamefont {Zhao}},\ }\bibfield  {title} {\enquote {\bibinfo {title} {{Valence parton distribution of the pion from lattice QCD: Approaching the continuum limit}},}\ }\href {\doibase 10.1103/PhysRevD.102.094513} {\bibfield  {journal} {\bibinfo  {journal} {Phys. Rev. D}\ }\textbf {\bibinfo {volume} {102}},\ \bibinfo {pages} {094513} (\bibinfo {year} {2020})},\ \Eprint {http://arxiv.org/abs/2007.06590}
  {arXiv:2007.06590 [hep-lat]} \BibitemShut {NoStop}%
\bibitem [{\citenamefont {Bringewatt}\ \emph {et~al.}(2021)\citenamefont {Bringewatt}, \citenamefont {Sato}, \citenamefont {Melnitchouk}, \citenamefont {Qiu}, \citenamefont {Steffens},\ and\ \citenamefont {Constantinou}}]{Bringewatt:2020ixn}%
  \BibitemOpen
  \bibfield  {author} {\bibinfo {author} {\bibfnamefont {J.}~\bibnamefont {Bringewatt}}, \bibinfo {author} {\bibfnamefont {N.}~\bibnamefont {Sato}}, \bibinfo {author} {\bibfnamefont {W.}~\bibnamefont {Melnitchouk}}, \bibinfo {author} {\bibfnamefont {Jian-Wei}\ \bibnamefont {Qiu}}, \bibinfo {author} {\bibfnamefont {F.}~\bibnamefont {Steffens}}, \ and\ \bibinfo {author} {\bibfnamefont {M.}~\bibnamefont {Constantinou}},\ }\bibfield  {title} {\enquote {\bibinfo {title} {{Confronting lattice parton distributions with global QCD analysis}},}\ }\href {\doibase 10.1103/PhysRevD.103.016003} {\bibfield  {journal} {\bibinfo  {journal} {Phys. Rev. D}\ }\textbf {\bibinfo {volume} {103}},\ \bibinfo {pages} {016003} (\bibinfo {year} {2021})},\ \Eprint {http://arxiv.org/abs/2010.00548} {arXiv:2010.00548 [hep-ph]} \BibitemShut {NoStop}%
\bibitem [{\citenamefont {Hua}\ \emph {et~al.}(2021)\citenamefont {Hua}, \citenamefont {Chu}, \citenamefont {Sun}, \citenamefont {Wang}, \citenamefont {Xu}, \citenamefont {Yang}, \citenamefont {Zhang},\ and\ \citenamefont {Zhang}}]{Hua:2020gnw}%
  \BibitemOpen
  \bibfield  {author} {\bibinfo {author} {\bibfnamefont {Jun}\ \bibnamefont {Hua}}, \bibinfo {author} {\bibfnamefont {Min-Huan}\ \bibnamefont {Chu}}, \bibinfo {author} {\bibfnamefont {Peng}\ \bibnamefont {Sun}}, \bibinfo {author} {\bibfnamefont {Wei}\ \bibnamefont {Wang}}, \bibinfo {author} {\bibfnamefont {Ji}~\bibnamefont {Xu}}, \bibinfo {author} {\bibfnamefont {Yi-Bo}\ \bibnamefont {Yang}}, \bibinfo {author} {\bibfnamefont {Jian-Hui}\ \bibnamefont {Zhang}}, \ and\ \bibinfo {author} {\bibfnamefont {Qi-An}\ \bibnamefont {Zhang}} (\bibinfo {collaboration} {Lattice Parton}),\ }\bibfield  {title} {\enquote {\bibinfo {title} {{Distribution Amplitudes of K* and \ensuremath{\phi} at the Physical Pion Mass from Lattice QCD}},}\ }\href {\doibase 10.1103/PhysRevLett.127.062002} {\bibfield  {journal} {\bibinfo  {journal} {Phys. Rev. Lett.}\ }\textbf {\bibinfo {volume} {127}},\ \bibinfo {pages} {062002} (\bibinfo {year} {2021})},\ \Eprint {http://arxiv.org/abs/2011.09788} {arXiv:2011.09788 [hep-lat]} \BibitemShut
  {NoStop}%
\bibitem [{\citenamefont {Alexandrou}\ \emph {et~al.}(2021{\natexlab{c}})\citenamefont {Alexandrou}, \citenamefont {Constantinou}, \citenamefont {Hadjiyiannakou}, \citenamefont {Jansen},\ and\ \citenamefont {Manigrasso}}]{Alexandrou:2021oih}%
  \BibitemOpen
  \bibfield  {author} {\bibinfo {author} {\bibfnamefont {Constantia}\ \bibnamefont {Alexandrou}}, \bibinfo {author} {\bibfnamefont {Martha}\ \bibnamefont {Constantinou}}, \bibinfo {author} {\bibfnamefont {Kyriakos}\ \bibnamefont {Hadjiyiannakou}}, \bibinfo {author} {\bibfnamefont {Karl}\ \bibnamefont {Jansen}}, \ and\ \bibinfo {author} {\bibfnamefont {Floriano}\ \bibnamefont {Manigrasso}},\ }\bibfield  {title} {\enquote {\bibinfo {title} {{Flavor decomposition of the nucleon unpolarized, helicity, and transversity parton distribution functions from lattice QCD simulations}},}\ }\href {\doibase 10.1103/PhysRevD.104.054503} {\bibfield  {journal} {\bibinfo  {journal} {Phys. Rev. D}\ }\textbf {\bibinfo {volume} {104}},\ \bibinfo {pages} {054503} (\bibinfo {year} {2021}{\natexlab{c}})},\ \Eprint {http://arxiv.org/abs/2106.16065} {arXiv:2106.16065 [hep-lat]} \BibitemShut {NoStop}%
\bibitem [{\citenamefont {Alexandrou}\ \emph {et~al.}(2022)\citenamefont {Alexandrou}, \citenamefont {Cichy}, \citenamefont {Constantinou}, \citenamefont {Hadjiyiannakou}, \citenamefont {Jansen}, \citenamefont {Scapellato},\ and\ \citenamefont {Steffens}}]{Alexandrou:2021bbo}%
  \BibitemOpen
  \bibfield  {author} {\bibinfo {author} {\bibfnamefont {Constantia}\ \bibnamefont {Alexandrou}}, \bibinfo {author} {\bibfnamefont {Krzysztof}\ \bibnamefont {Cichy}}, \bibinfo {author} {\bibfnamefont {Martha}\ \bibnamefont {Constantinou}}, \bibinfo {author} {\bibfnamefont {Kyriakos}\ \bibnamefont {Hadjiyiannakou}}, \bibinfo {author} {\bibfnamefont {Karl}\ \bibnamefont {Jansen}}, \bibinfo {author} {\bibfnamefont {Aurora}\ \bibnamefont {Scapellato}}, \ and\ \bibinfo {author} {\bibfnamefont {Fernanda}\ \bibnamefont {Steffens}},\ }\bibfield  {title} {\enquote {\bibinfo {title} {{Transversity GPDs of the proton from lattice QCD}},}\ }\href {\doibase 10.1103/PhysRevD.105.034501} {\bibfield  {journal} {\bibinfo  {journal} {Phys. Rev. D}\ }\textbf {\bibinfo {volume} {105}},\ \bibinfo {pages} {034501} (\bibinfo {year} {2022})},\ \Eprint {http://arxiv.org/abs/2108.10789} {arXiv:2108.10789 [hep-lat]} \BibitemShut {NoStop}%
\bibitem [{\citenamefont {Bhattacharya}\ \emph {et~al.}(2021)\citenamefont {Bhattacharya}, \citenamefont {Cichy}, \citenamefont {Constantinou}, \citenamefont {Metz}, \citenamefont {Scapellato},\ and\ \citenamefont {Steffens}}]{Bhattacharya:2021moj}%
  \BibitemOpen
  \bibfield  {author} {\bibinfo {author} {\bibfnamefont {Shohini}\ \bibnamefont {Bhattacharya}}, \bibinfo {author} {\bibfnamefont {Krzysztof}\ \bibnamefont {Cichy}}, \bibinfo {author} {\bibfnamefont {Martha}\ \bibnamefont {Constantinou}}, \bibinfo {author} {\bibfnamefont {Andreas}\ \bibnamefont {Metz}}, \bibinfo {author} {\bibfnamefont {Aurora}\ \bibnamefont {Scapellato}}, \ and\ \bibinfo {author} {\bibfnamefont {Fernanda}\ \bibnamefont {Steffens}},\ }\bibfield  {title} {\enquote {\bibinfo {title} {{Parton distribution functions beyond leading twist from lattice QCD: The hL(x) case}},}\ }\href {\doibase 10.1103/PhysRevD.104.114510} {\bibfield  {journal} {\bibinfo  {journal} {Phys. Rev. D}\ }\textbf {\bibinfo {volume} {104}},\ \bibinfo {pages} {114510} (\bibinfo {year} {2021})},\ \Eprint {http://arxiv.org/abs/2107.02574} {arXiv:2107.02574 [hep-lat]} \BibitemShut {NoStop}%
\bibitem [{\citenamefont {Gao}\ \emph {et~al.}(2022{\natexlab{a}})\citenamefont {Gao}, \citenamefont {Hanlon}, \citenamefont {Mukherjee}, \citenamefont {Petreczky}, \citenamefont {Scior}, \citenamefont {Syritsyn},\ and\ \citenamefont {Zhao}}]{Gao:2021dbh}%
  \BibitemOpen
  \bibfield  {author} {\bibinfo {author} {\bibfnamefont {Xiang}\ \bibnamefont {Gao}}, \bibinfo {author} {\bibfnamefont {Andrew~D.}\ \bibnamefont {Hanlon}}, \bibinfo {author} {\bibfnamefont {Swagato}\ \bibnamefont {Mukherjee}}, \bibinfo {author} {\bibfnamefont {Peter}\ \bibnamefont {Petreczky}}, \bibinfo {author} {\bibfnamefont {Philipp}\ \bibnamefont {Scior}}, \bibinfo {author} {\bibfnamefont {Sergey}\ \bibnamefont {Syritsyn}}, \ and\ \bibinfo {author} {\bibfnamefont {Yong}\ \bibnamefont {Zhao}},\ }\bibfield  {title} {\enquote {\bibinfo {title} {{Lattice QCD Determination of the Bjorken-x Dependence of Parton Distribution Functions at Next-to-Next-to-Leading Order}},}\ }\href {\doibase 10.1103/PhysRevLett.128.142003} {\bibfield  {journal} {\bibinfo  {journal} {Phys. Rev. Lett.}\ }\textbf {\bibinfo {volume} {128}},\ \bibinfo {pages} {142003} (\bibinfo {year} {2022}{\natexlab{a}})},\ \Eprint {http://arxiv.org/abs/2112.02208} {arXiv:2112.02208 [hep-lat]} \BibitemShut {NoStop}%
\bibitem [{\citenamefont {Hua}\ \emph {et~al.}(2022)\citenamefont {Hua} \emph {et~al.}}]{Hua:2022kcm}%
  \BibitemOpen
  \bibfield  {author} {\bibinfo {author} {\bibfnamefont {Jun}\ \bibnamefont {Hua}} \emph {et~al.} (\bibinfo {collaboration} {Lattice Parton}),\ }\bibfield  {title} {\enquote {\bibinfo {title} {{Pion and Kaon Distribution Amplitudes from Lattice QCD}},}\ }\href {\doibase 10.1103/PhysRevLett.129.132001} {\bibfield  {journal} {\bibinfo  {journal} {Phys. Rev. Lett.}\ }\textbf {\bibinfo {volume} {129}},\ \bibinfo {pages} {132001} (\bibinfo {year} {2022})},\ \Eprint {http://arxiv.org/abs/2201.09173} {arXiv:2201.09173 [hep-lat]} \BibitemShut {NoStop}%
\bibitem [{\citenamefont {Gao}\ \emph {et~al.}(2022{\natexlab{b}})\citenamefont {Gao}, \citenamefont {Hanlon}, \citenamefont {Karthik}, \citenamefont {Mukherjee}, \citenamefont {Petreczky}, \citenamefont {Scior}, \citenamefont {Shi}, \citenamefont {Syritsyn}, \citenamefont {Zhao},\ and\ \citenamefont {Zhou}}]{Gao:2022iex}%
  \BibitemOpen
  \bibfield  {author} {\bibinfo {author} {\bibfnamefont {Xiang}\ \bibnamefont {Gao}}, \bibinfo {author} {\bibfnamefont {Andrew~D.}\ \bibnamefont {Hanlon}}, \bibinfo {author} {\bibfnamefont {Nikhil}\ \bibnamefont {Karthik}}, \bibinfo {author} {\bibfnamefont {Swagato}\ \bibnamefont {Mukherjee}}, \bibinfo {author} {\bibfnamefont {Peter}\ \bibnamefont {Petreczky}}, \bibinfo {author} {\bibfnamefont {Philipp}\ \bibnamefont {Scior}}, \bibinfo {author} {\bibfnamefont {Shuzhe}\ \bibnamefont {Shi}}, \bibinfo {author} {\bibfnamefont {Sergey}\ \bibnamefont {Syritsyn}}, \bibinfo {author} {\bibfnamefont {Yong}\ \bibnamefont {Zhao}}, \ and\ \bibinfo {author} {\bibfnamefont {Kai}\ \bibnamefont {Zhou}},\ }\bibfield  {title} {\enquote {\bibinfo {title} {{Continuum-extrapolated NNLO valence PDF of the pion at the physical point}},}\ }\href {\doibase 10.1103/PhysRevD.106.114510} {\bibfield  {journal} {\bibinfo  {journal} {Phys. Rev. D}\ }\textbf {\bibinfo {volume} {106}},\ \bibinfo {pages} {114510} (\bibinfo {year}
  {2022}{\natexlab{b}})},\ \Eprint {http://arxiv.org/abs/2208.02297} {arXiv:2208.02297 [hep-lat]} \BibitemShut {NoStop}%
\bibitem [{\citenamefont {Bhattacharya}\ \emph {et~al.}(2022)\citenamefont {Bhattacharya}, \citenamefont {Cichy}, \citenamefont {Constantinou}, \citenamefont {Dodson}, \citenamefont {Gao}, \citenamefont {Metz}, \citenamefont {Mukherjee}, \citenamefont {Scapellato}, \citenamefont {Steffens},\ and\ \citenamefont {Zhao}}]{Bhattacharya:2022aob}%
  \BibitemOpen
  \bibfield  {author} {\bibinfo {author} {\bibfnamefont {Shohini}\ \bibnamefont {Bhattacharya}}, \bibinfo {author} {\bibfnamefont {Krzysztof}\ \bibnamefont {Cichy}}, \bibinfo {author} {\bibfnamefont {Martha}\ \bibnamefont {Constantinou}}, \bibinfo {author} {\bibfnamefont {Jack}\ \bibnamefont {Dodson}}, \bibinfo {author} {\bibfnamefont {Xiang}\ \bibnamefont {Gao}}, \bibinfo {author} {\bibfnamefont {Andreas}\ \bibnamefont {Metz}}, \bibinfo {author} {\bibfnamefont {Swagato}\ \bibnamefont {Mukherjee}}, \bibinfo {author} {\bibfnamefont {Aurora}\ \bibnamefont {Scapellato}}, \bibinfo {author} {\bibfnamefont {Fernanda}\ \bibnamefont {Steffens}}, \ and\ \bibinfo {author} {\bibfnamefont {Yong}\ \bibnamefont {Zhao}},\ }\bibfield  {title} {\enquote {\bibinfo {title} {{Generalized parton distributions from lattice QCD with asymmetric momentum transfer: Unpolarized quarks}},}\ }\href {\doibase 10.1103/PhysRevD.106.114512} {\bibfield  {journal} {\bibinfo  {journal} {Phys. Rev. D}\ }\textbf {\bibinfo {volume} {106}},\
  \bibinfo {pages} {114512} (\bibinfo {year} {2022})},\ \Eprint {http://arxiv.org/abs/2209.05373} {arXiv:2209.05373 [hep-lat]} \BibitemShut {NoStop}%
\bibitem [{\citenamefont {Yao}\ \emph {et~al.}(2023)\citenamefont {Yao} \emph {et~al.}}]{LatticeParton:2022xsd}%
  \BibitemOpen
  \bibfield  {author} {\bibinfo {author} {\bibfnamefont {Fei}\ \bibnamefont {Yao}} \emph {et~al.} (\bibinfo {collaboration} {Lattice Parton}),\ }\bibfield  {title} {\enquote {\bibinfo {title} {{Nucleon Transversity Distribution in the Continuum and Physical Mass Limit from Lattice QCD}},}\ }\href {\doibase 10.1103/PhysRevLett.131.261901} {\bibfield  {journal} {\bibinfo  {journal} {Phys. Rev. Lett.}\ }\textbf {\bibinfo {volume} {131}},\ \bibinfo {pages} {261901} (\bibinfo {year} {2023})},\ \Eprint {http://arxiv.org/abs/2208.08008} {arXiv:2208.08008 [hep-lat]} \BibitemShut {NoStop}%
\bibitem [{\citenamefont {Bhattacharya}\ \emph {et~al.}(2023)\citenamefont {Bhattacharya}, \citenamefont {Cichy}, \citenamefont {Constantinou}, \citenamefont {Dodson}, \citenamefont {Metz}, \citenamefont {Scapellato},\ and\ \citenamefont {Steffens}}]{Bhattacharya:2023nmv}%
  \BibitemOpen
  \bibfield  {author} {\bibinfo {author} {\bibfnamefont {Shohini}\ \bibnamefont {Bhattacharya}}, \bibinfo {author} {\bibfnamefont {Krzysztof}\ \bibnamefont {Cichy}}, \bibinfo {author} {\bibfnamefont {Martha}\ \bibnamefont {Constantinou}}, \bibinfo {author} {\bibfnamefont {Jack}\ \bibnamefont {Dodson}}, \bibinfo {author} {\bibfnamefont {Andreas}\ \bibnamefont {Metz}}, \bibinfo {author} {\bibfnamefont {Aurora}\ \bibnamefont {Scapellato}}, \ and\ \bibinfo {author} {\bibfnamefont {Fernanda}\ \bibnamefont {Steffens}},\ }\bibfield  {title} {\enquote {\bibinfo {title} {{Chiral-even axial twist-3 GPDs of the proton from lattice QCD}},}\ }\href {\doibase 10.1103/PhysRevD.108.054501} {\bibfield  {journal} {\bibinfo  {journal} {Phys. Rev. D}\ }\textbf {\bibinfo {volume} {108}},\ \bibinfo {pages} {054501} (\bibinfo {year} {2023})},\ \Eprint {http://arxiv.org/abs/2306.05533} {arXiv:2306.05533 [hep-lat]} \BibitemShut {NoStop}%
\bibitem [{\citenamefont {Orginos}\ \emph {et~al.}(2017)\citenamefont {Orginos}, \citenamefont {Radyushkin}, \citenamefont {Karpie},\ and\ \citenamefont {Zafeiropoulos}}]{Orginos:2017kos}%
  \BibitemOpen
  \bibfield  {author} {\bibinfo {author} {\bibfnamefont {Kostas}\ \bibnamefont {Orginos}}, \bibinfo {author} {\bibfnamefont {Anatoly}\ \bibnamefont {Radyushkin}}, \bibinfo {author} {\bibfnamefont {Joseph}\ \bibnamefont {Karpie}}, \ and\ \bibinfo {author} {\bibfnamefont {Savvas}\ \bibnamefont {Zafeiropoulos}},\ }\bibfield  {title} {\enquote {\bibinfo {title} {{Lattice QCD exploration of parton pseudo-distribution functions}},}\ }\href {\doibase 10.1103/PhysRevD.96.094503} {\bibfield  {journal} {\bibinfo  {journal} {Phys. Rev. D}\ }\textbf {\bibinfo {volume} {96}},\ \bibinfo {pages} {094503} (\bibinfo {year} {2017})},\ \Eprint {http://arxiv.org/abs/1706.05373} {arXiv:1706.05373 [hep-ph]} \BibitemShut {NoStop}%
\bibitem [{\citenamefont {Karpie}\ \emph {et~al.}(2018)\citenamefont {Karpie}, \citenamefont {Orginos},\ and\ \citenamefont {Zafeiropoulos}}]{Karpie:2018zaz}%
  \BibitemOpen
  \bibfield  {author} {\bibinfo {author} {\bibfnamefont {Joseph}\ \bibnamefont {Karpie}}, \bibinfo {author} {\bibfnamefont {Kostas}\ \bibnamefont {Orginos}}, \ and\ \bibinfo {author} {\bibfnamefont {Savvas}\ \bibnamefont {Zafeiropoulos}},\ }\bibfield  {title} {\enquote {\bibinfo {title} {{Moments of Ioffe time parton distribution functions from non-local matrix elements}},}\ }\href {\doibase 10.1007/JHEP11(2018)178} {\bibfield  {journal} {\bibinfo  {journal} {JHEP}\ }\textbf {\bibinfo {volume} {11}},\ \bibinfo {pages} {178} (\bibinfo {year} {2018})},\ \Eprint {http://arxiv.org/abs/1807.10933} {arXiv:1807.10933 [hep-lat]} \BibitemShut {NoStop}%
\bibitem [{\citenamefont {Karpie}\ \emph {et~al.}(2019)\citenamefont {Karpie}, \citenamefont {Orginos}, \citenamefont {Rothkopf},\ and\ \citenamefont {Zafeiropoulos}}]{Karpie:2019eiq}%
  \BibitemOpen
  \bibfield  {author} {\bibinfo {author} {\bibfnamefont {Joseph}\ \bibnamefont {Karpie}}, \bibinfo {author} {\bibfnamefont {Kostas}\ \bibnamefont {Orginos}}, \bibinfo {author} {\bibfnamefont {Alexander}\ \bibnamefont {Rothkopf}}, \ and\ \bibinfo {author} {\bibfnamefont {Savvas}\ \bibnamefont {Zafeiropoulos}},\ }\bibfield  {title} {\enquote {\bibinfo {title} {{Reconstructing parton distribution functions from Ioffe time data: from Bayesian methods to Neural Networks}},}\ }\href {\doibase 10.1007/JHEP04(2019)057} {\bibfield  {journal} {\bibinfo  {journal} {JHEP}\ }\textbf {\bibinfo {volume} {04}},\ \bibinfo {pages} {057} (\bibinfo {year} {2019})},\ \Eprint {http://arxiv.org/abs/1901.05408} {arXiv:1901.05408 [hep-lat]} \BibitemShut {NoStop}%
\bibitem [{\citenamefont {Jo\'o}\ \emph {et~al.}(2019{\natexlab{a}})\citenamefont {Jo\'o}, \citenamefont {Karpie}, \citenamefont {Orginos}, \citenamefont {Radyushkin}, \citenamefont {Richards},\ and\ \citenamefont {Zafeiropoulos}}]{Joo:2019jct}%
  \BibitemOpen
  \bibfield  {author} {\bibinfo {author} {\bibfnamefont {B\'alint}\ \bibnamefont {Jo\'o}}, \bibinfo {author} {\bibfnamefont {Joseph}\ \bibnamefont {Karpie}}, \bibinfo {author} {\bibfnamefont {Kostas}\ \bibnamefont {Orginos}}, \bibinfo {author} {\bibfnamefont {Anatoly}\ \bibnamefont {Radyushkin}}, \bibinfo {author} {\bibfnamefont {David}\ \bibnamefont {Richards}}, \ and\ \bibinfo {author} {\bibfnamefont {Savvas}\ \bibnamefont {Zafeiropoulos}},\ }\bibfield  {title} {\enquote {\bibinfo {title} {{Parton Distribution Functions from Ioffe time pseudo-distributions}},}\ }\href {\doibase 10.1007/JHEP12(2019)081} {\bibfield  {journal} {\bibinfo  {journal} {JHEP}\ }\textbf {\bibinfo {volume} {12}},\ \bibinfo {pages} {081} (\bibinfo {year} {2019}{\natexlab{a}})},\ \Eprint {http://arxiv.org/abs/1908.09771} {arXiv:1908.09771 [hep-lat]} \BibitemShut {NoStop}%
\bibitem [{\citenamefont {Jo\'o}\ \emph {et~al.}(2019{\natexlab{b}})\citenamefont {Jo\'o}, \citenamefont {Karpie}, \citenamefont {Orginos}, \citenamefont {Radyushkin}, \citenamefont {Richards}, \citenamefont {Sufian},\ and\ \citenamefont {Zafeiropoulos}}]{Joo:2019bzr}%
  \BibitemOpen
  \bibfield  {author} {\bibinfo {author} {\bibfnamefont {B\'alint}\ \bibnamefont {Jo\'o}}, \bibinfo {author} {\bibfnamefont {Joseph}\ \bibnamefont {Karpie}}, \bibinfo {author} {\bibfnamefont {Kostas}\ \bibnamefont {Orginos}}, \bibinfo {author} {\bibfnamefont {Anatoly~V.}\ \bibnamefont {Radyushkin}}, \bibinfo {author} {\bibfnamefont {David~G.}\ \bibnamefont {Richards}}, \bibinfo {author} {\bibfnamefont {Raza~Sabbir}\ \bibnamefont {Sufian}}, \ and\ \bibinfo {author} {\bibfnamefont {Savvas}\ \bibnamefont {Zafeiropoulos}},\ }\bibfield  {title} {\enquote {\bibinfo {title} {{Pion valence structure from Ioffe-time parton pseudodistribution functions}},}\ }\href {\doibase 10.1103/PhysRevD.100.114512} {\bibfield  {journal} {\bibinfo  {journal} {Phys. Rev. D}\ }\textbf {\bibinfo {volume} {100}},\ \bibinfo {pages} {114512} (\bibinfo {year} {2019}{\natexlab{b}})},\ \Eprint {http://arxiv.org/abs/1909.08517} {arXiv:1909.08517 [hep-lat]} \BibitemShut {NoStop}%
\bibitem [{\citenamefont {Jo\'o}\ \emph {et~al.}(2020)\citenamefont {Jo\'o}, \citenamefont {Karpie}, \citenamefont {Orginos}, \citenamefont {Radyushkin}, \citenamefont {Richards},\ and\ \citenamefont {Zafeiropoulos}}]{Joo:2020spy}%
  \BibitemOpen
  \bibfield  {author} {\bibinfo {author} {\bibfnamefont {B\'alint}\ \bibnamefont {Jo\'o}}, \bibinfo {author} {\bibfnamefont {Joseph}\ \bibnamefont {Karpie}}, \bibinfo {author} {\bibfnamefont {Kostas}\ \bibnamefont {Orginos}}, \bibinfo {author} {\bibfnamefont {Anatoly~V.}\ \bibnamefont {Radyushkin}}, \bibinfo {author} {\bibfnamefont {David~G.}\ \bibnamefont {Richards}}, \ and\ \bibinfo {author} {\bibfnamefont {Savvas}\ \bibnamefont {Zafeiropoulos}},\ }\bibfield  {title} {\enquote {\bibinfo {title} {{Parton Distribution Functions from Ioffe Time Pseudodistributions from Lattice Calculations: Approaching the Physical Point}},}\ }\href {\doibase 10.1103/PhysRevLett.125.232003} {\bibfield  {journal} {\bibinfo  {journal} {Phys. Rev. Lett.}\ }\textbf {\bibinfo {volume} {125}},\ \bibinfo {pages} {232003} (\bibinfo {year} {2020})},\ \Eprint {http://arxiv.org/abs/2004.01687} {arXiv:2004.01687 [hep-lat]} \BibitemShut {NoStop}%
\bibitem [{\citenamefont {Bhat}\ \emph {et~al.}(2021)\citenamefont {Bhat}, \citenamefont {Cichy}, \citenamefont {Constantinou},\ and\ \citenamefont {Scapellato}}]{Bhat:2020ktg}%
  \BibitemOpen
  \bibfield  {author} {\bibinfo {author} {\bibfnamefont {Manjunath}\ \bibnamefont {Bhat}}, \bibinfo {author} {\bibfnamefont {Krzysztof}\ \bibnamefont {Cichy}}, \bibinfo {author} {\bibfnamefont {Martha}\ \bibnamefont {Constantinou}}, \ and\ \bibinfo {author} {\bibfnamefont {Aurora}\ \bibnamefont {Scapellato}},\ }\bibfield  {title} {\enquote {\bibinfo {title} {{Flavor nonsinglet parton distribution functions from lattice QCD at physical quark masses via the pseudodistribution approach}},}\ }\href {\doibase 10.1103/PhysRevD.103.034510} {\bibfield  {journal} {\bibinfo  {journal} {Phys. Rev. D}\ }\textbf {\bibinfo {volume} {103}},\ \bibinfo {pages} {034510} (\bibinfo {year} {2021})},\ \Eprint {http://arxiv.org/abs/2005.02102} {arXiv:2005.02102 [hep-lat]} \BibitemShut {NoStop}%
\bibitem [{\citenamefont {Del~Debbio}\ \emph {et~al.}(2021)\citenamefont {Del~Debbio}, \citenamefont {Giani}, \citenamefont {Karpie}, \citenamefont {Orginos}, \citenamefont {Radyushkin},\ and\ \citenamefont {Zafeiropoulos}}]{DelDebbio:2020rgv}%
  \BibitemOpen
  \bibfield  {author} {\bibinfo {author} {\bibfnamefont {Luigi}\ \bibnamefont {Del~Debbio}}, \bibinfo {author} {\bibfnamefont {Tommaso}\ \bibnamefont {Giani}}, \bibinfo {author} {\bibfnamefont {Joseph}\ \bibnamefont {Karpie}}, \bibinfo {author} {\bibfnamefont {Kostas}\ \bibnamefont {Orginos}}, \bibinfo {author} {\bibfnamefont {Anatoly}\ \bibnamefont {Radyushkin}}, \ and\ \bibinfo {author} {\bibfnamefont {Savvas}\ \bibnamefont {Zafeiropoulos}},\ }\bibfield  {title} {\enquote {\bibinfo {title} {{Neural-network analysis of Parton Distribution Functions from Ioffe-time pseudodistributions}},}\ }\href {\doibase 10.1007/JHEP02(2021)138} {\bibfield  {journal} {\bibinfo  {journal} {JHEP}\ }\textbf {\bibinfo {volume} {02}},\ \bibinfo {pages} {138} (\bibinfo {year} {2021})},\ \Eprint {http://arxiv.org/abs/2010.03996} {arXiv:2010.03996 [hep-ph]} \BibitemShut {NoStop}%
\bibitem [{\citenamefont {Karpie}\ \emph {et~al.}(2021)\citenamefont {Karpie}, \citenamefont {Orginos}, \citenamefont {Radyushkin},\ and\ \citenamefont {Zafeiropoulos}}]{Karpie:2021pap}%
  \BibitemOpen
  \bibfield  {author} {\bibinfo {author} {\bibfnamefont {Joseph}\ \bibnamefont {Karpie}}, \bibinfo {author} {\bibfnamefont {Kostas}\ \bibnamefont {Orginos}}, \bibinfo {author} {\bibfnamefont {Anatoly}\ \bibnamefont {Radyushkin}}, \ and\ \bibinfo {author} {\bibfnamefont {Savvas}\ \bibnamefont {Zafeiropoulos}} (\bibinfo {collaboration} {HadStruc}),\ }\bibfield  {title} {\enquote {\bibinfo {title} {{The continuum and leading twist limits of parton distribution functions in lattice QCD}},}\ }\href {\doibase 10.1007/JHEP11(2021)024} {\bibfield  {journal} {\bibinfo  {journal} {JHEP}\ }\textbf {\bibinfo {volume} {11}},\ \bibinfo {pages} {024} (\bibinfo {year} {2021})},\ \Eprint {http://arxiv.org/abs/2105.13313} {arXiv:2105.13313 [hep-lat]} \BibitemShut {NoStop}%
\bibitem [{\citenamefont {Egerer}\ \emph {et~al.}(2021)\citenamefont {Egerer}, \citenamefont {Edwards}, \citenamefont {Kallidonis}, \citenamefont {Orginos}, \citenamefont {Radyushkin}, \citenamefont {Richards}, \citenamefont {Romero},\ and\ \citenamefont {Zafeiropoulos}}]{Egerer:2021ymv}%
  \BibitemOpen
  \bibfield  {author} {\bibinfo {author} {\bibfnamefont {Colin}\ \bibnamefont {Egerer}}, \bibinfo {author} {\bibfnamefont {Robert~G.}\ \bibnamefont {Edwards}}, \bibinfo {author} {\bibfnamefont {Christos}\ \bibnamefont {Kallidonis}}, \bibinfo {author} {\bibfnamefont {Kostas}\ \bibnamefont {Orginos}}, \bibinfo {author} {\bibfnamefont {Anatoly~V.}\ \bibnamefont {Radyushkin}}, \bibinfo {author} {\bibfnamefont {David~G.}\ \bibnamefont {Richards}}, \bibinfo {author} {\bibfnamefont {Eloy}\ \bibnamefont {Romero}}, \ and\ \bibinfo {author} {\bibfnamefont {Savvas}\ \bibnamefont {Zafeiropoulos}} (\bibinfo {collaboration} {HadStruc}),\ }\bibfield  {title} {\enquote {\bibinfo {title} {{Towards high-precision parton distributions from lattice QCD via distillation}},}\ }\href {\doibase 10.1007/JHEP11(2021)148} {\bibfield  {journal} {\bibinfo  {journal} {JHEP}\ }\textbf {\bibinfo {volume} {11}},\ \bibinfo {pages} {148} (\bibinfo {year} {2021})},\ \Eprint {http://arxiv.org/abs/2107.05199} {arXiv:2107.05199 [hep-lat]}
  \BibitemShut {NoStop}%
\bibitem [{\citenamefont {Egerer}\ \emph {et~al.}(2022{\natexlab{a}})\citenamefont {Egerer} \emph {et~al.}}]{HadStruc:2021qdf}%
  \BibitemOpen
  \bibfield  {author} {\bibinfo {author} {\bibfnamefont {Colin}\ \bibnamefont {Egerer}} \emph {et~al.} (\bibinfo {collaboration} {HadStruc}),\ }\bibfield  {title} {\enquote {\bibinfo {title} {{Transversity parton distribution function of the nucleon using the pseudodistribution approach}},}\ }\href {\doibase 10.1103/PhysRevD.105.034507} {\bibfield  {journal} {\bibinfo  {journal} {Phys. Rev. D}\ }\textbf {\bibinfo {volume} {105}},\ \bibinfo {pages} {034507} (\bibinfo {year} {2022}{\natexlab{a}})},\ \Eprint {http://arxiv.org/abs/2111.01808} {arXiv:2111.01808 [hep-lat]} \BibitemShut {NoStop}%
\bibitem [{\citenamefont {Bhat}\ \emph {et~al.}(2022)\citenamefont {Bhat}, \citenamefont {Chomicki}, \citenamefont {Cichy}, \citenamefont {Constantinou}, \citenamefont {Green},\ and\ \citenamefont {Scapellato}}]{Bhat:2022zrw}%
  \BibitemOpen
  \bibfield  {author} {\bibinfo {author} {\bibfnamefont {Manjunath}\ \bibnamefont {Bhat}}, \bibinfo {author} {\bibfnamefont {Wojciech}\ \bibnamefont {Chomicki}}, \bibinfo {author} {\bibfnamefont {Krzysztof}\ \bibnamefont {Cichy}}, \bibinfo {author} {\bibfnamefont {Martha}\ \bibnamefont {Constantinou}}, \bibinfo {author} {\bibfnamefont {Jeremy~R.}\ \bibnamefont {Green}}, \ and\ \bibinfo {author} {\bibfnamefont {Aurora}\ \bibnamefont {Scapellato}},\ }\bibfield  {title} {\enquote {\bibinfo {title} {{Continuum limit of parton distribution functions from the pseudodistribution approach on the lattice}},}\ }\href {\doibase 10.1103/PhysRevD.106.054504} {\bibfield  {journal} {\bibinfo  {journal} {Phys. Rev. D}\ }\textbf {\bibinfo {volume} {106}},\ \bibinfo {pages} {054504} (\bibinfo {year} {2022})},\ \Eprint {http://arxiv.org/abs/2205.07585} {arXiv:2205.07585 [hep-lat]} \BibitemShut {NoStop}%
\bibitem [{\citenamefont {Edwards}\ \emph {et~al.}(2023)\citenamefont {Edwards} \emph {et~al.}}]{HadStruc:2022nay}%
  \BibitemOpen
  \bibfield  {author} {\bibinfo {author} {\bibfnamefont {Robert~G.}\ \bibnamefont {Edwards}} \emph {et~al.} (\bibinfo {collaboration} {HadStruc}),\ }\bibfield  {title} {\enquote {\bibinfo {title} {{Non-singlet quark helicity PDFs of the nucleon from pseudo-distributions}},}\ }\href {\doibase 10.1007/JHEP03(2023)086} {\bibfield  {journal} {\bibinfo  {journal} {JHEP}\ }\textbf {\bibinfo {volume} {03}},\ \bibinfo {pages} {086} (\bibinfo {year} {2023})},\ \Eprint {http://arxiv.org/abs/2211.04434} {arXiv:2211.04434 [hep-lat]} \BibitemShut {NoStop}%
\bibitem [{\citenamefont {Ji}\ \emph {et~al.}(2015)\citenamefont {Ji}, \citenamefont {Sun}, \citenamefont {Xiong},\ and\ \citenamefont {Yuan}}]{Ji:2014hxa}%
  \BibitemOpen
  \bibfield  {author} {\bibinfo {author} {\bibfnamefont {Xiangdong}\ \bibnamefont {Ji}}, \bibinfo {author} {\bibfnamefont {Peng}\ \bibnamefont {Sun}}, \bibinfo {author} {\bibfnamefont {Xiaonu}\ \bibnamefont {Xiong}}, \ and\ \bibinfo {author} {\bibfnamefont {Feng}\ \bibnamefont {Yuan}},\ }\bibfield  {title} {\enquote {\bibinfo {title} {{Soft factor subtraction and transverse momentum dependent parton distributions on the lattice}},}\ }\href {\doibase 10.1103/PhysRevD.91.074009} {\bibfield  {journal} {\bibinfo  {journal} {Phys. Rev. D}\ }\textbf {\bibinfo {volume} {91}},\ \bibinfo {pages} {074009} (\bibinfo {year} {2015})},\ \Eprint {http://arxiv.org/abs/1405.7640} {arXiv:1405.7640 [hep-ph]} \BibitemShut {NoStop}%
\bibitem [{\citenamefont {Engelhardt}\ \emph {et~al.}(2016)\citenamefont {Engelhardt}, \citenamefont {H\"agler}, \citenamefont {Musch}, \citenamefont {Negele},\ and\ \citenamefont {Sch\"afer}}]{Engelhardt:2015xja}%
  \BibitemOpen
  \bibfield  {author} {\bibinfo {author} {\bibfnamefont {M.}~\bibnamefont {Engelhardt}}, \bibinfo {author} {\bibfnamefont {P.}~\bibnamefont {H\"agler}}, \bibinfo {author} {\bibfnamefont {B.}~\bibnamefont {Musch}}, \bibinfo {author} {\bibfnamefont {J.}~\bibnamefont {Negele}}, \ and\ \bibinfo {author} {\bibfnamefont {A.}~\bibnamefont {Sch\"afer}},\ }\bibfield  {title} {\enquote {\bibinfo {title} {{Lattice QCD study of the Boer-Mulders effect in a pion}},}\ }\href {\doibase 10.1103/PhysRevD.93.054501} {\bibfield  {journal} {\bibinfo  {journal} {Phys. Rev. D}\ }\textbf {\bibinfo {volume} {93}},\ \bibinfo {pages} {054501} (\bibinfo {year} {2016})},\ \Eprint {http://arxiv.org/abs/1506.07826} {arXiv:1506.07826 [hep-lat]} \BibitemShut {NoStop}%
\bibitem [{\citenamefont {Radyushkin}(2017{\natexlab{a}})}]{Radyushkin:2016hsy}%
  \BibitemOpen
  \bibfield  {author} {\bibinfo {author} {\bibfnamefont {Anatoly}\ \bibnamefont {Radyushkin}},\ }\bibfield  {title} {\enquote {\bibinfo {title} {{Nonperturbative Evolution of Parton Quasi-Distributions}},}\ }\href@noop {} {\bibfield  {journal} {\bibinfo  {journal} {Phys. Lett. B}\ }\textbf {\bibinfo {volume} {767}},\ \bibinfo {pages} {314--320} (\bibinfo {year} {2017}{\natexlab{a}})},\ \Eprint {http://arxiv.org/abs/1612.05170} {arXiv:1612.05170 [hep-ph]} \BibitemShut {NoStop}%
\bibitem [{\citenamefont {Radyushkin}(2017{\natexlab{b}})}]{Radyushkin:2017ffo}%
  \BibitemOpen
  \bibfield  {author} {\bibinfo {author} {\bibfnamefont {Anatoly}\ \bibnamefont {Radyushkin}},\ }\bibfield  {title} {\enquote {\bibinfo {title} {{Target Mass Effects in Parton Quasi-Distributions}},}\ }\href {\doibase 10.1016/j.physletb.2017.05.024} {\bibfield  {journal} {\bibinfo  {journal} {Phys. Lett. B}\ }\textbf {\bibinfo {volume} {770}},\ \bibinfo {pages} {514--522} (\bibinfo {year} {2017}{\natexlab{b}})},\ \Eprint {http://arxiv.org/abs/1702.01726} {arXiv:1702.01726 [hep-ph]} \BibitemShut {NoStop}%
\bibitem [{\citenamefont {Yoon}\ \emph {et~al.}(2017)\citenamefont {Yoon}, \citenamefont {Engelhardt}, \citenamefont {Gupta}, \citenamefont {Bhattacharya}, \citenamefont {Green}, \citenamefont {Musch}, \citenamefont {Negele}, \citenamefont {Pochinsky}, \citenamefont {Sch\"afer},\ and\ \citenamefont {Syritsyn}}]{Yoon:2017qzo}%
  \BibitemOpen
  \bibfield  {author} {\bibinfo {author} {\bibfnamefont {Boram}\ \bibnamefont {Yoon}}, \bibinfo {author} {\bibfnamefont {Michael}\ \bibnamefont {Engelhardt}}, \bibinfo {author} {\bibfnamefont {Rajan}\ \bibnamefont {Gupta}}, \bibinfo {author} {\bibfnamefont {Tanmoy}\ \bibnamefont {Bhattacharya}}, \bibinfo {author} {\bibfnamefont {Jeremy~R.}\ \bibnamefont {Green}}, \bibinfo {author} {\bibfnamefont {Bernhard~U.}\ \bibnamefont {Musch}}, \bibinfo {author} {\bibfnamefont {John~W.}\ \bibnamefont {Negele}}, \bibinfo {author} {\bibfnamefont {Andrew~V.}\ \bibnamefont {Pochinsky}}, \bibinfo {author} {\bibfnamefont {Andreas}\ \bibnamefont {Sch\"afer}}, \ and\ \bibinfo {author} {\bibfnamefont {Sergey~N.}\ \bibnamefont {Syritsyn}},\ }\bibfield  {title} {\enquote {\bibinfo {title} {{Nucleon Transverse Momentum-dependent Parton Distributions in Lattice QCD: Renormalization Patterns and Discretization Effects}},}\ }\href {\doibase 10.1103/PhysRevD.96.094508} {\bibfield  {journal} {\bibinfo  {journal} {Phys. Rev. D}\ }\textbf
  {\bibinfo {volume} {96}},\ \bibinfo {pages} {094508} (\bibinfo {year} {2017})},\ \Eprint {http://arxiv.org/abs/1706.03406} {arXiv:1706.03406 [hep-lat]} \BibitemShut {NoStop}%
\bibitem [{\citenamefont {Broniowski}\ and\ \citenamefont {Ruiz~Arriola}(2018)}]{Broniowski:2017gfp}%
  \BibitemOpen
  \bibfield  {author} {\bibinfo {author} {\bibfnamefont {Wojciech}\ \bibnamefont {Broniowski}}\ and\ \bibinfo {author} {\bibfnamefont {Enrique}\ \bibnamefont {Ruiz~Arriola}},\ }\bibfield  {title} {\enquote {\bibinfo {title} {{Partonic quasidistributions of the proton and pion from transverse-momentum distributions}},}\ }\href {\doibase 10.1103/PhysRevD.97.034031} {\bibfield  {journal} {\bibinfo  {journal} {Phys. Rev. D}\ }\textbf {\bibinfo {volume} {97}},\ \bibinfo {pages} {034031} (\bibinfo {year} {2018})},\ \Eprint {http://arxiv.org/abs/1711.03377} {arXiv:1711.03377 [hep-ph]} \BibitemShut {NoStop}%
\bibitem [{\citenamefont {Ji}\ \emph {et~al.}(2019)\citenamefont {Ji}, \citenamefont {Jin}, \citenamefont {Yuan}, \citenamefont {Zhang},\ and\ \citenamefont {Zhao}}]{Ji:2018hvs}%
  \BibitemOpen
  \bibfield  {author} {\bibinfo {author} {\bibfnamefont {Xiangdong}\ \bibnamefont {Ji}}, \bibinfo {author} {\bibfnamefont {Lu-Chang}\ \bibnamefont {Jin}}, \bibinfo {author} {\bibfnamefont {Feng}\ \bibnamefont {Yuan}}, \bibinfo {author} {\bibfnamefont {Jian-Hui}\ \bibnamefont {Zhang}}, \ and\ \bibinfo {author} {\bibfnamefont {Yong}\ \bibnamefont {Zhao}},\ }\bibfield  {title} {\enquote {\bibinfo {title} {{Transverse momentum dependent parton quasidistributions}},}\ }\href {\doibase 10.1103/PhysRevD.99.114006} {\bibfield  {journal} {\bibinfo  {journal} {Phys. Rev. D}\ }\textbf {\bibinfo {volume} {99}},\ \bibinfo {pages} {114006} (\bibinfo {year} {2019})},\ \Eprint {http://arxiv.org/abs/1801.05930} {arXiv:1801.05930 [hep-ph]} \BibitemShut {NoStop}%
\bibitem [{\citenamefont {Shanahan}\ \emph {et~al.}(2020{\natexlab{a}})\citenamefont {Shanahan}, \citenamefont {Wagman},\ and\ \citenamefont {Zhao}}]{Shanahan:2019zcq}%
  \BibitemOpen
  \bibfield  {author} {\bibinfo {author} {\bibfnamefont {Phiala}\ \bibnamefont {Shanahan}}, \bibinfo {author} {\bibfnamefont {Michael~L.}\ \bibnamefont {Wagman}}, \ and\ \bibinfo {author} {\bibfnamefont {Yong}\ \bibnamefont {Zhao}},\ }\bibfield  {title} {\enquote {\bibinfo {title} {{Nonperturbative renormalization of staple-shaped Wilson line operators in lattice QCD}},}\ }\href {\doibase 10.1103/PhysRevD.101.074505} {\bibfield  {journal} {\bibinfo  {journal} {Phys. Rev. D}\ }\textbf {\bibinfo {volume} {101}},\ \bibinfo {pages} {074505} (\bibinfo {year} {2020}{\natexlab{a}})},\ \Eprint {http://arxiv.org/abs/1911.00800} {arXiv:1911.00800 [hep-lat]} \BibitemShut {NoStop}%
\bibitem [{\citenamefont {Ebert}\ \emph {et~al.}(2022)\citenamefont {Ebert}, \citenamefont {Schindler}, \citenamefont {Stewart},\ and\ \citenamefont {Zhao}}]{Ebert:2022fmh}%
  \BibitemOpen
  \bibfield  {author} {\bibinfo {author} {\bibfnamefont {Markus~A.}\ \bibnamefont {Ebert}}, \bibinfo {author} {\bibfnamefont {Stella~T.}\ \bibnamefont {Schindler}}, \bibinfo {author} {\bibfnamefont {Iain~W.}\ \bibnamefont {Stewart}}, \ and\ \bibinfo {author} {\bibfnamefont {Yong}\ \bibnamefont {Zhao}},\ }\bibfield  {title} {\enquote {\bibinfo {title} {{Factorization connecting continuum \& lattice TMDs}},}\ }\href {\doibase 10.1007/JHEP04(2022)178} {\bibfield  {journal} {\bibinfo  {journal} {JHEP}\ }\textbf {\bibinfo {volume} {04}},\ \bibinfo {pages} {178} (\bibinfo {year} {2022})},\ \Eprint {http://arxiv.org/abs/2201.08401} {arXiv:2201.08401 [hep-ph]} \BibitemShut {NoStop}%
\bibitem [{\citenamefont {Zhang}\ \emph {et~al.}(2020{\natexlab{b}})\citenamefont {Zhang} \emph {et~al.}}]{LatticeParton:2020uhz}%
  \BibitemOpen
  \bibfield  {author} {\bibinfo {author} {\bibfnamefont {Qi-An}\ \bibnamefont {Zhang}} \emph {et~al.} (\bibinfo {collaboration} {Lattice Parton}),\ }\bibfield  {title} {\enquote {\bibinfo {title} {{Lattice-QCD Calculations of TMD Soft Function Through Large-Momentum Effective Theory}},}\ }\href {\doibase 10.22323/1.396.0477} {\bibfield  {journal} {\bibinfo  {journal} {Phys. Rev. Lett.}\ }\textbf {\bibinfo {volume} {125}},\ \bibinfo {pages} {192001} (\bibinfo {year} {2020}{\natexlab{b}})},\ \Eprint {http://arxiv.org/abs/2005.14572} {arXiv:2005.14572 [hep-lat]} \BibitemShut {NoStop}%
\bibitem [{\citenamefont {Shanahan}\ \emph {et~al.}(2020{\natexlab{b}})\citenamefont {Shanahan}, \citenamefont {Wagman},\ and\ \citenamefont {Zhao}}]{Shanahan:2020zxr}%
  \BibitemOpen
  \bibfield  {author} {\bibinfo {author} {\bibfnamefont {Phiala}\ \bibnamefont {Shanahan}}, \bibinfo {author} {\bibfnamefont {Michael}\ \bibnamefont {Wagman}}, \ and\ \bibinfo {author} {\bibfnamefont {Yong}\ \bibnamefont {Zhao}},\ }\bibfield  {title} {\enquote {\bibinfo {title} {{Collins-Soper kernel for TMD evolution from lattice QCD}},}\ }\href {\doibase 10.1103/PhysRevD.102.014511} {\bibfield  {journal} {\bibinfo  {journal} {Phys. Rev. D}\ }\textbf {\bibinfo {volume} {102}},\ \bibinfo {pages} {014511} (\bibinfo {year} {2020}{\natexlab{b}})},\ \Eprint {http://arxiv.org/abs/2003.06063} {arXiv:2003.06063 [hep-lat]} \BibitemShut {NoStop}%
\bibitem [{\citenamefont {Li}\ \emph {et~al.}(2022)\citenamefont {Li} \emph {et~al.}}]{Li:2021wvl}%
  \BibitemOpen
  \bibfield  {author} {\bibinfo {author} {\bibfnamefont {Yuan}\ \bibnamefont {Li}} \emph {et~al.},\ }\bibfield  {title} {\enquote {\bibinfo {title} {{Lattice QCD Study of Transverse-Momentum Dependent Soft Function}},}\ }\href {\doibase 10.1103/PhysRevLett.128.062002} {\bibfield  {journal} {\bibinfo  {journal} {Phys. Rev. Lett.}\ }\textbf {\bibinfo {volume} {128}},\ \bibinfo {pages} {062002} (\bibinfo {year} {2022})},\ \Eprint {http://arxiv.org/abs/2106.13027} {arXiv:2106.13027 [hep-lat]} \BibitemShut {NoStop}%
\bibitem [{\citenamefont {Schlemmer}\ \emph {et~al.}(2021)\citenamefont {Schlemmer}, \citenamefont {Vladimirov}, \citenamefont {Zimmermann}, \citenamefont {Engelhardt},\ and\ \citenamefont {Sch\"afer}}]{Schlemmer:2021aij}%
  \BibitemOpen
  \bibfield  {author} {\bibinfo {author} {\bibfnamefont {Maximilian}\ \bibnamefont {Schlemmer}}, \bibinfo {author} {\bibfnamefont {Alexey}\ \bibnamefont {Vladimirov}}, \bibinfo {author} {\bibfnamefont {Christian}\ \bibnamefont {Zimmermann}}, \bibinfo {author} {\bibfnamefont {Michael}\ \bibnamefont {Engelhardt}}, \ and\ \bibinfo {author} {\bibfnamefont {Andreas}\ \bibnamefont {Sch\"afer}},\ }\bibfield  {title} {\enquote {\bibinfo {title} {{Determination of the Collins-Soper Kernel from Lattice QCD}},}\ }\href {\doibase 10.1007/JHEP08(2021)004} {\bibfield  {journal} {\bibinfo  {journal} {JHEP}\ }\textbf {\bibinfo {volume} {08}},\ \bibinfo {pages} {004} (\bibinfo {year} {2021})},\ \Eprint {http://arxiv.org/abs/2103.16991} {arXiv:2103.16991 [hep-lat]} \BibitemShut {NoStop}%
\bibitem [{\citenamefont {Shanahan}\ \emph {et~al.}(2021)\citenamefont {Shanahan}, \citenamefont {Wagman},\ and\ \citenamefont {Zhao}}]{Shanahan:2021tst}%
  \BibitemOpen
  \bibfield  {author} {\bibinfo {author} {\bibfnamefont {Phiala}\ \bibnamefont {Shanahan}}, \bibinfo {author} {\bibfnamefont {Michael}\ \bibnamefont {Wagman}}, \ and\ \bibinfo {author} {\bibfnamefont {Yong}\ \bibnamefont {Zhao}},\ }\bibfield  {title} {\enquote {\bibinfo {title} {{Lattice QCD calculation of the Collins-Soper kernel from quasi-TMDPDFs}},}\ }\href {\doibase 10.1103/PhysRevD.104.114502} {\bibfield  {journal} {\bibinfo  {journal} {Phys. Rev. D}\ }\textbf {\bibinfo {volume} {104}},\ \bibinfo {pages} {114502} (\bibinfo {year} {2021})},\ \Eprint {http://arxiv.org/abs/2107.11930} {arXiv:2107.11930 [hep-lat]} \BibitemShut {NoStop}%
\bibitem [{\citenamefont {Ji}\ \emph {et~al.}(2021{\natexlab{a}})\citenamefont {Ji}, \citenamefont {Zhang}, \citenamefont {Zhao},\ and\ \citenamefont {Zhu}}]{Ji:2021uvr}%
  \BibitemOpen
  \bibfield  {author} {\bibinfo {author} {\bibfnamefont {Yao}\ \bibnamefont {Ji}}, \bibinfo {author} {\bibfnamefont {Jian-Hui}\ \bibnamefont {Zhang}}, \bibinfo {author} {\bibfnamefont {Shuai}\ \bibnamefont {Zhao}}, \ and\ \bibinfo {author} {\bibfnamefont {Ruilin}\ \bibnamefont {Zhu}},\ }\bibfield  {title} {\enquote {\bibinfo {title} {{Renormalization and mixing of staple-shaped Wilson line operators on the lattice revisited}},}\ }\href {\doibase 10.1103/PhysRevD.104.094510} {\bibfield  {journal} {\bibinfo  {journal} {Phys. Rev. D}\ }\textbf {\bibinfo {volume} {104}},\ \bibinfo {pages} {094510} (\bibinfo {year} {2021}{\natexlab{a}})},\ \Eprint {http://arxiv.org/abs/2104.13345} {arXiv:2104.13345 [hep-ph]} \BibitemShut {NoStop}%
\bibitem [{\citenamefont {Zhang}\ \emph {et~al.}(2022)\citenamefont {Zhang}, \citenamefont {Ji}, \citenamefont {Yang}, \citenamefont {Yao},\ and\ \citenamefont {Zhang}}]{Zhang:2022xuw}%
  \BibitemOpen
  \bibfield  {author} {\bibinfo {author} {\bibfnamefont {Kuan}\ \bibnamefont {Zhang}}, \bibinfo {author} {\bibfnamefont {Xiangdong}\ \bibnamefont {Ji}}, \bibinfo {author} {\bibfnamefont {Yi-Bo}\ \bibnamefont {Yang}}, \bibinfo {author} {\bibfnamefont {Fei}\ \bibnamefont {Yao}}, \ and\ \bibinfo {author} {\bibfnamefont {Jian-Hui}\ \bibnamefont {Zhang}} (\bibinfo {collaboration} {[Lattice Parton Collaboration (LPC)]}),\ }\bibfield  {title} {\enquote {\bibinfo {title} {{Renormalization of Transverse-Momentum-Dependent Parton Distribution on the Lattice}},}\ }\href {\doibase 10.1103/PhysRevLett.129.082002} {\bibfield  {journal} {\bibinfo  {journal} {Phys. Rev. Lett.}\ }\textbf {\bibinfo {volume} {129}},\ \bibinfo {pages} {082002} (\bibinfo {year} {2022})},\ \Eprint {http://arxiv.org/abs/2205.13402} {arXiv:2205.13402 [hep-lat]} \BibitemShut {NoStop}%
\bibitem [{\citenamefont {Chu}\ \emph {et~al.}(2023)\citenamefont {Chu} \emph {et~al.}}]{LatticePartonLPC:2023pdv}%
  \BibitemOpen
  \bibfield  {author} {\bibinfo {author} {\bibfnamefont {Min-Huan}\ \bibnamefont {Chu}} \emph {et~al.} (\bibinfo {collaboration} {Lattice Parton (LPC)}),\ }\bibfield  {title} {\enquote {\bibinfo {title} {{Lattice calculation of the intrinsic soft function and the Collins-Soper kernel}},}\ }\href {\doibase 10.1007/JHEP08(2023)172} {\bibfield  {journal} {\bibinfo  {journal} {JHEP}\ }\textbf {\bibinfo {volume} {08}},\ \bibinfo {pages} {172} (\bibinfo {year} {2023})},\ \Eprint {http://arxiv.org/abs/2306.06488} {arXiv:2306.06488 [hep-lat]} \BibitemShut {NoStop}%
\bibitem [{\citenamefont {Alexandrou}\ \emph {et~al.}(2023)\citenamefont {Alexandrou} \emph {et~al.}}]{Alexandrou:2023ucc}%
  \BibitemOpen
  \bibfield  {author} {\bibinfo {author} {\bibfnamefont {Constantia}\ \bibnamefont {Alexandrou}} \emph {et~al.},\ }\bibfield  {title} {\enquote {\bibinfo {title} {{Nonperturbative renormalization of asymmetric staple-shaped operators in twisted mass lattice QCD}},}\ }\href {\doibase 10.1103/PhysRevD.108.114503} {\bibfield  {journal} {\bibinfo  {journal} {Phys. Rev. D}\ }\textbf {\bibinfo {volume} {108}},\ \bibinfo {pages} {114503} (\bibinfo {year} {2023})},\ \Eprint {http://arxiv.org/abs/2305.11824} {arXiv:2305.11824 [hep-lat]} \BibitemShut {NoStop}%
\bibitem [{\citenamefont {Fan}\ \emph {et~al.}(2018)\citenamefont {Fan}, \citenamefont {Yang}, \citenamefont {Anthony}, \citenamefont {Lin},\ and\ \citenamefont {Liu}}]{Fan:2018dxu}%
  \BibitemOpen
  \bibfield  {author} {\bibinfo {author} {\bibfnamefont {Zhou-You}\ \bibnamefont {Fan}}, \bibinfo {author} {\bibfnamefont {Yi-Bo}\ \bibnamefont {Yang}}, \bibinfo {author} {\bibfnamefont {Adam}\ \bibnamefont {Anthony}}, \bibinfo {author} {\bibfnamefont {Huey-Wen}\ \bibnamefont {Lin}}, \ and\ \bibinfo {author} {\bibfnamefont {Keh-Fei}\ \bibnamefont {Liu}},\ }\bibfield  {title} {\enquote {\bibinfo {title} {{Gluon Quasi-Parton-Distribution Functions from Lattice QCD}},}\ }\href {\doibase 10.1103/PhysRevLett.121.242001} {\bibfield  {journal} {\bibinfo  {journal} {Phys. Rev. Lett.}\ }\textbf {\bibinfo {volume} {121}},\ \bibinfo {pages} {242001} (\bibinfo {year} {2018})},\ \Eprint {http://arxiv.org/abs/1808.02077} {arXiv:1808.02077 [hep-lat]} \BibitemShut {NoStop}%
\bibitem [{\citenamefont {Zhang}\ \emph {et~al.}(2019{\natexlab{c}})\citenamefont {Zhang}, \citenamefont {Ji}, \citenamefont {Sch\"afer}, \citenamefont {Wang},\ and\ \citenamefont {Zhao}}]{Zhang:2018diq}%
  \BibitemOpen
  \bibfield  {author} {\bibinfo {author} {\bibfnamefont {Jian-Hui}\ \bibnamefont {Zhang}}, \bibinfo {author} {\bibfnamefont {Xiangdong}\ \bibnamefont {Ji}}, \bibinfo {author} {\bibfnamefont {Andreas}\ \bibnamefont {Sch\"afer}}, \bibinfo {author} {\bibfnamefont {Wei}\ \bibnamefont {Wang}}, \ and\ \bibinfo {author} {\bibfnamefont {Shuai}\ \bibnamefont {Zhao}},\ }\bibfield  {title} {\enquote {\bibinfo {title} {{Accessing Gluon Parton Distributions in Large Momentum Effective Theory}},}\ }\href {\doibase 10.1103/PhysRevLett.122.142001} {\bibfield  {journal} {\bibinfo  {journal} {Phys. Rev. Lett.}\ }\textbf {\bibinfo {volume} {122}},\ \bibinfo {pages} {142001} (\bibinfo {year} {2019}{\natexlab{c}})},\ \Eprint {http://arxiv.org/abs/1808.10824} {arXiv:1808.10824 [hep-ph]} \BibitemShut {NoStop}%
\bibitem [{\citenamefont {Fan}\ \emph {et~al.}(2021)\citenamefont {Fan}, \citenamefont {Zhang},\ and\ \citenamefont {Lin}}]{Fan:2020cpa}%
  \BibitemOpen
  \bibfield  {author} {\bibinfo {author} {\bibfnamefont {Zhouyou}\ \bibnamefont {Fan}}, \bibinfo {author} {\bibfnamefont {Rui}\ \bibnamefont {Zhang}}, \ and\ \bibinfo {author} {\bibfnamefont {Huey-Wen}\ \bibnamefont {Lin}},\ }\bibfield  {title} {\enquote {\bibinfo {title} {{Nucleon gluon distribution function from 2 + 1 + 1-flavor lattice QCD}},}\ }\href {\doibase 10.1142/S0217751X21500809} {\bibfield  {journal} {\bibinfo  {journal} {Int. J. Mod. Phys. A}\ }\textbf {\bibinfo {volume} {36}},\ \bibinfo {pages} {2150080} (\bibinfo {year} {2021})},\ \Eprint {http://arxiv.org/abs/2007.16113} {arXiv:2007.16113 [hep-lat]} \BibitemShut {NoStop}%
\bibitem [{\citenamefont {Fan}\ and\ \citenamefont {Lin}(2021)}]{Fan:2021bcr}%
  \BibitemOpen
  \bibfield  {author} {\bibinfo {author} {\bibfnamefont {Zhouyou}\ \bibnamefont {Fan}}\ and\ \bibinfo {author} {\bibfnamefont {Huey-Wen}\ \bibnamefont {Lin}},\ }\bibfield  {title} {\enquote {\bibinfo {title} {{Gluon parton distribution of the pion from lattice QCD}},}\ }\href {\doibase 10.1016/j.physletb.2021.136778} {\bibfield  {journal} {\bibinfo  {journal} {Phys. Lett. B}\ }\textbf {\bibinfo {volume} {823}},\ \bibinfo {pages} {136778} (\bibinfo {year} {2021})},\ \Eprint {http://arxiv.org/abs/2104.06372} {arXiv:2104.06372 [hep-lat]} \BibitemShut {NoStop}%
\bibitem [{\citenamefont {Salas-Chavira}\ \emph {et~al.}(2022)\citenamefont {Salas-Chavira}, \citenamefont {Fan},\ and\ \citenamefont {Lin}}]{Salas-Chavira:2021wui}%
  \BibitemOpen
  \bibfield  {author} {\bibinfo {author} {\bibfnamefont {Alejandro}\ \bibnamefont {Salas-Chavira}}, \bibinfo {author} {\bibfnamefont {Zhouyou}\ \bibnamefont {Fan}}, \ and\ \bibinfo {author} {\bibfnamefont {Huey-Wen}\ \bibnamefont {Lin}},\ }\bibfield  {title} {\enquote {\bibinfo {title} {{First glimpse into the kaon gluon parton distribution using lattice QCD}},}\ }\href {\doibase 10.1103/PhysRevD.106.094510} {\bibfield  {journal} {\bibinfo  {journal} {Phys. Rev. D}\ }\textbf {\bibinfo {volume} {106}},\ \bibinfo {pages} {094510} (\bibinfo {year} {2022})},\ \Eprint {http://arxiv.org/abs/2112.03124} {arXiv:2112.03124 [hep-lat]} \BibitemShut {NoStop}%
\bibitem [{\citenamefont {Khan}\ \emph {et~al.}(2021)\citenamefont {Khan} \emph {et~al.}}]{HadStruc:2021wmh}%
  \BibitemOpen
  \bibfield  {author} {\bibinfo {author} {\bibfnamefont {Tanjib}\ \bibnamefont {Khan}} \emph {et~al.} (\bibinfo {collaboration} {HadStruc}),\ }\bibfield  {title} {\enquote {\bibinfo {title} {{Unpolarized gluon distribution in the nucleon from lattice quantum chromodynamics}},}\ }\href {\doibase 10.1103/PhysRevD.104.094516} {\bibfield  {journal} {\bibinfo  {journal} {Phys. Rev. D}\ }\textbf {\bibinfo {volume} {104}},\ \bibinfo {pages} {094516} (\bibinfo {year} {2021})},\ \Eprint {http://arxiv.org/abs/2107.08960} {arXiv:2107.08960 [hep-lat]} \BibitemShut {NoStop}%
\bibitem [{\citenamefont {Egerer}\ \emph {et~al.}(2022{\natexlab{b}})\citenamefont {Egerer} \emph {et~al.}}]{HadStruc:2022yaw}%
  \BibitemOpen
  \bibfield  {author} {\bibinfo {author} {\bibfnamefont {Colin}\ \bibnamefont {Egerer}} \emph {et~al.} (\bibinfo {collaboration} {HadStruc}),\ }\bibfield  {title} {\enquote {\bibinfo {title} {{Toward the determination of the gluon helicity distribution in the nucleon from lattice quantum chromodynamics}},}\ }\href {\doibase 10.1103/PhysRevD.106.094511} {\bibfield  {journal} {\bibinfo  {journal} {Phys. Rev. D}\ }\textbf {\bibinfo {volume} {106}},\ \bibinfo {pages} {094511} (\bibinfo {year} {2022}{\natexlab{b}})},\ \Eprint {http://arxiv.org/abs/2207.08733} {arXiv:2207.08733 [hep-lat]} \BibitemShut {NoStop}%
\bibitem [{\citenamefont {Khan}\ \emph {et~al.}(2023)\citenamefont {Khan}, \citenamefont {Liu},\ and\ \citenamefont {Sufian}}]{Khan:2022vot}%
  \BibitemOpen
  \bibfield  {author} {\bibinfo {author} {\bibfnamefont {Tanjib}\ \bibnamefont {Khan}}, \bibinfo {author} {\bibfnamefont {Tianbo}\ \bibnamefont {Liu}}, \ and\ \bibinfo {author} {\bibfnamefont {Raza~Sabbir}\ \bibnamefont {Sufian}},\ }\bibfield  {title} {\enquote {\bibinfo {title} {{Gluon helicity in the nucleon from lattice QCD and machine learning}},}\ }\href {\doibase 10.1103/PhysRevD.108.074502} {\bibfield  {journal} {\bibinfo  {journal} {Phys. Rev. D}\ }\textbf {\bibinfo {volume} {108}},\ \bibinfo {pages} {074502} (\bibinfo {year} {2023})},\ \Eprint {http://arxiv.org/abs/2211.15587} {arXiv:2211.15587 [hep-lat]} \BibitemShut {NoStop}%
\bibitem [{\citenamefont {Fan}\ \emph {et~al.}(2023)\citenamefont {Fan}, \citenamefont {Good},\ and\ \citenamefont {Lin}}]{Fan:2022kcb}%
  \BibitemOpen
  \bibfield  {author} {\bibinfo {author} {\bibfnamefont {Zhouyou}\ \bibnamefont {Fan}}, \bibinfo {author} {\bibfnamefont {William}\ \bibnamefont {Good}}, \ and\ \bibinfo {author} {\bibfnamefont {Huey-Wen}\ \bibnamefont {Lin}},\ }\bibfield  {title} {\enquote {\bibinfo {title} {{Gluon parton distribution of the nucleon from (2+1+1)-flavor lattice QCD in the physical-continuum limit}},}\ }\href {\doibase 10.1103/PhysRevD.108.014508} {\bibfield  {journal} {\bibinfo  {journal} {Phys. Rev. D}\ }\textbf {\bibinfo {volume} {108}},\ \bibinfo {pages} {014508} (\bibinfo {year} {2023})},\ \Eprint {http://arxiv.org/abs/2210.09985} {arXiv:2210.09985 [hep-lat]} \BibitemShut {NoStop}%
\bibitem [{\citenamefont {Delmar}\ \emph {et~al.}(2023)\citenamefont {Delmar}, \citenamefont {Alexandrou}, \citenamefont {Cichy}, \citenamefont {Constantinou},\ and\ \citenamefont {Hadjiyiannakou}}]{Delmar:2023agv}%
  \BibitemOpen
  \bibfield  {author} {\bibinfo {author} {\bibfnamefont {Joseph}\ \bibnamefont {Delmar}}, \bibinfo {author} {\bibfnamefont {Constantia}\ \bibnamefont {Alexandrou}}, \bibinfo {author} {\bibfnamefont {Krzysztof}\ \bibnamefont {Cichy}}, \bibinfo {author} {\bibfnamefont {Martha}\ \bibnamefont {Constantinou}}, \ and\ \bibinfo {author} {\bibfnamefont {Kyriakos}\ \bibnamefont {Hadjiyiannakou}},\ }\bibfield  {title} {\enquote {\bibinfo {title} {{Gluon PDF of the proton using twisted mass fermions}},}\ }\href {\doibase 10.1103/PhysRevD.108.094515} {\bibfield  {journal} {\bibinfo  {journal} {Phys. Rev. D}\ }\textbf {\bibinfo {volume} {108}},\ \bibinfo {pages} {094515} (\bibinfo {year} {2023})},\ \Eprint {http://arxiv.org/abs/2310.01389} {arXiv:2310.01389 [hep-lat]} \BibitemShut {NoStop}%
\bibitem [{\citenamefont {Jia}\ and\ \citenamefont {Xiong}(2016)}]{Jia:2015pxx}%
  \BibitemOpen
  \bibfield  {author} {\bibinfo {author} {\bibfnamefont {Yu}~\bibnamefont {Jia}}\ and\ \bibinfo {author} {\bibfnamefont {Xiaonu}\ \bibnamefont {Xiong}},\ }\bibfield  {title} {\enquote {\bibinfo {title} {{Quasidistribution amplitude of heavy quarkonia}},}\ }\href {\doibase 10.1103/PhysRevD.94.094005} {\bibfield  {journal} {\bibinfo  {journal} {Phys. Rev. D}\ }\textbf {\bibinfo {volume} {94}},\ \bibinfo {pages} {094005} (\bibinfo {year} {2016})},\ \Eprint {http://arxiv.org/abs/1511.04430} {arXiv:1511.04430 [hep-ph]} \BibitemShut {NoStop}%
\bibitem [{\citenamefont {Radyushkin}(2017{\natexlab{c}})}]{Radyushkin:2017gjd}%
  \BibitemOpen
  \bibfield  {author} {\bibinfo {author} {\bibfnamefont {Anatoly~V.}\ \bibnamefont {Radyushkin}},\ }\bibfield  {title} {\enquote {\bibinfo {title} {{Pion Distribution Amplitude and Quasi-Distributions}},}\ }\href {\doibase 10.1103/PhysRevD.95.056020} {\bibfield  {journal} {\bibinfo  {journal} {Phys. Rev. D}\ }\textbf {\bibinfo {volume} {95}},\ \bibinfo {pages} {056020} (\bibinfo {year} {2017}{\natexlab{c}})},\ \Eprint {http://arxiv.org/abs/1701.02688} {arXiv:1701.02688 [hep-ph]} \BibitemShut {NoStop}%
\bibitem [{\citenamefont {Broniowski}\ and\ \citenamefont {Ruiz~Arriola}(2017)}]{Broniowski:2017wbr}%
  \BibitemOpen
  \bibfield  {author} {\bibinfo {author} {\bibfnamefont {Wojciech}\ \bibnamefont {Broniowski}}\ and\ \bibinfo {author} {\bibfnamefont {Enrique}\ \bibnamefont {Ruiz~Arriola}},\ }\bibfield  {title} {\enquote {\bibinfo {title} {{Nonperturbative partonic quasidistributions of the pion from chiral quark models}},}\ }\href {\doibase 10.1016/j.physletb.2017.08.055} {\bibfield  {journal} {\bibinfo  {journal} {Phys. Lett. B}\ }\textbf {\bibinfo {volume} {773}},\ \bibinfo {pages} {385--390} (\bibinfo {year} {2017})},\ \Eprint {http://arxiv.org/abs/1707.09588} {arXiv:1707.09588 [hep-ph]} \BibitemShut {NoStop}%
\bibitem [{\citenamefont {Zhang}\ \emph {et~al.}(2019{\natexlab{d}})\citenamefont {Zhang}, \citenamefont {Jin}, \citenamefont {Lin}, \citenamefont {Sch\"afer}, \citenamefont {Sun}, \citenamefont {Yang}, \citenamefont {Zhang}, \citenamefont {Zhao},\ and\ \citenamefont {Chen}}]{Chen:2017gck}%
  \BibitemOpen
  \bibfield  {author} {\bibinfo {author} {\bibfnamefont {Jian-Hui}\ \bibnamefont {Zhang}}, \bibinfo {author} {\bibfnamefont {Luchang}\ \bibnamefont {Jin}}, \bibinfo {author} {\bibfnamefont {Huey-Wen}\ \bibnamefont {Lin}}, \bibinfo {author} {\bibfnamefont {Andreas}\ \bibnamefont {Sch\"afer}}, \bibinfo {author} {\bibfnamefont {Peng}\ \bibnamefont {Sun}}, \bibinfo {author} {\bibfnamefont {Yi-Bo}\ \bibnamefont {Yang}}, \bibinfo {author} {\bibfnamefont {Rui}\ \bibnamefont {Zhang}}, \bibinfo {author} {\bibfnamefont {Yong}\ \bibnamefont {Zhao}}, \ and\ \bibinfo {author} {\bibfnamefont {Jiunn-Wei}\ \bibnamefont {Chen}} (\bibinfo {collaboration} {LP3}),\ }\bibfield  {title} {\enquote {\bibinfo {title} {{Kaon Distribution Amplitude from Lattice QCD and the Flavor SU(3) Symmetry}},}\ }\href {\doibase 10.1016/j.nuclphysb.2018.12.020} {\bibfield  {journal} {\bibinfo  {journal} {Nucl. Phys. B}\ }\textbf {\bibinfo {volume} {939}},\ \bibinfo {pages} {429--446} (\bibinfo {year} {2019}{\natexlab{d}})},\ \Eprint
  {http://arxiv.org/abs/1712.10025} {arXiv:1712.10025 [hep-ph]} \BibitemShut {NoStop}%
\bibitem [{\citenamefont {Cichy}\ and\ \citenamefont {Constantinou}(2019)}]{Cichy:2018mum}%
  \BibitemOpen
  \bibfield  {author} {\bibinfo {author} {\bibfnamefont {Krzysztof}\ \bibnamefont {Cichy}}\ and\ \bibinfo {author} {\bibfnamefont {Martha}\ \bibnamefont {Constantinou}},\ }\bibfield  {title} {\enquote {\bibinfo {title} {{A guide to light-cone PDFs from Lattice QCD: an overview of approaches, techniques and results}},}\ }\href {\doibase 10.1155/2019/3036904} {\bibfield  {journal} {\bibinfo  {journal} {Adv. High Energy Phys.}\ }\textbf {\bibinfo {volume} {2019}},\ \bibinfo {pages} {3036904} (\bibinfo {year} {2019})},\ \Eprint {http://arxiv.org/abs/1811.07248} {arXiv:1811.07248 [hep-lat]} \BibitemShut {NoStop}%
\bibitem [{\citenamefont {Ji}\ \emph {et~al.}(2021{\natexlab{b}})\citenamefont {Ji}, \citenamefont {Liu}, \citenamefont {Liu}, \citenamefont {Zhang},\ and\ \citenamefont {Zhao}}]{Ji:2020ect}%
  \BibitemOpen
  \bibfield  {author} {\bibinfo {author} {\bibfnamefont {Xiangdong}\ \bibnamefont {Ji}}, \bibinfo {author} {\bibfnamefont {Yu-Sheng}\ \bibnamefont {Liu}}, \bibinfo {author} {\bibfnamefont {Yizhuang}\ \bibnamefont {Liu}}, \bibinfo {author} {\bibfnamefont {Jian-Hui}\ \bibnamefont {Zhang}}, \ and\ \bibinfo {author} {\bibfnamefont {Yong}\ \bibnamefont {Zhao}},\ }\bibfield  {title} {\enquote {\bibinfo {title} {{Large-momentum effective theory}},}\ }\href {\doibase 10.1103/RevModPhys.93.035005} {\bibfield  {journal} {\bibinfo  {journal} {Rev. Mod. Phys.}\ }\textbf {\bibinfo {volume} {93}},\ \bibinfo {pages} {035005} (\bibinfo {year} {2021}{\natexlab{b}})},\ \Eprint {http://arxiv.org/abs/2004.03543} {arXiv:2004.03543 [hep-ph]} \BibitemShut {NoStop}%
\bibitem [{\citenamefont {Constantinou}(2021)}]{Constantinou:2020pek}%
  \BibitemOpen
  \bibfield  {author} {\bibinfo {author} {\bibfnamefont {Martha}\ \bibnamefont {Constantinou}},\ }\bibfield  {title} {\enquote {\bibinfo {title} {{The x-dependence of hadronic parton distributions: A review on the progress of lattice QCD}},}\ }\href {\doibase 10.1140/epja/s10050-021-00353-7} {\bibfield  {journal} {\bibinfo  {journal} {Eur. Phys. J. A}\ }\textbf {\bibinfo {volume} {57}},\ \bibinfo {pages} {77} (\bibinfo {year} {2021})},\ \Eprint {http://arxiv.org/abs/2010.02445} {arXiv:2010.02445 [hep-lat]} \BibitemShut {NoStop}%
\bibitem [{\citenamefont {Cichy}(2022{\natexlab{a}})}]{Cichy:2021lih}%
  \BibitemOpen
  \bibfield  {author} {\bibinfo {author} {\bibfnamefont {Krzysztof}\ \bibnamefont {Cichy}},\ }\bibfield  {title} {\enquote {\bibinfo {title} {{Progress in $x$-dependent partonic distributions from lattice QCD}},}\ }\href {\doibase 10.22323/1.396.0017} {\bibfield  {journal} {\bibinfo  {journal} {PoS}\ }\textbf {\bibinfo {volume} {LATTICE2021}},\ \bibinfo {pages} {017} (\bibinfo {year} {2022}{\natexlab{a}})},\ \Eprint {http://arxiv.org/abs/2110.07440} {arXiv:2110.07440 [hep-lat]} \BibitemShut {NoStop}%
\bibitem [{\citenamefont {Cichy}(2022{\natexlab{b}})}]{Cichy:2021ewm}%
  \BibitemOpen
  \bibfield  {author} {\bibinfo {author} {\bibfnamefont {Krzysztof}\ \bibnamefont {Cichy}},\ }\bibfield  {title} {\enquote {\bibinfo {title} {{Overview of lattice calculations of the x-dependence of PDFs, GPDs and TMDs}},}\ }\href {\doibase 10.1051/epjconf/202225801005} {\bibfield  {journal} {\bibinfo  {journal} {EPJ Web Conf.}\ }\textbf {\bibinfo {volume} {258}},\ \bibinfo {pages} {01005} (\bibinfo {year} {2022}{\natexlab{b}})},\ \Eprint {http://arxiv.org/abs/2111.04552} {arXiv:2111.04552 [hep-lat]} \BibitemShut {NoStop}%
\bibitem [{\citenamefont {Accardi}\ \emph {et~al.}(2016)\citenamefont {Accardi} \emph {et~al.}}]{Accardi:2012qut}%
  \BibitemOpen
  \bibfield  {author} {\bibinfo {author} {\bibfnamefont {A.}~\bibnamefont {Accardi}} \emph {et~al.},\ }\bibfield  {title} {\enquote {\bibinfo {title} {{Electron Ion Collider: The Next QCD Frontier}: {Understanding the glue that binds us all}},}\ }\href {\doibase 10.1140/epja/i2016-16268-9} {\bibfield  {journal} {\bibinfo  {journal} {Eur. Phys. J. A}\ }\textbf {\bibinfo {volume} {52}},\ \bibinfo {pages} {268} (\bibinfo {year} {2016})},\ \Eprint {http://arxiv.org/abs/1212.1701} {arXiv:1212.1701 [nucl-ex]} \BibitemShut {NoStop}%
\bibitem [{\citenamefont {Abdul~Khalek}\ \emph {et~al.}(2022)\citenamefont {Abdul~Khalek} \emph {et~al.}}]{AbdulKhalek:2021gbh}%
  \BibitemOpen
  \bibfield  {author} {\bibinfo {author} {\bibfnamefont {R.}~\bibnamefont {Abdul~Khalek}} \emph {et~al.},\ }\bibfield  {title} {\enquote {\bibinfo {title} {{Science Requirements and Detector Concepts for the Electron-Ion Collider}: {EIC Yellow Report}},}\ }\href {\doibase 10.1016/j.nuclphysa.2022.122447} {\bibfield  {journal} {\bibinfo  {journal} {Nucl. Phys. A}\ }\textbf {\bibinfo {volume} {1026}},\ \bibinfo {pages} {122447} (\bibinfo {year} {2022})},\ \Eprint {http://arxiv.org/abs/2103.05419} {arXiv:2103.05419 [physics.ins-det]} \BibitemShut {NoStop}%
\bibitem [{\citenamefont {Anderle}\ \emph {et~al.}(2021)\citenamefont {Anderle} \emph {et~al.}}]{Anderle:2021wcy}%
  \BibitemOpen
  \bibfield  {author} {\bibinfo {author} {\bibfnamefont {Daniele~P.}\ \bibnamefont {Anderle}} \emph {et~al.},\ }\bibfield  {title} {\enquote {\bibinfo {title} {{Electron-ion collider in China}},}\ }\href {\doibase 10.1007/s11467-021-1062-0} {\bibfield  {journal} {\bibinfo  {journal} {Front. Phys. (Beijing)}\ }\textbf {\bibinfo {volume} {16}},\ \bibinfo {pages} {64701} (\bibinfo {year} {2021})},\ \Eprint {http://arxiv.org/abs/2102.09222} {arXiv:2102.09222 [nucl-ex]} \BibitemShut {NoStop}%
\bibitem [{\citenamefont {Wang}\ \emph {et~al.}(2019)\citenamefont {Wang}, \citenamefont {Zhang}, \citenamefont {Zhao},\ and\ \citenamefont {Zhu}}]{Wang:2019tgg}%
  \BibitemOpen
  \bibfield  {author} {\bibinfo {author} {\bibfnamefont {Wei}\ \bibnamefont {Wang}}, \bibinfo {author} {\bibfnamefont {Jian-Hui}\ \bibnamefont {Zhang}}, \bibinfo {author} {\bibfnamefont {Shuai}\ \bibnamefont {Zhao}}, \ and\ \bibinfo {author} {\bibfnamefont {Ruilin}\ \bibnamefont {Zhu}},\ }\bibfield  {title} {\enquote {\bibinfo {title} {{Complete matching for quasidistribution functions in large momentum effective theory}},}\ }\href {\doibase 10.1103/PhysRevD.100.074509} {\bibfield  {journal} {\bibinfo  {journal} {Phys. Rev. D}\ }\textbf {\bibinfo {volume} {100}},\ \bibinfo {pages} {074509} (\bibinfo {year} {2019})},\ \Eprint {http://arxiv.org/abs/1904.00978} {arXiv:1904.00978 [hep-ph]} \BibitemShut {NoStop}%
\bibitem [{\citenamefont {Wang}\ \emph {et~al.}(2018)\citenamefont {Wang}, \citenamefont {Zhao},\ and\ \citenamefont {Zhu}}]{Wang:2017qyg}%
  \BibitemOpen
  \bibfield  {author} {\bibinfo {author} {\bibfnamefont {Wei}\ \bibnamefont {Wang}}, \bibinfo {author} {\bibfnamefont {Shuai}\ \bibnamefont {Zhao}}, \ and\ \bibinfo {author} {\bibfnamefont {Ruilin}\ \bibnamefont {Zhu}},\ }\bibfield  {title} {\enquote {\bibinfo {title} {{Gluon quasidistribution function at one loop}},}\ }\href {\doibase 10.1140/epjc/s10052-018-5617-3} {\bibfield  {journal} {\bibinfo  {journal} {Eur. Phys. J. C}\ }\textbf {\bibinfo {volume} {78}},\ \bibinfo {pages} {147} (\bibinfo {year} {2018})},\ \Eprint {http://arxiv.org/abs/1708.02458} {arXiv:1708.02458 [hep-ph]} \BibitemShut {NoStop}%
\bibitem [{\citenamefont {Chen}\ \emph {et~al.}(2020)\citenamefont {Chen}, \citenamefont {Wang},\ and\ \citenamefont {Zhu}}]{Chen:2020arf}%
  \BibitemOpen
  \bibfield  {author} {\bibinfo {author} {\bibfnamefont {Long-Bin}\ \bibnamefont {Chen}}, \bibinfo {author} {\bibfnamefont {Wei}\ \bibnamefont {Wang}}, \ and\ \bibinfo {author} {\bibfnamefont {Ruilin}\ \bibnamefont {Zhu}},\ }\bibfield  {title} {\enquote {\bibinfo {title} {{Quasi parton distribution functions at NNLO: flavor non-diagonal quark contributions}},}\ }\href {\doibase 10.1103/PhysRevD.102.011503} {\bibfield  {journal} {\bibinfo  {journal} {Phys. Rev. D}\ }\textbf {\bibinfo {volume} {102}},\ \bibinfo {pages} {011503} (\bibinfo {year} {2020})},\ \Eprint {http://arxiv.org/abs/2005.13757} {arXiv:2005.13757 [hep-ph]} \BibitemShut {NoStop}%
\bibitem [{\citenamefont {Li}\ \emph {et~al.}(2021)\citenamefont {Li}, \citenamefont {Ma},\ and\ \citenamefont {Qiu}}]{Li:2020xml}%
  \BibitemOpen
  \bibfield  {author} {\bibinfo {author} {\bibfnamefont {Zheng-Yang}\ \bibnamefont {Li}}, \bibinfo {author} {\bibfnamefont {Yan-Qing}\ \bibnamefont {Ma}}, \ and\ \bibinfo {author} {\bibfnamefont {Jian-Wei}\ \bibnamefont {Qiu}},\ }\bibfield  {title} {\enquote {\bibinfo {title} {{Extraction of Next-to-Next-to-Leading-Order Parton Distribution Functions from Lattice QCD Calculations}},}\ }\href {\doibase 10.1103/PhysRevLett.126.072001} {\bibfield  {journal} {\bibinfo  {journal} {Phys. Rev. Lett.}\ }\textbf {\bibinfo {volume} {126}},\ \bibinfo {pages} {072001} (\bibinfo {year} {2021})},\ \Eprint {http://arxiv.org/abs/2006.12370} {arXiv:2006.12370 [hep-ph]} \BibitemShut {NoStop}%
\bibitem [{\citenamefont {Mandelstam}(1968)}]{Mandelstam:1968hz}%
  \BibitemOpen
  \bibfield  {author} {\bibinfo {author} {\bibfnamefont {Stanley}\ \bibnamefont {Mandelstam}},\ }\bibfield  {title} {\enquote {\bibinfo {title} {{Feynman rules for electromagnetic and Yang-Mills fields from the gauge independent field theoretic formalism}},}\ }\href {\doibase 10.1103/PhysRev.175.1580} {\bibfield  {journal} {\bibinfo  {journal} {Phys. Rev.}\ }\textbf {\bibinfo {volume} {175}},\ \bibinfo {pages} {1580--1623} (\bibinfo {year} {1968})}\BibitemShut {NoStop}%
\bibitem [{\citenamefont {Polyakov}(1979)}]{Polyakov:1979gp}%
  \BibitemOpen
  \bibfield  {author} {\bibinfo {author} {\bibfnamefont {Alexander~M.}\ \bibnamefont {Polyakov}},\ }\bibfield  {title} {\enquote {\bibinfo {title} {{String Representations and Hidden Symmetries for Gauge Fields}},}\ }\href {\doibase 10.1016/0370-2693(79)90747-0} {\bibfield  {journal} {\bibinfo  {journal} {Phys. Lett.}\ }\textbf {\bibinfo {volume} {82B}},\ \bibinfo {pages} {247--250} (\bibinfo {year} {1979})}\BibitemShut {NoStop}%
\bibitem [{\citenamefont {Makeenko}\ and\ \citenamefont {Migdal}(1979)}]{Makeenko:1979pb}%
  \BibitemOpen
  \bibfield  {author} {\bibinfo {author} {\bibfnamefont {{\relax Yu}.~M.}\ \bibnamefont {Makeenko}}\ and\ \bibinfo {author} {\bibfnamefont {Alexander~A.}\ \bibnamefont {Migdal}},\ }\bibfield  {title} {\enquote {\bibinfo {title} {{Exact Equation for the Loop Average in Multicolor QCD}},}\ }\href {\doibase 10.1016/0370-2693(79)90131-X} {\bibfield  {journal} {\bibinfo  {journal} {Phys. Lett.}\ }\textbf {\bibinfo {volume} {88B}},\ \bibinfo {pages} {135} (\bibinfo {year} {1979})}\BibitemShut {NoStop}%
\bibitem [{\citenamefont {Dotsenko}\ and\ \citenamefont {Vergeles}(1980)}]{Dotsenko:1979wb}%
  \BibitemOpen
  \bibfield  {author} {\bibinfo {author} {\bibfnamefont {V.~S.}\ \bibnamefont {Dotsenko}}\ and\ \bibinfo {author} {\bibfnamefont {S.~N.}\ \bibnamefont {Vergeles}},\ }\bibfield  {title} {\enquote {\bibinfo {title} {{Renormalizability of Phase Factors in the Nonabelian Gauge Theory}},}\ }\href {\doibase 10.1016/0550-3213(80)90103-0} {\bibfield  {journal} {\bibinfo  {journal} {Nucl. Phys.}\ }\textbf {\bibinfo {volume} {B169}},\ \bibinfo {pages} {527--546} (\bibinfo {year} {1980})}\BibitemShut {NoStop}%
\bibitem [{\citenamefont {Craigie}\ and\ \citenamefont {Dorn}(1981)}]{Craigie:1980qs}%
  \BibitemOpen
  \bibfield  {author} {\bibinfo {author} {\bibfnamefont {N.~S.}\ \bibnamefont {Craigie}}\ and\ \bibinfo {author} {\bibfnamefont {Harald}\ \bibnamefont {Dorn}},\ }\bibfield  {title} {\enquote {\bibinfo {title} {{On the Renormalization and Short Distance Properties of Hadronic Operators in {QCD}}},}\ }\href {\doibase 10.1016/0550-3213(81)90372-2} {\bibfield  {journal} {\bibinfo  {journal} {Nucl. Phys.}\ }\textbf {\bibinfo {volume} {B185}},\ \bibinfo {pages} {204--220} (\bibinfo {year} {1981})}\BibitemShut {NoStop}%
\bibitem [{\citenamefont {Brandt}\ \emph {et~al.}(1981)\citenamefont {Brandt}, \citenamefont {Neri},\ and\ \citenamefont {Sato}}]{Brandt:1981kf}%
  \BibitemOpen
  \bibfield  {author} {\bibinfo {author} {\bibfnamefont {Richard~A.}\ \bibnamefont {Brandt}}, \bibinfo {author} {\bibfnamefont {Filippo}\ \bibnamefont {Neri}}, \ and\ \bibinfo {author} {\bibfnamefont {Masa-aki}\ \bibnamefont {Sato}},\ }\bibfield  {title} {\enquote {\bibinfo {title} {{Renormalization of Loop Functions for All Loops}},}\ }\href {\doibase 10.1103/PhysRevD.24.879} {\bibfield  {journal} {\bibinfo  {journal} {Phys. Rev.}\ }\textbf {\bibinfo {volume} {D24}},\ \bibinfo {pages} {879} (\bibinfo {year} {1981})}\BibitemShut {NoStop}%
\bibitem [{\citenamefont {Stefanis}(1984)}]{Stefanis:1983ke}%
  \BibitemOpen
  \bibfield  {author} {\bibinfo {author} {\bibfnamefont {N.~G.}\ \bibnamefont {Stefanis}},\ }\bibfield  {title} {\enquote {\bibinfo {title} {{Gauge-invariant quark two-point Green's function through connector insertion to $\mathcal{O} (\alpha_s)$}},}\ }\href {\doibase 10.1007/BF02902597} {\bibfield  {journal} {\bibinfo  {journal} {Nuovo Cim.}\ }\textbf {\bibinfo {volume} {A83}},\ \bibinfo {pages} {205} (\bibinfo {year} {1984})}\BibitemShut {NoStop}%
\bibitem [{\citenamefont {Knauss}\ and\ \citenamefont {Scharnhorst}(1984)}]{Knauss:1984rx}%
  \BibitemOpen
  \bibfield  {author} {\bibinfo {author} {\bibfnamefont {D.}~\bibnamefont {Knauss}}\ and\ \bibinfo {author} {\bibfnamefont {K.}~\bibnamefont {Scharnhorst}},\ }\bibfield  {title} {\enquote {\bibinfo {title} {{Two Loop Renormalization of Nonsmooth String Operators in {Yang-Mills} Theory}},}\ }\href {\doibase 10.1002/andp.19844960413} {\bibfield  {journal} {\bibinfo  {journal} {Annalen Phys.}\ }\textbf {\bibinfo {volume} {496}},\ \bibinfo {pages} {331--344} (\bibinfo {year} {1984})}\BibitemShut {NoStop}%
\bibitem [{\citenamefont {Dorn}(1986)}]{Dorn:1986dt}%
  \BibitemOpen
  \bibfield  {author} {\bibinfo {author} {\bibfnamefont {Harald}\ \bibnamefont {Dorn}},\ }\bibfield  {title} {\enquote {\bibinfo {title} {{Renormalization of Path Ordered Phase Factors and Related Hadron Operators in Gauge Field Theories}},}\ }\href {\doibase 10.1002/prop.19860340104} {\bibfield  {journal} {\bibinfo  {journal} {Fortsch. Phys.}\ }\textbf {\bibinfo {volume} {34}},\ \bibinfo {pages} {11--56} (\bibinfo {year} {1986})}\BibitemShut {NoStop}%
\bibitem [{\citenamefont {Korchemsky}\ and\ \citenamefont {Radyushkin}(1987)}]{Korchemsky:1987wg}%
  \BibitemOpen
  \bibfield  {author} {\bibinfo {author} {\bibfnamefont {G.~P.}\ \bibnamefont {Korchemsky}}\ and\ \bibinfo {author} {\bibfnamefont {A.~V.}\ \bibnamefont {Radyushkin}},\ }\bibfield  {title} {\enquote {\bibinfo {title} {{Renormalization of the Wilson Loops Beyond the Leading Order}},}\ }\href {\doibase 10.1016/0550-3213(87)90277-X} {\bibfield  {journal} {\bibinfo  {journal} {Nucl. Phys.}\ }\textbf {\bibinfo {volume} {B283}},\ \bibinfo {pages} {342--364} (\bibinfo {year} {1987})}\BibitemShut {NoStop}%
\bibitem [{\citenamefont {Constantinou}\ and\ \citenamefont {Panagopoulos}(2017)}]{Constantinou:2017sej}%
  \BibitemOpen
  \bibfield  {author} {\bibinfo {author} {\bibfnamefont {Martha}\ \bibnamefont {Constantinou}}\ and\ \bibinfo {author} {\bibfnamefont {Haralambos}\ \bibnamefont {Panagopoulos}},\ }\bibfield  {title} {\enquote {\bibinfo {title} {{Perturbative renormalization of quasi-parton distribution functions}},}\ }\href {\doibase 10.1103/PhysRevD.96.054506} {\bibfield  {journal} {\bibinfo  {journal} {Phys. Rev. D}\ }\textbf {\bibinfo {volume} {96}},\ \bibinfo {pages} {054506} (\bibinfo {year} {2017})},\ \Eprint {http://arxiv.org/abs/1705.11193} {arXiv:1705.11193 [hep-lat]} \BibitemShut {NoStop}%
\bibitem [{\citenamefont {Spanoudes}\ and\ \citenamefont {Panagopoulos}(2018)}]{Spanoudes:2018zya}%
  \BibitemOpen
  \bibfield  {author} {\bibinfo {author} {\bibfnamefont {Gregoris}\ \bibnamefont {Spanoudes}}\ and\ \bibinfo {author} {\bibfnamefont {Haralambos}\ \bibnamefont {Panagopoulos}},\ }\bibfield  {title} {\enquote {\bibinfo {title} {{Renormalization of Wilson-line operators in the presence of nonzero quark masses}},}\ }\href {\doibase 10.1103/PhysRevD.98.014509} {\bibfield  {journal} {\bibinfo  {journal} {Phys. Rev. D}\ }\textbf {\bibinfo {volume} {98}},\ \bibinfo {pages} {014509} (\bibinfo {year} {2018})},\ \Eprint {http://arxiv.org/abs/1805.01164} {arXiv:1805.01164 [hep-lat]} \BibitemShut {NoStop}%
\bibitem [{\citenamefont {Constantinou}\ and\ \citenamefont {Panagopoulos}(2023)}]{Constantinou:2022aij}%
  \BibitemOpen
  \bibfield  {author} {\bibinfo {author} {\bibfnamefont {Martha}\ \bibnamefont {Constantinou}}\ and\ \bibinfo {author} {\bibfnamefont {Haralambos}\ \bibnamefont {Panagopoulos}},\ }\bibfield  {title} {\enquote {\bibinfo {title} {{Improved renormalization scheme for nonlocal operators}},}\ }\href {\doibase 10.1103/PhysRevD.107.014503} {\bibfield  {journal} {\bibinfo  {journal} {Phys. Rev. D}\ }\textbf {\bibinfo {volume} {107}},\ \bibinfo {pages} {014503} (\bibinfo {year} {2023})},\ \Eprint {http://arxiv.org/abs/2207.09977} {arXiv:2207.09977 [hep-lat]} \BibitemShut {NoStop}%
\bibitem [{\citenamefont {Constantinou}\ \emph {et~al.}(2019)\citenamefont {Constantinou}, \citenamefont {Panagopoulos},\ and\ \citenamefont {Spanoudes}}]{Constantinou:2019vyb}%
  \BibitemOpen
  \bibfield  {author} {\bibinfo {author} {\bibfnamefont {Martha}\ \bibnamefont {Constantinou}}, \bibinfo {author} {\bibfnamefont {Haralambos}\ \bibnamefont {Panagopoulos}}, \ and\ \bibinfo {author} {\bibfnamefont {Gregoris}\ \bibnamefont {Spanoudes}},\ }\bibfield  {title} {\enquote {\bibinfo {title} {{One-loop renormalization of staple-shaped operators in continuum and lattice regularizations}},}\ }\href {\doibase 10.1103/PhysRevD.99.074508} {\bibfield  {journal} {\bibinfo  {journal} {Phys. Rev. D}\ }\textbf {\bibinfo {volume} {99}},\ \bibinfo {pages} {074508} (\bibinfo {year} {2019})},\ \Eprint {http://arxiv.org/abs/1901.03862} {arXiv:1901.03862 [hep-lat]} \BibitemShut {NoStop}%
\bibitem [{\citenamefont {Chen}\ \emph {et~al.}(2017)\citenamefont {Chen}, \citenamefont {Ji},\ and\ \citenamefont {Zhang}}]{Chen:2016fxx}%
  \BibitemOpen
  \bibfield  {author} {\bibinfo {author} {\bibfnamefont {Jiunn-Wei}\ \bibnamefont {Chen}}, \bibinfo {author} {\bibfnamefont {Xiangdong}\ \bibnamefont {Ji}}, \ and\ \bibinfo {author} {\bibfnamefont {Jian-Hui}\ \bibnamefont {Zhang}},\ }\bibfield  {title} {\enquote {\bibinfo {title} {{Improved quasi parton distribution through Wilson line renormalization}},}\ }\href {\doibase 10.1016/j.nuclphysb.2016.12.004} {\bibfield  {journal} {\bibinfo  {journal} {Nucl. Phys. B}\ }\textbf {\bibinfo {volume} {915}},\ \bibinfo {pages} {1--9} (\bibinfo {year} {2017})},\ \Eprint {http://arxiv.org/abs/1609.08102} {arXiv:1609.08102 [hep-ph]} \BibitemShut {NoStop}%
\bibitem [{\citenamefont {Ishikawa}\ \emph {et~al.}(2016)\citenamefont {Ishikawa}, \citenamefont {Ma}, \citenamefont {Qiu},\ and\ \citenamefont {Yoshida}}]{Ishikawa:2016znu}%
  \BibitemOpen
  \bibfield  {author} {\bibinfo {author} {\bibfnamefont {Tomomi}\ \bibnamefont {Ishikawa}}, \bibinfo {author} {\bibfnamefont {Yan-Qing}\ \bibnamefont {Ma}}, \bibinfo {author} {\bibfnamefont {Jian-Wei}\ \bibnamefont {Qiu}}, \ and\ \bibinfo {author} {\bibfnamefont {Shinsuke}\ \bibnamefont {Yoshida}},\ }\bibfield  {title} {\enquote {\bibinfo {title} {{Practical quasi parton distribution functions}},}\ }\href@noop {} {\  (\bibinfo {year} {2016})},\ \Eprint {http://arxiv.org/abs/1609.02018} {arXiv:1609.02018 [hep-lat]} \BibitemShut {NoStop}%
\bibitem [{\citenamefont {Carlson}\ and\ \citenamefont {Freid}(2017)}]{Carlson:2017gpk}%
  \BibitemOpen
  \bibfield  {author} {\bibinfo {author} {\bibfnamefont {Carl~E.}\ \bibnamefont {Carlson}}\ and\ \bibinfo {author} {\bibfnamefont {Michael}\ \bibnamefont {Freid}},\ }\bibfield  {title} {\enquote {\bibinfo {title} {{Lattice corrections to the quark quasidistribution at one-loop}},}\ }\href {\doibase 10.1103/PhysRevD.95.094504} {\bibfield  {journal} {\bibinfo  {journal} {Phys. Rev. D}\ }\textbf {\bibinfo {volume} {95}},\ \bibinfo {pages} {094504} (\bibinfo {year} {2017})},\ \Eprint {http://arxiv.org/abs/1702.05775} {arXiv:1702.05775 [hep-ph]} \BibitemShut {NoStop}%
\bibitem [{\citenamefont {Xiong}\ \emph {et~al.}(2017)\citenamefont {Xiong}, \citenamefont {Luu},\ and\ \citenamefont {Mei\ss{}ner}}]{Xiong:2017jtn}%
  \BibitemOpen
  \bibfield  {author} {\bibinfo {author} {\bibfnamefont {Xiaonu}\ \bibnamefont {Xiong}}, \bibinfo {author} {\bibfnamefont {Thomas}\ \bibnamefont {Luu}}, \ and\ \bibinfo {author} {\bibfnamefont {Ulf-G.}\ \bibnamefont {Mei\ss{}ner}},\ }\bibfield  {title} {\enquote {\bibinfo {title} {{Quasi-Parton Distribution Function in Lattice Perturbation Theory}},}\ }\href@noop {} {\  (\bibinfo {year} {2017})},\ \Eprint {http://arxiv.org/abs/1705.00246} {arXiv:1705.00246 [hep-ph]} \BibitemShut {NoStop}%
\bibitem [{\citenamefont {Bomhof}\ \emph {et~al.}(2006)\citenamefont {Bomhof}, \citenamefont {Mulders},\ and\ \citenamefont {Pijlman}}]{Bomhof:2006dp}%
  \BibitemOpen
  \bibfield  {author} {\bibinfo {author} {\bibfnamefont {C.~J.}\ \bibnamefont {Bomhof}}, \bibinfo {author} {\bibfnamefont {P.~J.}\ \bibnamefont {Mulders}}, \ and\ \bibinfo {author} {\bibfnamefont {F.}~\bibnamefont {Pijlman}},\ }\bibfield  {title} {\enquote {\bibinfo {title} {{The Construction of gauge-links in arbitrary hard processes}},}\ }\href {\doibase 10.1140/epjc/s2006-02554-2} {\bibfield  {journal} {\bibinfo  {journal} {Eur. Phys. J. C}\ }\textbf {\bibinfo {volume} {47}},\ \bibinfo {pages} {147--162} (\bibinfo {year} {2006})},\ \Eprint {http://arxiv.org/abs/hep-ph/0601171} {arXiv:hep-ph/0601171} \BibitemShut {NoStop}%
\bibitem [{\citenamefont {Chen}\ \emph {et~al.}(2019)\citenamefont {Chen}, \citenamefont {Ishikawa}, \citenamefont {Jin}, \citenamefont {Lin}, \citenamefont {Zhang},\ and\ \citenamefont {Zhao}}]{Chen:2017mie}%
  \BibitemOpen
  \bibfield  {author} {\bibinfo {author} {\bibfnamefont {Jiunn-Wei}\ \bibnamefont {Chen}}, \bibinfo {author} {\bibfnamefont {Tomomi}\ \bibnamefont {Ishikawa}}, \bibinfo {author} {\bibfnamefont {Luchang}\ \bibnamefont {Jin}}, \bibinfo {author} {\bibfnamefont {Huey-Wen}\ \bibnamefont {Lin}}, \bibinfo {author} {\bibfnamefont {Jian-Hui}\ \bibnamefont {Zhang}}, \ and\ \bibinfo {author} {\bibfnamefont {Yong}\ \bibnamefont {Zhao}} (\bibinfo {collaboration} {LP3}),\ }\bibfield  {title} {\enquote {\bibinfo {title} {{Symmetry properties of nonlocal quark bilinear operators on a Lattice}},}\ }\href {\doibase 10.1088/1674-1137/43/10/103101} {\bibfield  {journal} {\bibinfo  {journal} {Chin. Phys. C}\ }\textbf {\bibinfo {volume} {43}},\ \bibinfo {pages} {103101} (\bibinfo {year} {2019})},\ \Eprint {http://arxiv.org/abs/1710.01089} {arXiv:1710.01089 [hep-lat]} \BibitemShut {NoStop}%
\bibitem [{\citenamefont {Constantinou}(Feb 1-3, 2017)}]{GHP}%
  \BibitemOpen
  \bibfield  {author} {\bibinfo {author} {\bibfnamefont {M}~\bibnamefont {Constantinou}},\ }\bibfield  {title} {\enquote {\bibinfo {title} {{Renormalization Issues on Long-Link Operators}},}\ }in\ \href@noop {} {\emph {\bibinfo {booktitle} {7th Workshop of the APS Topical Group on Hadronic Physics}}}\ (\bibinfo {year} {Feb 1-3, 2017})\BibitemShut {NoStop}%
\bibitem [{\citenamefont {Chen}\ \emph {et~al.}(2018)\citenamefont {Chen}, \citenamefont {Ishikawa}, \citenamefont {Jin}, \citenamefont {Lin}, \citenamefont {Yang}, \citenamefont {Zhang},\ and\ \citenamefont {Zhao}}]{Chen:2017mzz}%
  \BibitemOpen
  \bibfield  {author} {\bibinfo {author} {\bibfnamefont {Jiunn-Wei}\ \bibnamefont {Chen}}, \bibinfo {author} {\bibfnamefont {Tomomi}\ \bibnamefont {Ishikawa}}, \bibinfo {author} {\bibfnamefont {Luchang}\ \bibnamefont {Jin}}, \bibinfo {author} {\bibfnamefont {Huey-Wen}\ \bibnamefont {Lin}}, \bibinfo {author} {\bibfnamefont {Yi-Bo}\ \bibnamefont {Yang}}, \bibinfo {author} {\bibfnamefont {Jian-Hui}\ \bibnamefont {Zhang}}, \ and\ \bibinfo {author} {\bibfnamefont {Yong}\ \bibnamefont {Zhao}},\ }\bibfield  {title} {\enquote {\bibinfo {title} {{Parton distribution function with nonperturbative renormalization from lattice QCD}},}\ }\href {\doibase 10.1103/PhysRevD.97.014505} {\bibfield  {journal} {\bibinfo  {journal} {Phys. Rev. D}\ }\textbf {\bibinfo {volume} {97}},\ \bibinfo {pages} {014505} (\bibinfo {year} {2018})},\ \Eprint {http://arxiv.org/abs/1706.01295} {arXiv:1706.01295 [hep-lat]} \BibitemShut {NoStop}%
\bibitem [{\citenamefont {Musch}\ \emph {et~al.}(2011)\citenamefont {Musch}, \citenamefont {Hagler}, \citenamefont {Negele},\ and\ \citenamefont {Schafer}}]{Musch:2010ka}%
  \BibitemOpen
  \bibfield  {author} {\bibinfo {author} {\bibfnamefont {Bernhard~U.}\ \bibnamefont {Musch}}, \bibinfo {author} {\bibfnamefont {Philipp}\ \bibnamefont {Hagler}}, \bibinfo {author} {\bibfnamefont {John~W.}\ \bibnamefont {Negele}}, \ and\ \bibinfo {author} {\bibfnamefont {Andreas}\ \bibnamefont {Schafer}},\ }\bibfield  {title} {\enquote {\bibinfo {title} {{Exploring quark transverse momentum distributions with lattice QCD}},}\ }\href {\doibase 10.1103/PhysRevD.83.094507} {\bibfield  {journal} {\bibinfo  {journal} {Phys. Rev. D}\ }\textbf {\bibinfo {volume} {83}},\ \bibinfo {pages} {094507} (\bibinfo {year} {2011})},\ \Eprint {http://arxiv.org/abs/1011.1213} {arXiv:1011.1213 [hep-lat]} \BibitemShut {NoStop}%
\bibitem [{\citenamefont {Ji}\ \emph {et~al.}(2018)\citenamefont {Ji}, \citenamefont {Zhang},\ and\ \citenamefont {Zhao}}]{Ji:2017oey}%
  \BibitemOpen
  \bibfield  {author} {\bibinfo {author} {\bibfnamefont {Xiangdong}\ \bibnamefont {Ji}}, \bibinfo {author} {\bibfnamefont {Jian-Hui}\ \bibnamefont {Zhang}}, \ and\ \bibinfo {author} {\bibfnamefont {Yong}\ \bibnamefont {Zhao}},\ }\bibfield  {title} {\enquote {\bibinfo {title} {{Renormalization in Large Momentum Effective Theory of Parton Physics}},}\ }\href {\doibase 10.1103/PhysRevLett.120.112001} {\bibfield  {journal} {\bibinfo  {journal} {Phys. Rev. Lett.}\ }\textbf {\bibinfo {volume} {120}},\ \bibinfo {pages} {112001} (\bibinfo {year} {2018})},\ \Eprint {http://arxiv.org/abs/1706.08962} {arXiv:1706.08962 [hep-ph]} \BibitemShut {NoStop}%
\bibitem [{\citenamefont {Green}\ \emph {et~al.}(2018)\citenamefont {Green}, \citenamefont {Jansen},\ and\ \citenamefont {Steffens}}]{Green:2017xeu}%
  \BibitemOpen
  \bibfield  {author} {\bibinfo {author} {\bibfnamefont {Jeremy}\ \bibnamefont {Green}}, \bibinfo {author} {\bibfnamefont {Karl}\ \bibnamefont {Jansen}}, \ and\ \bibinfo {author} {\bibfnamefont {Fernanda}\ \bibnamefont {Steffens}},\ }\bibfield  {title} {\enquote {\bibinfo {title} {{Nonperturbative Renormalization of Nonlocal Quark Bilinears for Parton Quasidistribution Functions on the Lattice Using an Auxiliary Field}},}\ }\href {\doibase 10.1103/PhysRevLett.121.022004} {\bibfield  {journal} {\bibinfo  {journal} {Phys. Rev. Lett.}\ }\textbf {\bibinfo {volume} {121}},\ \bibinfo {pages} {022004} (\bibinfo {year} {2018})},\ \Eprint {http://arxiv.org/abs/1707.07152} {arXiv:1707.07152 [hep-lat]} \BibitemShut {NoStop}%
\bibitem [{\citenamefont {Wang}\ and\ \citenamefont {Zhao}(2018)}]{Wang:2017eel}%
  \BibitemOpen
  \bibfield  {author} {\bibinfo {author} {\bibfnamefont {Wei}\ \bibnamefont {Wang}}\ and\ \bibinfo {author} {\bibfnamefont {Shuai}\ \bibnamefont {Zhao}},\ }\bibfield  {title} {\enquote {\bibinfo {title} {{On the power divergence in quasi gluon distribution function}},}\ }\href {\doibase 10.1007/JHEP05(2018)142} {\bibfield  {journal} {\bibinfo  {journal} {JHEP}\ }\textbf {\bibinfo {volume} {05}},\ \bibinfo {pages} {142} (\bibinfo {year} {2018})},\ \Eprint {http://arxiv.org/abs/1712.09247} {arXiv:1712.09247 [hep-ph]} \BibitemShut {NoStop}%
\bibitem [{\citenamefont {Ji}\ \emph {et~al.}(2021{\natexlab{c}})\citenamefont {Ji}, \citenamefont {Liu}, \citenamefont {Sch\"afer}, \citenamefont {Wang}, \citenamefont {Yang}, \citenamefont {Zhang},\ and\ \citenamefont {Zhao}}]{Ji:2020brr}%
  \BibitemOpen
  \bibfield  {author} {\bibinfo {author} {\bibfnamefont {Xiangdong}\ \bibnamefont {Ji}}, \bibinfo {author} {\bibfnamefont {Yizhuang}\ \bibnamefont {Liu}}, \bibinfo {author} {\bibfnamefont {Andreas}\ \bibnamefont {Sch\"afer}}, \bibinfo {author} {\bibfnamefont {Wei}\ \bibnamefont {Wang}}, \bibinfo {author} {\bibfnamefont {Yi-Bo}\ \bibnamefont {Yang}}, \bibinfo {author} {\bibfnamefont {Jian-Hui}\ \bibnamefont {Zhang}}, \ and\ \bibinfo {author} {\bibfnamefont {Yong}\ \bibnamefont {Zhao}},\ }\bibfield  {title} {\enquote {\bibinfo {title} {{A Hybrid Renormalization Scheme for Quasi Light-Front Correlations in Large-Momentum Effective Theory}},}\ }\href {\doibase 10.1016/j.nuclphysb.2021.115311} {\bibfield  {journal} {\bibinfo  {journal} {Nucl. Phys. B}\ }\textbf {\bibinfo {volume} {964}},\ \bibinfo {pages} {115311} (\bibinfo {year} {2021}{\natexlab{c}})},\ \Eprint {http://arxiv.org/abs/2008.03886} {arXiv:2008.03886 [hep-ph]} \BibitemShut {NoStop}%
\bibitem [{\citenamefont {Zhang}\ \emph {et~al.}(2021)\citenamefont {Zhang}, \citenamefont {Li}, \citenamefont {Huo}, \citenamefont {Sch\"afer}, \citenamefont {Sun},\ and\ \citenamefont {Yang}}]{Zhang:2020rsx}%
  \BibitemOpen
  \bibfield  {author} {\bibinfo {author} {\bibfnamefont {Kuan}\ \bibnamefont {Zhang}}, \bibinfo {author} {\bibfnamefont {Yuan-Yuan}\ \bibnamefont {Li}}, \bibinfo {author} {\bibfnamefont {Yi-Kai}\ \bibnamefont {Huo}}, \bibinfo {author} {\bibfnamefont {Andreas}\ \bibnamefont {Sch\"afer}}, \bibinfo {author} {\bibfnamefont {Peng}\ \bibnamefont {Sun}}, \ and\ \bibinfo {author} {\bibfnamefont {Yi-Bo}\ \bibnamefont {Yang}} (\bibinfo {collaboration} {\ensuremath{\chi}QCD}),\ }\bibfield  {title} {\enquote {\bibinfo {title} {{RI/MOM renormalization of the parton quasidistribution functions in lattice regularization}},}\ }\href {\doibase 10.1103/PhysRevD.104.074501} {\bibfield  {journal} {\bibinfo  {journal} {Phys. Rev. D}\ }\textbf {\bibinfo {volume} {104}},\ \bibinfo {pages} {074501} (\bibinfo {year} {2021})},\ \Eprint {http://arxiv.org/abs/2012.05448} {arXiv:2012.05448 [hep-lat]} \BibitemShut {NoStop}%
\bibitem [{\citenamefont {Ebert}\ \emph {et~al.}(2020)\citenamefont {Ebert}, \citenamefont {Stewart},\ and\ \citenamefont {Zhao}}]{Ebert:2019tvc}%
  \BibitemOpen
  \bibfield  {author} {\bibinfo {author} {\bibfnamefont {Markus~A.}\ \bibnamefont {Ebert}}, \bibinfo {author} {\bibfnamefont {Iain~W.}\ \bibnamefont {Stewart}}, \ and\ \bibinfo {author} {\bibfnamefont {Yong}\ \bibnamefont {Zhao}},\ }\bibfield  {title} {\enquote {\bibinfo {title} {{Renormalization and Matching for the Collins-Soper Kernel from Lattice QCD}},}\ }\href {\doibase 10.1007/JHEP03(2020)099} {\bibfield  {journal} {\bibinfo  {journal} {JHEP}\ }\textbf {\bibinfo {volume} {03}},\ \bibinfo {pages} {099} (\bibinfo {year} {2020})},\ \Eprint {http://arxiv.org/abs/1910.08569} {arXiv:1910.08569 [hep-ph]} \BibitemShut {NoStop}%
\bibitem [{\citenamefont {Gattringer}\ and\ \citenamefont {Lang}(2010)}]{Gattringer:2010zz}%
  \BibitemOpen
  \bibfield  {author} {\bibinfo {author} {\bibfnamefont {Christof}\ \bibnamefont {Gattringer}}\ and\ \bibinfo {author} {\bibfnamefont {Christian~B.}\ \bibnamefont {Lang}},\ }\href {\doibase 10.1007/978-3-642-01850-3} {\emph {\bibinfo {title} {{Quantum chromodynamics on the lattice}}}},\ Vol.\ \bibinfo {volume} {788}\ (\bibinfo  {publisher} {Springer},\ \bibinfo {address} {Berlin},\ \bibinfo {year} {2010})\BibitemShut {NoStop}%
\bibitem [{\citenamefont {Constantinou}\ \emph {et~al.}(2009)\citenamefont {Constantinou}, \citenamefont {Lubicz}, \citenamefont {Panagopoulos},\ and\ \citenamefont {Stylianou}}]{Constantinou:2009tr}%
  \BibitemOpen
  \bibfield  {author} {\bibinfo {author} {\bibfnamefont {M.}~\bibnamefont {Constantinou}}, \bibinfo {author} {\bibfnamefont {V.}~\bibnamefont {Lubicz}}, \bibinfo {author} {\bibfnamefont {H.}~\bibnamefont {Panagopoulos}}, \ and\ \bibinfo {author} {\bibfnamefont {F.}~\bibnamefont {Stylianou}},\ }\bibfield  {title} {\enquote {\bibinfo {title} {{O(a**2) corrections to the one-loop propagator and bilinears of clover fermions with Symanzik improved gluons}},}\ }\href {\doibase 10.1088/1126-6708/2009/10/064} {\bibfield  {journal} {\bibinfo  {journal} {JHEP}\ }\textbf {\bibinfo {volume} {10}},\ \bibinfo {pages} {064} (\bibinfo {year} {2009})},\ \Eprint {http://arxiv.org/abs/0907.0381} {arXiv:0907.0381 [hep-lat]} \BibitemShut {NoStop}%
\bibitem [{\citenamefont {Constantinou}\ \emph {et~al.}(2013)\citenamefont {Constantinou}, \citenamefont {Costa}, \citenamefont {G\"ockeler}, \citenamefont {Horsley}, \citenamefont {Panagopoulos}, \citenamefont {Perlt}, \citenamefont {Rakow}, \citenamefont {Schierholz},\ and\ \citenamefont {Schiller}}]{Constantinou:2013ada}%
  \BibitemOpen
  \bibfield  {author} {\bibinfo {author} {\bibfnamefont {M.}~\bibnamefont {Constantinou}}, \bibinfo {author} {\bibfnamefont {M.}~\bibnamefont {Costa}}, \bibinfo {author} {\bibfnamefont {M.}~\bibnamefont {G\"ockeler}}, \bibinfo {author} {\bibfnamefont {R.}~\bibnamefont {Horsley}}, \bibinfo {author} {\bibfnamefont {H.}~\bibnamefont {Panagopoulos}}, \bibinfo {author} {\bibfnamefont {H.}~\bibnamefont {Perlt}}, \bibinfo {author} {\bibfnamefont {P.~E.~L.}\ \bibnamefont {Rakow}}, \bibinfo {author} {\bibfnamefont {G.}~\bibnamefont {Schierholz}}, \ and\ \bibinfo {author} {\bibfnamefont {A.}~\bibnamefont {Schiller}},\ }\bibfield  {title} {\enquote {\bibinfo {title} {{Perturbatively improving regularization-invariant momentum scheme renormalization constants}},}\ }\href {\doibase 10.1103/PhysRevD.87.096019} {\bibfield  {journal} {\bibinfo  {journal} {Phys. Rev. D}\ }\textbf {\bibinfo {volume} {87}},\ \bibinfo {pages} {096019} (\bibinfo {year} {2013})},\ \Eprint {http://arxiv.org/abs/1303.6776} {arXiv:1303.6776
  [hep-lat]} \BibitemShut {NoStop}%
\bibitem [{\citenamefont {Alexandrou}\ \emph {et~al.}(2017{\natexlab{c}})\citenamefont {Alexandrou}, \citenamefont {Constantinou},\ and\ \citenamefont {Panagopoulos}}]{Alexandrou:2015sea}%
  \BibitemOpen
  \bibfield  {author} {\bibinfo {author} {\bibfnamefont {Constantia}\ \bibnamefont {Alexandrou}}, \bibinfo {author} {\bibfnamefont {Martha}\ \bibnamefont {Constantinou}}, \ and\ \bibinfo {author} {\bibfnamefont {Haralambos}\ \bibnamefont {Panagopoulos}} (\bibinfo {collaboration} {ETM}),\ }\bibfield  {title} {\enquote {\bibinfo {title} {{Renormalization functions for Nf=2 and Nf=4 twisted mass fermions}},}\ }\href {\doibase 10.1103/PhysRevD.95.034505} {\bibfield  {journal} {\bibinfo  {journal} {Phys. Rev. D}\ }\textbf {\bibinfo {volume} {95}},\ \bibinfo {pages} {034505} (\bibinfo {year} {2017}{\natexlab{c}})},\ \Eprint {http://arxiv.org/abs/1509.00213} {arXiv:1509.00213 [hep-lat]} \BibitemShut {NoStop}%
\bibitem [{\citenamefont {Martinelli}\ \emph {et~al.}(1995)\citenamefont {Martinelli}, \citenamefont {Pittori}, \citenamefont {Sachrajda}, \citenamefont {Testa},\ and\ \citenamefont {Vladikas}}]{Martinelli:1994ty}%
  \BibitemOpen
  \bibfield  {author} {\bibinfo {author} {\bibfnamefont {G.}~\bibnamefont {Martinelli}}, \bibinfo {author} {\bibfnamefont {C.}~\bibnamefont {Pittori}}, \bibinfo {author} {\bibfnamefont {Christopher~T.}\ \bibnamefont {Sachrajda}}, \bibinfo {author} {\bibfnamefont {M.}~\bibnamefont {Testa}}, \ and\ \bibinfo {author} {\bibfnamefont {A.}~\bibnamefont {Vladikas}},\ }\bibfield  {title} {\enquote {\bibinfo {title} {{A General method for nonperturbative renormalization of lattice operators}},}\ }\href {\doibase 10.1016/0550-3213(95)00126-D} {\bibfield  {journal} {\bibinfo  {journal} {Nucl. Phys. B}\ }\textbf {\bibinfo {volume} {445}},\ \bibinfo {pages} {81--108} (\bibinfo {year} {1995})},\ \Eprint {http://arxiv.org/abs/hep-lat/9411010} {arXiv:hep-lat/9411010} \BibitemShut {NoStop}%
\bibitem [{\citenamefont {Martinelli}\ and\ \citenamefont {Sachrajda}(1999)}]{Martinelli:1998vt}%
  \BibitemOpen
  \bibfield  {author} {\bibinfo {author} {\bibfnamefont {G.}~\bibnamefont {Martinelli}}\ and\ \bibinfo {author} {\bibfnamefont {Christopher~T.}\ \bibnamefont {Sachrajda}},\ }\bibfield  {title} {\enquote {\bibinfo {title} {{Computation of the b quark mass with perturbative matching at the next-to-next-to-leading order}},}\ }\href {\doibase 10.1016/S0550-3213(99)00423-X} {\bibfield  {journal} {\bibinfo  {journal} {Nucl. Phys. B}\ }\textbf {\bibinfo {volume} {559}},\ \bibinfo {pages} {429--452} (\bibinfo {year} {1999})},\ \Eprint {http://arxiv.org/abs/hep-lat/9812001} {arXiv:hep-lat/9812001} \BibitemShut {NoStop}%
\bibitem [{\citenamefont {Sheikholeslami}\ and\ \citenamefont {Wohlert}(1985)}]{Sheikholeslami:1985ij}%
  \BibitemOpen
  \bibfield  {author} {\bibinfo {author} {\bibfnamefont {B.}~\bibnamefont {Sheikholeslami}}\ and\ \bibinfo {author} {\bibfnamefont {R.}~\bibnamefont {Wohlert}},\ }\bibfield  {title} {\enquote {\bibinfo {title} {{Improved Continuum Limit Lattice Action for QCD with Wilson Fermions}},}\ }\href {\doibase 10.1016/0550-3213(85)90002-1} {\bibfield  {journal} {\bibinfo  {journal} {Nucl. Phys. B}\ }\textbf {\bibinfo {volume} {259}},\ \bibinfo {pages} {572} (\bibinfo {year} {1985})}\BibitemShut {NoStop}%
\bibitem [{\citenamefont {Horsley}\ \emph {et~al.}(2004)\citenamefont {Horsley}, \citenamefont {Perlt}, \citenamefont {Rakow}, \citenamefont {Schierholz},\ and\ \citenamefont {Schiller}}]{Horsley:2004mx}%
  \BibitemOpen
  \bibfield  {author} {\bibinfo {author} {\bibfnamefont {R.}~\bibnamefont {Horsley}}, \bibinfo {author} {\bibfnamefont {H.}~\bibnamefont {Perlt}}, \bibinfo {author} {\bibfnamefont {Paul E.~L.}\ \bibnamefont {Rakow}}, \bibinfo {author} {\bibfnamefont {G.}~\bibnamefont {Schierholz}}, \ and\ \bibinfo {author} {\bibfnamefont {A.}~\bibnamefont {Schiller}} (\bibinfo {collaboration} {QCDSF}),\ }\bibfield  {title} {\enquote {\bibinfo {title} {{One-loop renormalisation of quark bilinears for overlap fermions with improved gauge actions}},}\ }\href {\doibase 10.1016/j.nuclphysb.2005.01.044} {\bibfield  {journal} {\bibinfo  {journal} {Nucl. Phys. B}\ }\textbf {\bibinfo {volume} {693}},\ \bibinfo {pages} {3--35} (\bibinfo {year} {2004})},\ \bibinfo {note} {[Erratum: Nucl.Phys.B 713, 601--606 (2005)]},\ \Eprint {http://arxiv.org/abs/hep-lat/0404007} {arXiv:hep-lat/0404007} \BibitemShut {NoStop}%
\bibitem [{\citenamefont {Shindler}(2008)}]{Shindler:2007vp}%
  \BibitemOpen
  \bibfield  {author} {\bibinfo {author} {\bibfnamefont {Andrea}\ \bibnamefont {Shindler}},\ }\bibfield  {title} {\enquote {\bibinfo {title} {{Twisted mass lattice QCD}},}\ }\href {\doibase 10.1016/j.physrep.2008.03.001} {\bibfield  {journal} {\bibinfo  {journal} {Phys. Rept.}\ }\textbf {\bibinfo {volume} {461}},\ \bibinfo {pages} {37--110} (\bibinfo {year} {2008})},\ \Eprint {http://arxiv.org/abs/0707.4093} {arXiv:0707.4093 [hep-lat]} \BibitemShut {NoStop}%
\bibitem [{\citenamefont {Luscher}\ and\ \citenamefont {Weisz}(1995)}]{Luscher:1995np}%
  \BibitemOpen
  \bibfield  {author} {\bibinfo {author} {\bibfnamefont {Martin}\ \bibnamefont {Luscher}}\ and\ \bibinfo {author} {\bibfnamefont {Peter}\ \bibnamefont {Weisz}},\ }\bibfield  {title} {\enquote {\bibinfo {title} {{Computation of the relation between the bare lattice coupling and the MS coupling in SU(N) gauge theories to two loops}},}\ }\href {\doibase 10.1016/0550-3213(95)00338-S} {\bibfield  {journal} {\bibinfo  {journal} {Nucl. Phys. B}\ }\textbf {\bibinfo {volume} {452}},\ \bibinfo {pages} {234--260} (\bibinfo {year} {1995})},\ \Eprint {http://arxiv.org/abs/hep-lat/9505011} {arXiv:hep-lat/9505011} \BibitemShut {NoStop}%
\bibitem [{\citenamefont {Boussarie}\ \emph {et~al.}(2023)\citenamefont {Boussarie} \emph {et~al.}}]{Boussarie:2023izj}%
  \BibitemOpen
  \bibfield  {author} {\bibinfo {author} {\bibfnamefont {Renaud}\ \bibnamefont {Boussarie}} \emph {et~al.},\ }\bibfield  {title} {\enquote {\bibinfo {title} {{TMD Handbook}},}\ }\href@noop {} {\  (\bibinfo {year} {2023})},\ \Eprint {http://arxiv.org/abs/2304.03302} {arXiv:2304.03302 [hep-ph]} \BibitemShut {NoStop}%
\bibitem [{\citenamefont {Chetyrkin}\ and\ \citenamefont {Tkachov}(1981)}]{Chetyrkin:1981qh}%
  \BibitemOpen
  \bibfield  {author} {\bibinfo {author} {\bibfnamefont {K.~G.}\ \bibnamefont {Chetyrkin}}\ and\ \bibinfo {author} {\bibfnamefont {F.~V.}\ \bibnamefont {Tkachov}},\ }\bibfield  {title} {\enquote {\bibinfo {title} {{Integration by parts: The algorithm to calculate $\beta$-functions in 4 loops}},}\ }\href {\doibase 10.1016/0550-3213(81)90199-1} {\bibfield  {journal} {\bibinfo  {journal} {Nucl. Phys. B}\ }\textbf {\bibinfo {volume} {192}},\ \bibinfo {pages} {159--204} (\bibinfo {year} {1981})}\BibitemShut {NoStop}%
\bibitem [{\citenamefont {Gracey}(2003)}]{Gracey:2003yr}%
  \BibitemOpen
  \bibfield  {author} {\bibinfo {author} {\bibfnamefont {J.~A.}\ \bibnamefont {Gracey}},\ }\bibfield  {title} {\enquote {\bibinfo {title} {{Three loop anomalous dimension of nonsinglet quark currents in the RI-prime scheme}},}\ }\href {\doibase 10.1016/S0550-3213(03)00335-3} {\bibfield  {journal} {\bibinfo  {journal} {Nucl. Phys. B}\ }\textbf {\bibinfo {volume} {662}},\ \bibinfo {pages} {247--278} (\bibinfo {year} {2003})},\ \Eprint {http://arxiv.org/abs/hep-ph/0304113} {arXiv:hep-ph/0304113} \BibitemShut {NoStop}%
\bibitem [{\citenamefont {Alexandrou}\ \emph {et~al.}(2012)\citenamefont {Alexandrou}, \citenamefont {Constantinou}, \citenamefont {Korzec}, \citenamefont {Panagopoulos},\ and\ \citenamefont {Stylianou}}]{Alexandrou:2012mt}%
  \BibitemOpen
  \bibfield  {author} {\bibinfo {author} {\bibfnamefont {C.}~\bibnamefont {Alexandrou}}, \bibinfo {author} {\bibfnamefont {M.}~\bibnamefont {Constantinou}}, \bibinfo {author} {\bibfnamefont {T.}~\bibnamefont {Korzec}}, \bibinfo {author} {\bibfnamefont {H.}~\bibnamefont {Panagopoulos}}, \ and\ \bibinfo {author} {\bibfnamefont {F.}~\bibnamefont {Stylianou}},\ }\bibfield  {title} {\enquote {\bibinfo {title} {{Renormalization constants of local operators for Wilson type improved fermions}},}\ }\href {\doibase 10.1103/PhysRevD.86.014505} {\bibfield  {journal} {\bibinfo  {journal} {Phys. Rev. D}\ }\textbf {\bibinfo {volume} {86}},\ \bibinfo {pages} {014505} (\bibinfo {year} {2012})},\ \Eprint {http://arxiv.org/abs/1201.5025} {arXiv:1201.5025 [hep-lat]} \BibitemShut {NoStop}%
\end{thebibliography}%

\end{document}